\documentclass[a4paper,11pt]{article}

\usepackage[utf8]{inputenc}
\usepackage[english]{babel}
\usepackage{amssymb,amsmath}
\usepackage{graphicx}
\usepackage{float}
\usepackage{jheppub}
\usepackage[font=footnotesize]{caption}
\usepackage[font=footnotesize]{subcaption}
\usepackage{dsfont} 
\usepackage{appendix}
\usepackage[table]{xcolor}

\makeatletter
\def\@fpheader{\bigskip\relax}
\makeatother

\def\be{\begin{equation}} 
\def\ee{\end{equation}} 
 
\newcommand{\dd}{\,\mathrm{d}}
\newcommand{\bx}{\mathbf{x}}
\newcommand{\sech}{\,\mathrm{sech}}

\newcommand{\acoth}{\,\mathrm{arccoth}}
\newcommand{\lp}{\left(}
\newcommand{\rp}{\right)}

\title{\boldmath Casting shadows on holographic reconstruction}
\author{Ben Freivogel, Robert A. Jefferson, Laurens Kabir, Benjamin Mosk, and I-Sheng Yang}
\affiliation{Universiteit van Amsterdam,\\Science Park 904, Amsterdam, the Netherlands}

\emailAdd{benfreivogel@gmail.com}
\emailAdd{rjefferson@uva.nl}
\emailAdd{l.r.kabir@uva.nl}
\emailAdd{b.mosk@uva.nl}
\emailAdd{i.s.yang@uva.nl}

\abstract{In the context of the AdS/CFT correspondence, we study several holographic probes that relate information about the bulk spacetime to CFT data. The best-known example is the relation between minimal surfaces in the bulk and entanglement entropy of a subregion in the CFT.  Building on earlier work, we identify  ``shadows'' in the bulk: regions that are not illuminated by any of the bulk probes we consider, in the sense that the bulk surfaces do not pass through these regions. We quantify the size of the shadow  in the near horizon region of a black hole and in the vicinity of a sufficiently dense star. The existence of shadows motivates further study of the bulk-boundary dictionary in order to identify CFT quantities that encode information about the shadow regions in the bulk. We speculate on the interpretation of our results from a dual field theory perspective.}

\begin{document}
\setcounter{tocdepth}{2}
\maketitle 
\flushbottom

\section{Introduction}
Despite many remarkable advances in our understanding of the AdS/CFT correspondence, significant obstacles remain in reconstructing local bulk physics from the CFT. These obstacles prevent us from answering elementary questions of enormous importance for our understanding of quantum gravity, such as whether an observer falling into an AdS-Schwarzschild black hole encounters a ``firewall'' \cite{AMPS, Mar-Pol, AMPSS, Sus-Mal}. A particularly important and difficult question is how to extract physics on scales short compared to the AdS radius near the black hole horizon. 

One powerful tool in reconstructing bulk physics comes from the Ryu-Takayanagi proposal \cite{RT1}. It directly links the area of minimal bulk surfaces to the entanglement entropy of spatial regions in the boundary field theory, and thereby provides a quantitative relationship between entanglement in holographic CFTs and spacetime geometry \cite{RT1,RT2,MvR_2010,Bianchi_Myers_2012,Swingle_2012,Faulkner_etal_2013,Lew-Mal}. In some cases, it has been shown that the entanglement entropy data alone is sufficient to completely determine the bulk solution \cite{Hubeny:2012wa}. This supports the ambitious claim that the spacetime is emergent and can be reconstructed from the boundary CFT \cite{Seiberg:2006wf,MvR_2010,ElShowk:2011ag,Bianchi_Myers_2012,Swingle_2012,BCCBH_2013}.

However, there is an obstacle to performing this type of reconstruction in more general geometries. In general, the bulk contains ``shadows'', or regions that are skipped over by the minimal surfaces. One reason for the existence of shadows is phase transition behavior: a given boundary region may have multiple bulk surfaces that are all local minima of the area. But the global minimum, with which the CFT quantity is associated, may switch from one branch of local minima to another, and thus the boundary dual skips over some bulk region \cite{Hartnoll:2014ppa,Czech:2014wka}. In asymptotically global AdS spacetime, it is possible that a region of the bulk is always skipped over no matter which boundary regions we choose. In \cite{BCCB_2014}, such regions into which no minimal area surface can probe were dubbed ``entanglement shadows.'' 

The most obvious way to overcome this obstacle is to find a better probe, i.e., one that reaches deeper into the bulk and penetrates the shadow. With this situation in mind, we present a generalized framework for determining the ``holographic shadows'' associated with extremal geometric objects.

Predictably, the interior of a static black hole lies within the entanglement shadow \cite{Hubeny_2012,Eng-Wal}, and is likely also part of the holographic shadow for any similar probe. Somewhat more surprising is the fact that, at least in all cases of which we are aware, holographic shadows always extend beyond the horizon. Furthermore, they are determined by the phase transition behaviour mentioned above, and are \emph{not} directly related to the presence of the black hole. Indeed, holographic probes can suffer shadows even in globally regular geometries, and we emphasize this by presenting an explicit example of an entanglement shadow in the case of a star in AdS$_{2+1}$.

In the case of singular spacetimes, the question ``how close to the black hole horizon can we probe?'' is both interesting and important. Thus, building on earlier work \cite{HMRT_2013}, we conduct a comparison of three distinct holographic probes in AdS-Schwarzschild geometries: minimal area surfaces, static Wilson loops, and causal information surfaces \cite{Hubeny:2012wa, Freivogel:2013zta, Kelly:2013aja}. Our results are summarized in table \ref{tab:results}. We find that in several cases, the causal information surfaces probe deepest into the bulk. In particular, for small black holes in higher dimensions, causal information surfaces get exponentially close to the horizon, while other probes remain of order one horizon distance away. 

\begin{table}[h!]
\centering
\begin{tabular}{|c|c|c|c|}
\hline
& Minimal Area & Wilson Loop & Causal \\ \hline
$d=2$, $r_H\ll l_{\rm AdS}$ & $\sim l_{\rm AdS}$ & $\sim l_{\rm AdS}$ & $\sim l_{\rm AdS}$ \\ \hline
$d=2$, $r_H\gg l_{\rm AdS}$ & $\sim e^{-\#r_H/l_{\rm AdS}}$ & $\sim r_H$ & $\sim e^{-\#r_H/l_{\rm AdS}}$ \\ \hline
$d>2$, $r_H\ll l_{\rm AdS}$ & $\sim r_H$ & $\sim (r_Hl_{\rm AdS})^{1/2}$ & $\sim e^{-\#l_{\rm AdS}/r_H}$ \\ \hline
$d>2$, $r_H\gg l_{\rm AdS}$ & $\sim e^{-\#r_H/l_{\rm AdS}}$ & $\sim r_H$ & $\sim e^{-\#r_H/l_{\rm AdS}}$ \\ \hline
\end{tabular}
\caption{Shadow summary for various probes of AdS-Schwarzschild; $d$ is the spatial dimension. The value listed is the distance from the black hole horizon $r_H$. The $\#$ symbol denotes an order one constant, which may depend on the spatial dimension; both this and the overall proportionality are determined explicitly in the main text.}\label{tab:results}
\end{table}

Our finding would seem to conflict the common impression that minimal area surfaces reach deeper than causal surfaces. To resolve this apparent conflict, it is important to distinguish local vs. global comparisons. The minimal area surface associated with a fixed boundary region does indeed reach deeper into the bulk than the corresponding causal surface \cite{Hubeny:2012wa}. However, the shadow is defined by the smallest radius accessible by \emph{any} bulk probe, i.e., the maximum depth among all possible boundary regions. In particular, the causal surfaces are not subject to the aforementioned switchover effect, which allows them to gain the advantage over minimal area surfaces despite being locally worse. It is in this second, global sense that we mean a given surface is ``better'' or reaches deeper, since having a smaller shadow is the more relevant standard for the purpose of holographic reconstruction. 

There is an additional, slightly more subtle consideration that may be important for bulk reconstruction. To retrieve complete information about a given bulk region, we might require a probe to not only reach every bulk point, but to do so with every possible orientation. Indeed, this is precisely the requirement of the hole-ographic construction of \cite{BCCBH_2013}. Thus, we also identify ``partial shadows''---regions of the bulk which are accessible by a given probe, but with only partial coverage of the tangent space. In this paper we present only some preliminary results regarding partial shadows, but we regard them as a potentially interesting aspect for future work.

Finally, we should emphasize that in higher than two spatial dimensions, our results strictly speaking do {\it not} prove the existence of holographic shadows. We have studied only boundary disks, rather than fully arbitrary boundary regions. Although it is natural to expect that more complicated boundary shapes cannot reduce the shadow size (since these tend to suffer from additional phase transition limitations), we have not succeeded in finding a general proof. We hope to return to this issue in the future.

The organization of this paper is as follows: In sec. \ref{sec:properties} we present the general framework for using extremal bulk surfaces as probes. We introduce and prove two ``coverage theorems'' in the interest of formalizing the conditions under which a spacetime exhibits holographic shadows. Then, in sec. \ref{sec:stellar}, we use these theorems to demonstrate the existence of entanglement shadows for globally well-defined geometries. In sections \ref{sub:BTZEE}, \ref{sub:BTZWILSON} and \ref{sub:BTZCAUSAL}, we extend our analysis to AdS-Schwarzschild geometries with three different probes: minimal area surfaces, static Wilson loops, and causal information surfaces. We present a comparison of these probes in the discussion, sec. \ref{sec:discussionMain}, and close with a summary and some comments on future directions. Appendix \ref{sec-lemma} contains proofs of some general properties of extremal surfaces. In appendix \ref{nonconstt}, we justify why we only consider static minimal area surfaces in $d=2$. Lastly, we include some detailed calculations of Wilson loops, in appendix \ref{wilsonhigherd}, including a proof that for $d\geq4$, shadows are due exclusively to the phase transition between different minima.

\section{Properties of minimal surfaces}\label{sec:properties}
In this section, we present some general properties, terminology, and theorems that will prove useful in the analysis of holographic shadows that follows. 

\subsection{Minimal area surfaces}
Let us first review the Ryu-Takayanagi proposal that relates bulk minimal surfaces to entanglement entropy on the boundary CFT \cite{RT1, RT2}. Consider a constant time slice in static, asymptotically AdS$_{d+1}$ spacetime. Let the set of all bulk points be $\mathcal{B}$, and let $\mathcal{A}$ be all points on the asymptotic boundary $S_{d-1}$. The proposal relates the entanglement entropy for a boundary region $a\subseteq\mathcal{A}$ to the area of a dual bulk surface $b\subset\mathcal{B}$ if (1) $b$ has the smallest area among all surfaces with $\partial b=\partial a$, and (2) $b$ can be continuously deformed to $a$. This proposal has many interesting aspects, but in this paper we will focus on one property with particular relevance for holographic reconstruction:

\bigskip

\noindent {\bf The Strong Coverage Property} (SCP): \\
$\forall x\in\mathcal{B},~\forall v\in T_x\mathcal{B}$, $\exists a\subset\mathcal{A}$ whose dual minimal surface $b$ intersects $x$ with tangent vector along $v$.

\bigskip

Intuitively, this says that the entire bulk and its tangent bundle are ``scanned over'' by the minimal surfaces $b$ of all possible boundary regions $a$. This is satisfied by empty AdS, and also holds up to small perturbations thereof. In $(2+1)$ dimensions, SCP is equivalent to the condition for boundary rigidity \cite{Pes-Uhl}, which means that knowing the entanglement entropy for every boundary region $a$ uniquely determines the bulk geometry. SCP is also a necessary condition for the ``hole-ographic'' reconstruction of \cite{BCCBH_2013} (see also \cite{BCCB_2014}). However, the requirement that one covers the entire tangent bundle is quite strong, and is not \emph{a priori} obviously necessary for a successful reconstruction scheme. We will therefore also consider a weaker property:

\bigskip

\noindent {\bf The Weak Coverage Property} (WCP): \\
$\forall x\in\mathcal{B}$, $\exists a\subset\mathcal{A}$ whose dual minimal surface $b$ intersects $x$.

\bigskip

This simply means that every bulk point is covered by the minimal surface $b$ of some boundary region $a$, but not necessarily scanning over all orientations in its tangent space. Note that this is not sufficient for boundary rigidity in 2 dimensions, nor for the aforementioned ``hole-ographic'' reconstruction. Nevertheless, this should be a minimal requirement for any attempt to reconstruct the bulk using this particular geometric dual.

\begin{figure}[h!]
\centering
\includegraphics[width=0.49\textwidth]{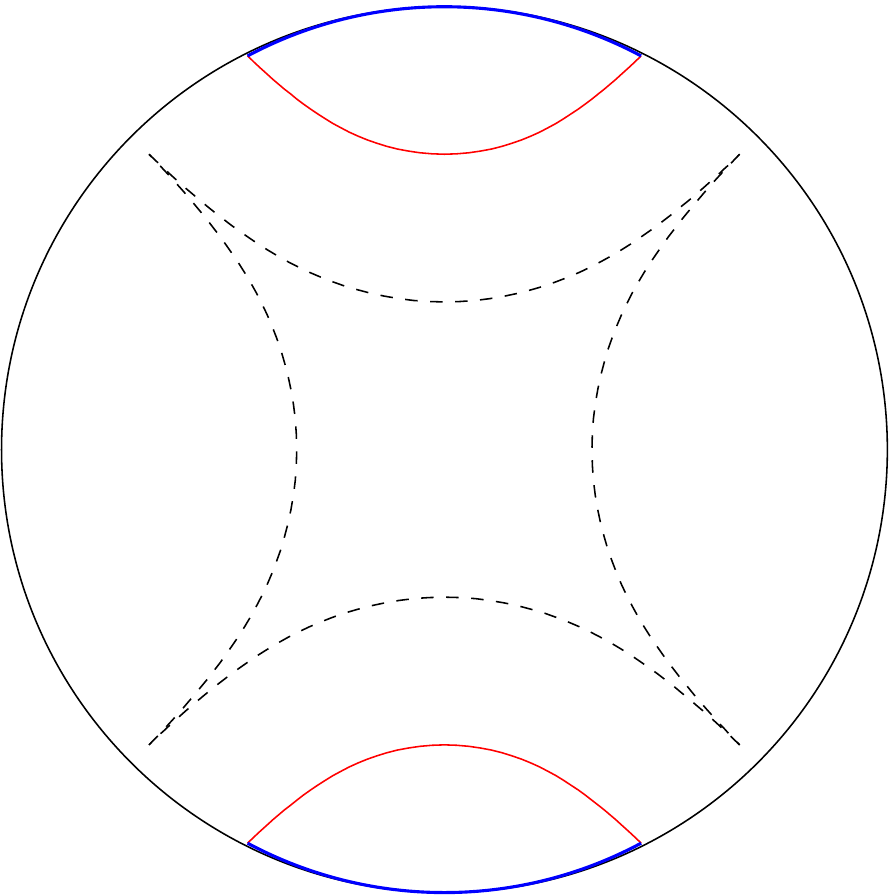}
\includegraphics[width=0.49\textwidth]{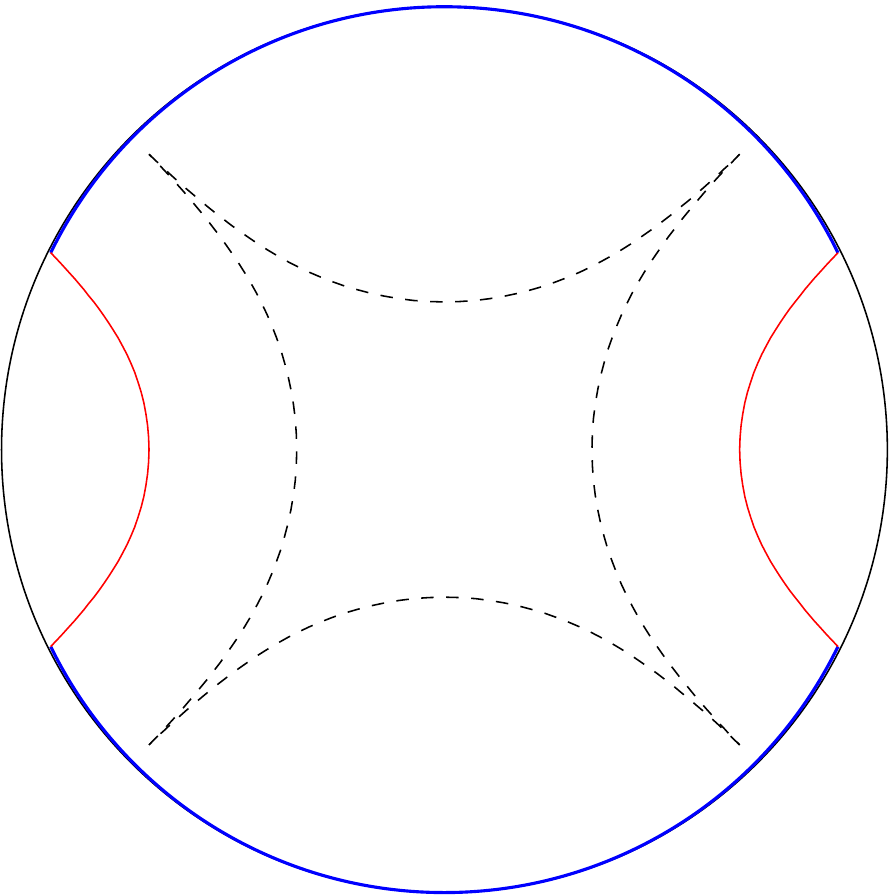}
\caption{The left figure shows a disconnected boundary region $a=\bigsqcup_i a_i$ (blue) and the corresponding disjoint minimal surface $b=\bigsqcup_jb_j$ in the bulk (red). As the boundary region is continuously increased, the bulk surfaces $b_j$ are pushed towards the dashed curve, at which point $b$ discontinuously switches to the new global minimum $b=\bigsqcup_jb_j'$ shown in the right figure. The region inside the dashed curves cannot be probed with this particular choice of bulk dual.\label{fig-seg2}}
\end{figure}

It is worth pointing out that in the case of a disjoint boundary region $a=\bigcup_ia_i$ with dual minimal surface $b=\bigcup_jb_j$, there need not be a direct correspondence between $a_i$ and $b_j$. This is illustrated in the case of two disconnected boundary subregions in figure \ref{fig-seg2}. There are two ways for the two bulk curves to end on the four boundary points that specify $\partial a$ without crossing, so there are (at least) two different local minima of their total area. Since the Ryu-Takayanagi proposal specifies $b$ as possessing the smallest area of all bulk surfaces with $\partial b=\partial a$, the choice of which of these two bulk possibilities to employ is determined by comparing their respective areas. 

As illustrated in figure \ref{fig-seg2}, as the boundary subregions $a_i$ are continuously increased, the bulk dual surfaces $b_j$ are pushed inwards until, at some critical point, $b$ switches over to the other possible combination of $b_j$, which are then pushed outwards towards the boundary as the $a_i$ continue to grow. This provides a simple example of a key concept underlying holographic shadows: rather than mirror the continuous deformation of the boundary, the bulk dual surface may undergo a discontinuous switchover in order to ascribe to the global minimum. This is a phase transition from the boundary point of view \cite{Hartnoll:2014ppa}, but here we will focus on the bulk implication. This switchover leaves out the middle region, and thereby limits the region of the bulk that can be probed.

Even without disconnected boundary regions, such switchovers can still occur. It has been examined in detail in the work of Hubeny in the context of AdS black holes \cite{HMRT_2013}, and also in geometries with a conical defect \cite{BCCB_2014}. In all of the above examples, one is tempted to ascribe this behaviour to nontrivial topology: either the boundary region is not simply connected, or the bulk has a horizon or a singularity. In fact, topology is not the real problem. Given globally well-defined manifolds and simply connected boundary regions, the weak coverage property can still be violated\footnote{Some have tried to establish that a globally regular,  WCP-violating geometry is unstable and should collapse into a black hole \cite{Nogueira_2013}. However, in this paper we will show that in $(3+1)$ dimensions, a star of radius 5 times its mass in Planck units -- e.g., neutron stars -- can already violate WCP. General stability issues are only a serious concern when the radius is near $2M$ \cite{Bondi_1963, Kleban_etal_2013}, which is the Schwarzschild radius. Hence we find no reason to doubt that stable, regular geometries can indeed violate WCP.}. Thus we will begin by studying the general behavior of SCP/WCP violation in spacetimes without horizons or singularities, and then proceed to analyze singular geometries. 

\subsection{Generalized minimal surfaces}
\label{sec-general}

Before proceeding, we shall first introduce a more general formulation of minimal bulk surfaces. In particular, one can formally take the Ryu-Takayanagi proposal as a special case of the following general prescription:
\begin{itemize}
\item Let $b\subset\mathcal{B}$ be an $n<d$ dimensional surface in the bulk, and define the geometric quantity
\begin{equation}
L(b) = \int_b \left|\dd^n\vec{B}\right|~F(g_{\mu\nu})~.\label{eq-gp}
\end{equation}
Over this surface, we integrate the area element and the function $F$ which only depends on the local geometry. This is then a very intuitive probe of the bulk geometry, as it does not care about the shape of $b$, but rather only about where $b$ reaches.
\item For an $n$ dimensional boundary region $a$ (or its boundary $\partial a$), one finds an observable $Q$ associated with the minimal value of the above geometric quantity:
\begin{equation}
Q(a) ={\rm Min}[L(b)]\bigg|_{\partial b=\partial a}.
\end{equation}
\end{itemize}
When $n=(d-1)$ and $F=1$, this reduces to the Ryu-Takayanagi proposal with $L=$ area and $Q=$ entanglement entropy. In addition, when $n=1$ and $F=g_{tt}$, this reduces to the action of certain Wilson loops. According to the form of eqn. \eqref{eq-gp}, one should always be interested in a minimum. A maximum is ill-defined as one can always arbitrarily deform the surface along the null directions. In this paper, we will also limit ourselves to quantities with $F>0$ and 
\begin{equation}
\lim_{b\rightarrow a}L(b) = \infty~.
\end{equation}
In other words, $L(b)$ is a positive definite quantity which diverges as one deforms $b$ toward the boundary. It is therefore very natural to expect the minimal surface to reach into the bulk. This is related to boundary observables which have UV divergences and needs to be regulated. 

We can now study the failure of the coverage properties above, and the consequent ``holographic shadows,'' in a more general manner not limited to minimal area surfaces vis-\`{a}-vis Ryu-Takayanagi. Other holographic duals can suffer from exactly the same obstacle, namely that the bulk probes fail to cover the entire manifold, thus placing a geometric limit on such reconstruction efforts. Our generalization makes it easier to compare different holographic probes and see which one is better, in the sense of which probe casts the smallest shadow.

\subsection{Seeking shadows}

In this paper, we will limit ourselves to $O(d)$ symmetric bulk geometries and $O(n)$ symmetric, simply connected boundary regions (disks). In such cases we can specify a bulk point $p$ by its radial distance to the origin, $r_{*}$. This point will be the $O(n)$ fixed point of a unique, $O(n)$ symmetric $n$ dimensional surface $b(r_{*})$ (modulo the remaining $SO(d-n)$ rotation) such that the first order variation of eqn.~(\ref{eq-gp}) is zero.\footnote{One might intuitively treat $r_{*}$ as the minimal radius reached by this critical surface, but there is no \emph{a priori} reason for this identification to hold for an arbitrary positive function $F$ in eqn. \eqref{eq-gp}. We will be very careful not to assume this identification in the proofs that follow.}

Proceeding from $r_{*}$, we follow the surface $b(r_{*})$ to the boundary at $r=\infty$ to find the $(n-1)$-dimensional boundary sphere $a$ on which it ends, $\partial a=\partial b$. We define the interior of $a$ to be the side closer to the initial bulk point $p$. In other words, one can deform from $b$ to $a$ without going through $r=0$. Denote the radius of this boundary ball $a$ as $\theta_{\infty}(r_{*})$.\footnote{There might be cases where some critical surfaces $b(r_{*})$ do not reach the boundary, so $\theta_{\infty}$ is not well-defined. This is exactly what happens when there is a horizon, but such cases may be more general.} We know two special values of this function: $\theta_\infty(\infty)=0$ and $\theta_\infty(0)=\pi/2$. The first is due to a surface $b(\infty)$ that effectively never leaves the boundary, while the second comes from symmetry: it is basically the surface that cuts the bulk into two halves.

This function is straightforward to compute (at least numerically), and possesses a number of useful properties. First of all, there is a condition which guarantees that a holographic reconstruction scheme will work:

\bigskip

\noindent {\bf Theorem 1:} 
{\it The set of all simply-connected, $O(n)$ symmetric boundary regions (balls) satisfies the Strong Coverage Property if $\theta_\infty\lp r_{*}\rp\in\lp0,\pi/2\rp$ is monotonic as $r_{*}$ goes from $0$ to $\infty$.}

\bigskip

Conversely, there is also a condition which guarantees that holographic reconstruction will fail: 

\bigskip

\noindent {\bf Theorem 2:}
{\it If $\dd\theta_\infty/\dd r_{*}>0$ as $r_{*}\rightarrow0$, then the weak coverage property fails for the set of all simply-connected, $O(n)$ symmetric boundary regions (balls).}

\bigskip

In this section, we will prove these two theorems using the following lemmas:

\bigskip

\noindent {\bf Lemma 1:} 
{\it For a boundary sphere $\partial a$, the bulk surface $b$ that minimizes $L$ in eqn. \eqref{eq-gp} with $\partial b=\partial a$ must be spherically symmetric.}\\
\\
{\bf Lemma 2:} 
{\it If the boundary anchors $\partial b$ and $\partial b'$ do not cross each other, but the corresponding bulk surfaces $b$ and $b'$ do, then $b$ and $b'$ cannot both be minimal surfaces.}

\bigskip

Proofs of these Lemmas will be given in appendix \ref{sec-lemma}.

\subsection*{Proof of Theorem 1}
\begin{figure}[h!]
\centering
\includegraphics[width=0.49\textwidth]{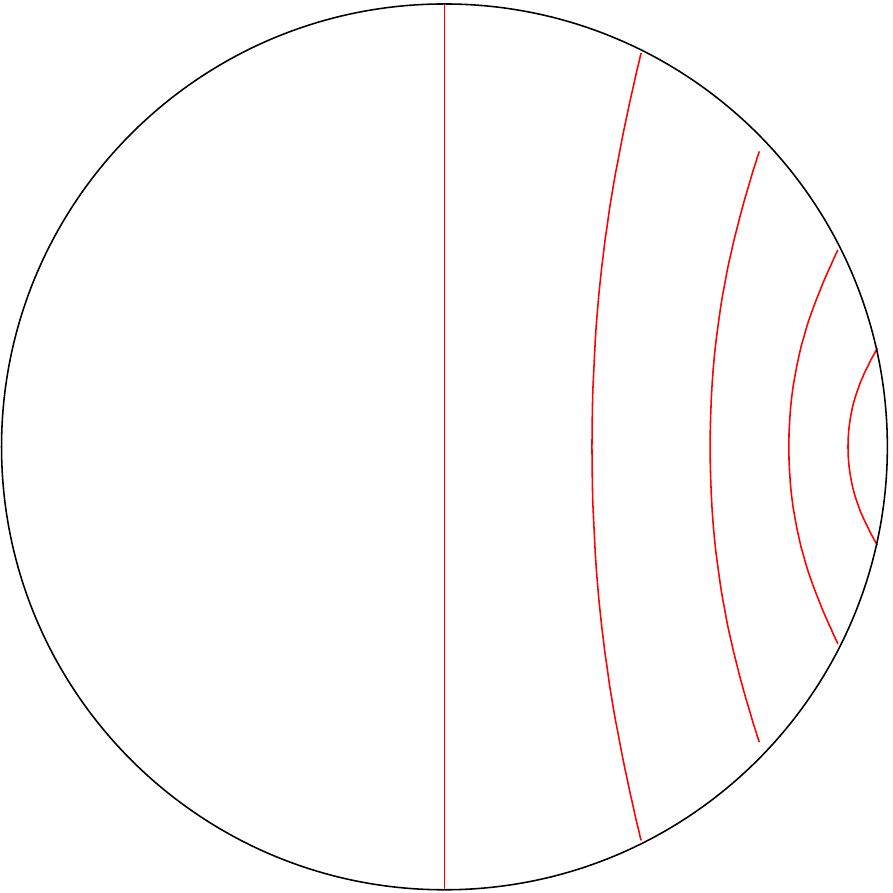}
\includegraphics[width=0.49\textwidth]{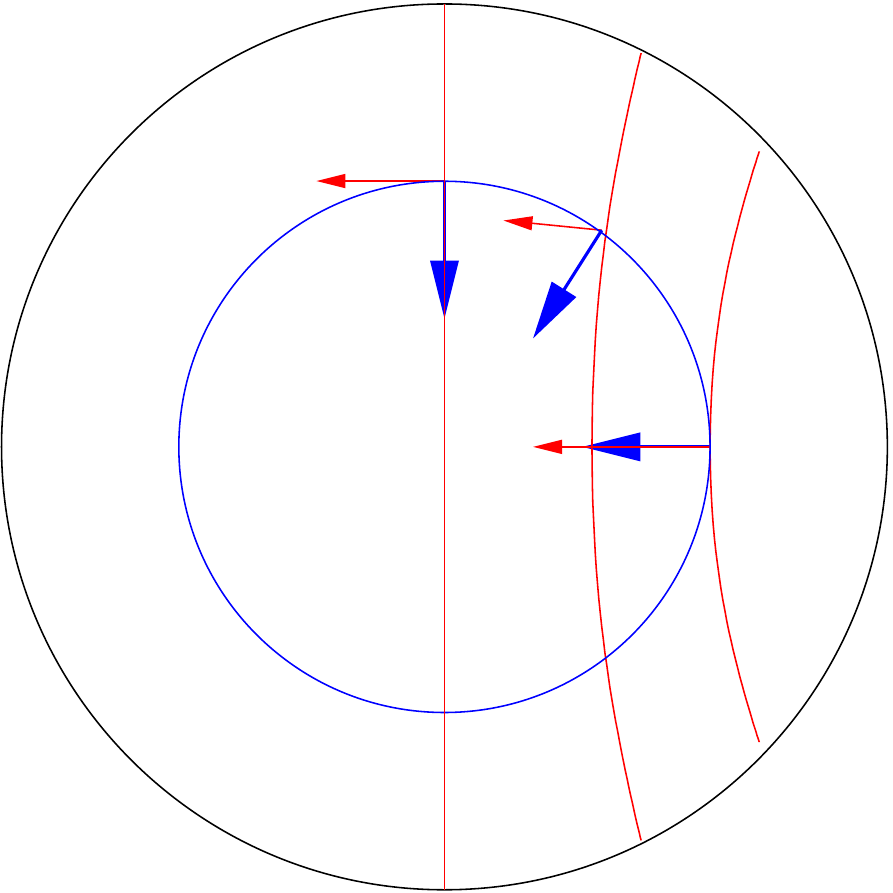}
\caption{The left figure shows a continuous foliation of minimal $n$-dimensional surfaces (red) on an $(n+1)$-dimensional equatorial slice of the bulk. The right figure shows how the angle between an $n$-sphere (blue circle) in the bulk and the foliation surfaces changes continuously from $0$ to $\pi/2$. Note that although the rightmost red surface is tangent to the blue circle at precisely $r_{*}$ in this plot, the proof does not rely on this.\label{fig-foliate}}
\end{figure}

Monotonicity of the boundary angle implies that every $b(r_{*})$ is the unique global minimum for the boundary ball $a$ of radius $\theta_\infty(r_{*})$. Lemma 1 then implies that the bulk can be foliated by a family of nonintersecting minimal surfaces anchored on the corresponding family of concentric boundary spheres, as illustrated in fig. \ref{fig-foliate}. Note that this is sufficient to satisfy WCP; for the strong coverage property, we need also demonstrate coverage of the bulk tangent bundle.

Consider a sphere with finite radius $R$ in the bulk. As shown in fig. \ref{fig-foliate}, it intersects $b(0)$ at an angle of $\pi/2$ between their normal vectors. As $r_{*}$ increases, $b(r_{*})$ will eventually stop intersecting this sphere. If we follow the intersection point during this process, the angle between the two normal vectors must continuously drop to $0$. Thus $b(r_{*})$ can cover the full tangent space of a point at radius $R$. Since $R$ is arbitrary, we have covered the full tangent bundle. QED

Note that the inverse of Theorem 1 is not generally true. That is, a non-monotonic $\theta_\infty(r_{*})$ does not guarantee the violation of SCP.\footnote{The inverse of Theorem 1 can be proven if we use the additional assumption that $r_{*}$ is the minimal radius reached by the surface $b(r_{*})$, which happens to be true in many examples.} But this is not so concerning. We have stipulated SCP as a \emph{sufficient} condition for a successful holographic reconstruction scheme; violating SCP does not necessarily imply that all schemes will fail. Thus, the more physically meaningful ``inverse'' statement is rather our Theorem 2, about the violation of WCP. Insofar as WCP is a \emph{necessary} condition, this indeed rules out holographic reconstruction (using the set of all boundary disks). Also note that Theorem 2 provides a sufficient condition to violate WCP. While WCP might be violated by other conditions, the condition Theorem 2 provides seems to be the most natural.

\subsection*{Proof of Theorem 2}
If $\dd\theta_\infty/\dd r_{*}>0$ when $r_{*}\rightarrow0$, then since $\theta_\infty(0)=\pi/2$ we can find some $r'>0$ such that $\theta_\infty(r_{*})\geq\pi/2$ for all $0\leq r_{*}\leq r'$. According to Lemma 2, none of the critical surfaces $b(r_{*})$ in this range can be the global minimum of the corresponding boundary sphere $\partial b$, because they always intersect their own mirror image.

\begin{figure}[h!]
\centering
\includegraphics[width=0.49\textwidth]{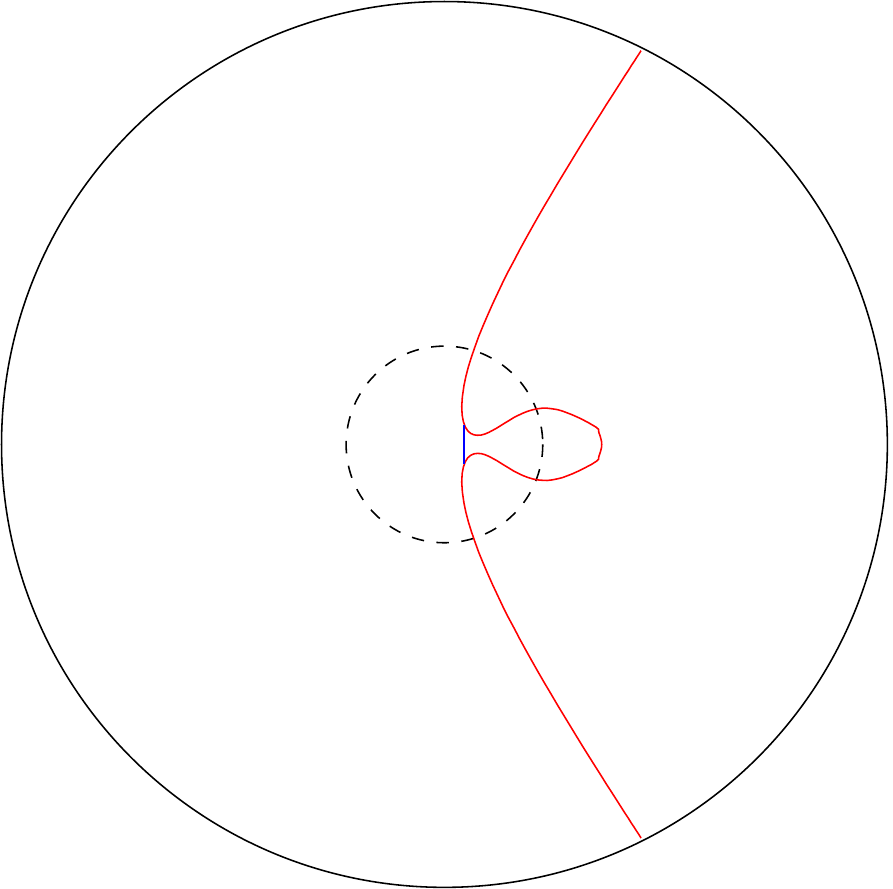}
\caption{A minimal surface (red) with its symmetric point sitting at a finite radius $r_{*}$ cannot have other points approach arbitrarily close to $r=0$. Otherwise, a pinched-off version (blue) will have even smaller area. \label{fig-rmin}}
\end{figure}

If for all minimal surfaces $b(r_{*})$, $r_{*}$ is the minimal radius reached, then no minimal surfaces can probe the region $r<r'$. On the other hand, if a point $p\in b(r_{*})$ with radius $r_p<r_{*}$ is allowed, one still cannot allow $r_p\rightarrow0$. As shown in fig. \ref{fig-rmin}, such a surface can be pinched-off to one with smaller $L$, which contradicts the assumption that the original surface is a global minimum. Thus in this case there must be a lower bound $r''$ with $0<r''<r'$ beyond which these minimal surfaces cannot probe. QED

In this paper, we will explore the simplest examples where $\dd\theta_\infty/\dd r_{*}>0$ for $r_{*}<r'$ and $\dd\theta_\infty/\dd r_{*}<0$ for $r_{*}>r'$. Additionally, in all the examples we study, we find that $r_{*}$ is the minimal radius reached by the surface $b(r_{*})$. Hence, in the rest of this paper we will adhere to the notation that $r_*$ refers to the minimal radius reached for a fixed boundary region, while $r_{\rm min}$ denotes the minimum $r_*$ among all possible boundary regions, i.e., the global minimum. Thus, $r_{\rm min}$ is also the size of the holographic shadow.

\section{Stellar shadows}\label{sec:stellar}
We begin our analysis by applying the above framework to identify shadows in globally regular geometries---namely, stars in AdS spacetimes. For our purposes, it is not necessary to specify the matter distribution; we assume only constant density. We first present analytical results for stars in AdS$_{2+1}$, and then numerically extend our analysis to AdS$_{3+1}$ as evidence for stability. 

\subsection{Analytical results in AdS$_{2+1}$}
In this section, we demonstrate an explicit example of a non-singular bulk geometry that nonetheless exhibits an entanglement shadow. The case we consider is that of an ideal (constant density) star of radius $R$ embedded in AdS$_{2+1}$, for which a physically reasonable metric is:
\begin{align}\label{eq:metricStar}
\dd s^2 = g_{tt}(r)\dd t^2+\frac{\dd r^2}{f(r)}+r^2 \dd\theta^2,\;\;\;\;
f(r)=
\begin{cases}
r^2+1-GM,&r>R\\
r^2+1-GM\frac{r^2}{R^2},&r\leq R 
\end{cases}
\end{align}
where the AdS radius $l_{\rm AdS}$ is set to 1, and $g_{tt}$ depends on the particular matter distribution. Since the metric admits the Killing vector $\partial_t$, we can analyze extremal surfaces associated to entanglement entropy on constant-time slices. We thus limit our example to entanglement surfaces, since an analysis of both Wilson loops and causal information surfaces would require explicit knowledge of the $g_{tt}$ component. Though a direct comparison of probes in this geometry would be interesting, the result for entanglement surfaces alone suffices to make our point: holographic shadows are general phenomena not limited to singular or topologically nontrivial geometries.

In what follows, we take $GM>1$, and try to solve for $\theta_\infty$ as a function of $r_*$\footnote{The $GM<1$ case corresponds to the conical defect geometry, for which the analysis proceeds along precisely similar lines.}. In the exterior region ($r > R$) the spatial part of the metric is identical to that of the BTZ metric (cf. \eqref{eq:BTZmetric}) with the identification $r_H^2\equiv GM-1$. Thus for $r_*>R$, $\theta_\infty(r_*)$ is identical to the BTZ solution as we will demonstrate later in eqn. \eqref{eq:BTZrmin}. For $r_*\leq R$, $\theta_\infty(r_*)$ is obtained by smoothly matching the $r<R$ segment and the $r>R$ segment. 

The length of a spacelike geodesic may be written:
\begin{equation}
 \begin{aligned}
L=\int\dd r\sqrt{\frac{1}{f(r)}+r^2{\theta'}^2}
\end{aligned}
\end{equation}
where the prime denotes differentiation with respect to $r$. Extremizing via Euler-Lagrange, we have
\begin{align*}
\frac{r^2\theta'}{\sqrt{f(r)^{-1}+r^2{\theta'}^2}}=\frac{\delta\mathcal{L}}{\delta\theta'}\equiv r_*
\end{align*}
where the minimum radius $r_*$ for this geodesic is, in our units, equivalent to the associated conserved angular momentum. Solving this expression for $\theta'$, we obtain
\begin{align}
\frac{\dd\theta}{\dd r}=\frac{r_*}{r\sqrt{f(r)(r^2-r_*^2)}}~.\label{eq:starIntegrand}
\end{align}
We may then perform an indefinite integral in the exterior ($r>R$), with $f(r)=r^2+1-GM$, to find
\begin{align}
\theta_E(r) =\frac{1}{2\sqrt{GM-1}}\cosh^{-1}\left(\frac{-2r_*^2(GM-1)+r^2(r_*^2+GM-1)}{r^2(r_*^2-GM+1)}\right)+g(r_*)\label{eq:starExt}
\end{align}
with constant of integration $g(r_*)$, and in the interior ($r<R$), with $f(r)=r^2+1-GMr^2/R^2$, to find
\begin{align}
\theta_I(r)= \frac{1}{2}\cos^{-1}\left(\frac{2r_*^2+r^2\lp-1+r_*^2\lp 1-\frac{GM}{R^2}\rp\rp}{r^2\lp 1+r_*^2\left(1-\frac{GM}{R^2}\right)\rp}\right)
\end{align}
where the subscripts $E$ and $I$ distinguish these functions as valid in the exterior and interior, respectively. For $\theta_I$, the constant of integration has been fixed to 0 by the symmetry assumption that demands that the minimum $r_*$ occurs at $\theta=0$, i.e. $\theta_I(r_*)=0$. To fix the constant of integration $g(r_*)$ in $\theta_E$, we demand continuity in both the function and its first derivative at the stellar boundary $r=R$. The latter condition is satisfied automatically by the conserved angular momentum $r_*$, thus we simply solve $\theta_I(R)=\theta_E(R)$ for $g(r_*)$:
\begin{equation}
\begin{aligned}
g(r_*)=&\frac{1}{2}\cos^{-1}\left(\frac{2r_*^2+R^2\lp-1+r_*^2\left(1-\frac{GM}{R^2}\right)\rp}{R^2+R^2r_*^2\left(1-\frac{GM}{R^2}\right)}\right)\\
&-\frac{1}{2\sqrt{GM-1}}\cosh^{-1}\left( \frac{-2r_*^2(GM-1)+R^2\lp r_*^2+GM-1\rp}{R^2\lp r_*^2-GM+1\rp}\right)
\end{aligned}
\end{equation}
which we may substitute into \eqref{eq:starExt}. The function $\theta_\infty(r_*)$ is then obtained by taking the $r\rightarrow\infty$ limit of the result. Dropping the subscript $E$, we at last obtain
\begin{eqnarray}
\theta_\infty(r_*)&=&\frac{1}{2\sqrt{GM-1}}\cosh^{-1}\left(\frac{r_*^2+GM-1}{r_*^2-GM+1}\right)
+\frac{1}{2}\cos^{-1}\left(\frac{2r_*^2+R^2\lp-1+r_*^2\left(1-\frac{GM}{R^2}\right)\rp}{R^2+R^2r_*^2\left(1-\frac{GM}{R^2}\right)}\right) \nonumber \\
&-&\frac{1}{2\sqrt{GM-1}}\cosh^{-1}\left( \frac{-2r_*^2(GM-1)+R^2\lp r_*^2+GM-1\rp}{R^2\lp r_*^2-GM+1\rp}\right)\label{solutionstar}
\end{eqnarray}
for the minimal geodesics extending from $r_*\leq R$ to the boundary at infinity. 

\begin{figure}[H]
\centering
\includegraphics[width=0.5\textwidth]{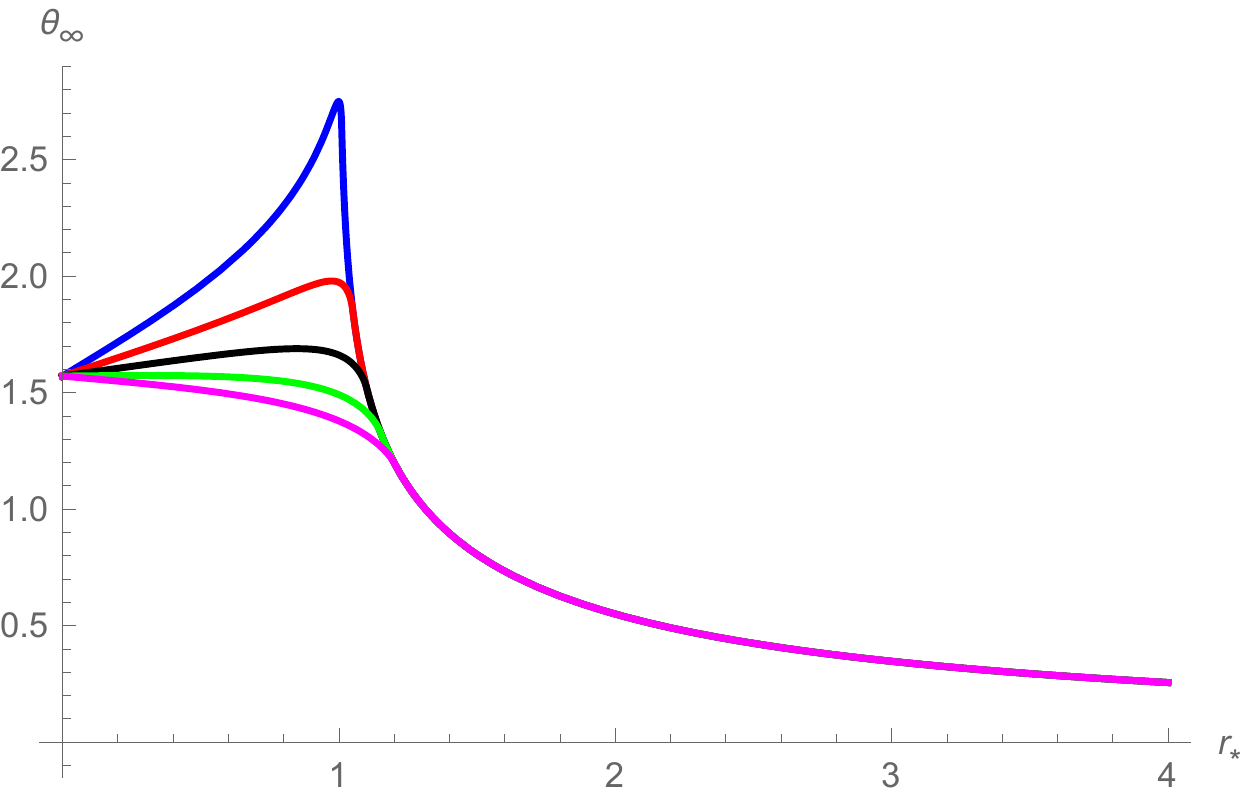}
\caption{$\theta_{\infty}(r_*)$ for $GM = 2$ and stellar radii $R = 1.01l_{\rm AdS}$ (blue), $1.05l_{\rm AdS}$ (red), $1.1l_{\rm AdS}$ (black), $1.15l_{\rm AdS}$ (green), and $1.2l_{\rm AdS}$ (magenta). The case $R=1.2l_{\rm AdS}$ is insufficiently dense, and hence exhibits a monotonic function with no shadows. But the other cases, with $R < \sqrt{4/3}~l_{\rm AdS}$ (cf. \eqref{range}), have a single maximum at finite radius $r_{\rm min}$, within which an entanglement shadow exists.}\label{StarINOUT}
\end{figure}

\begin{figure}[h!]
\centering
\begin{tabular}{cc}
\begin{subfigure}{0.49\textwidth}
	\includegraphics[width=\textwidth]{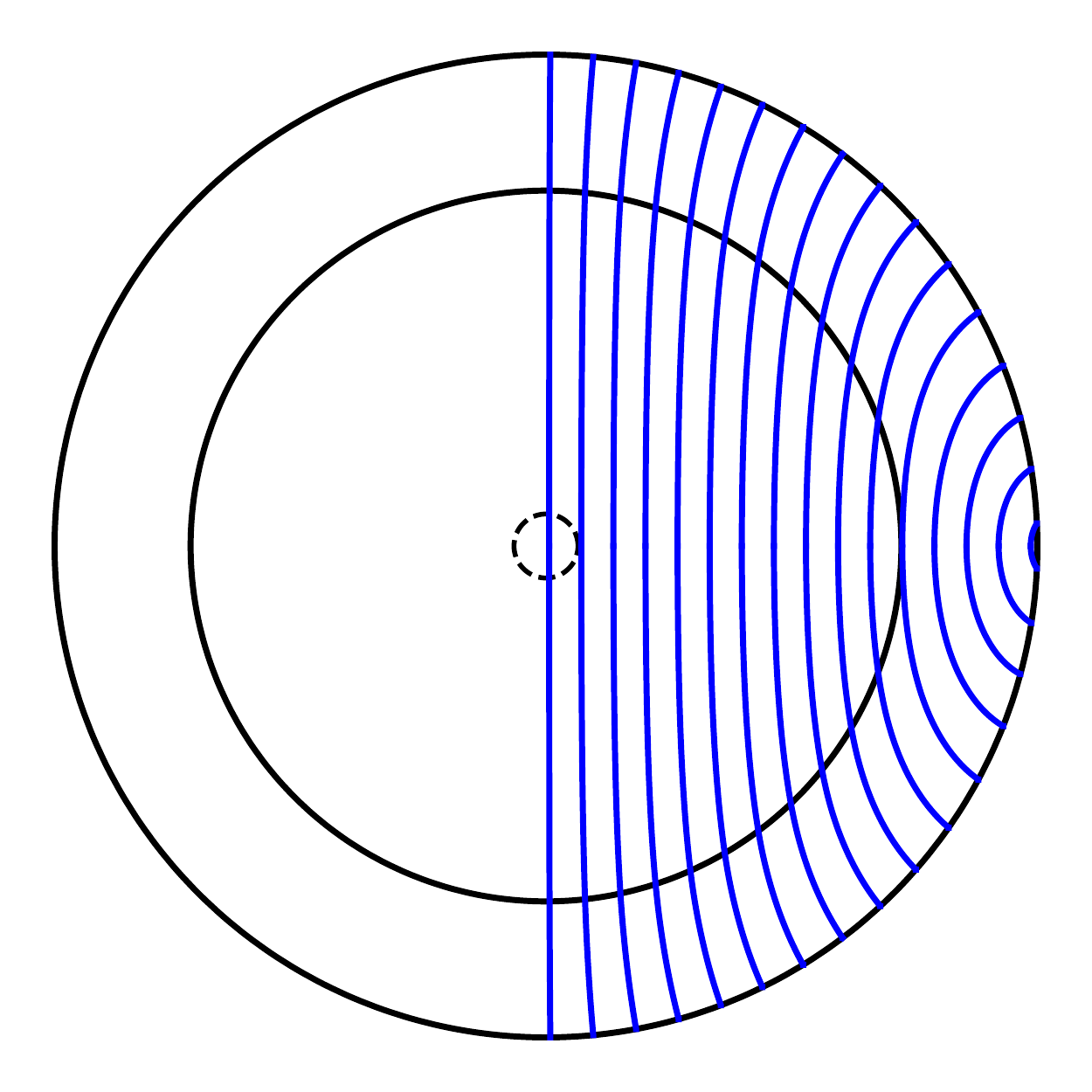}
	\caption{$GM = 1.01$, $R = 2 l_{\rm AdS}$: no shadow}
\end{subfigure} &
\begin{subfigure}{0.49\textwidth}
	\includegraphics[width=\textwidth]{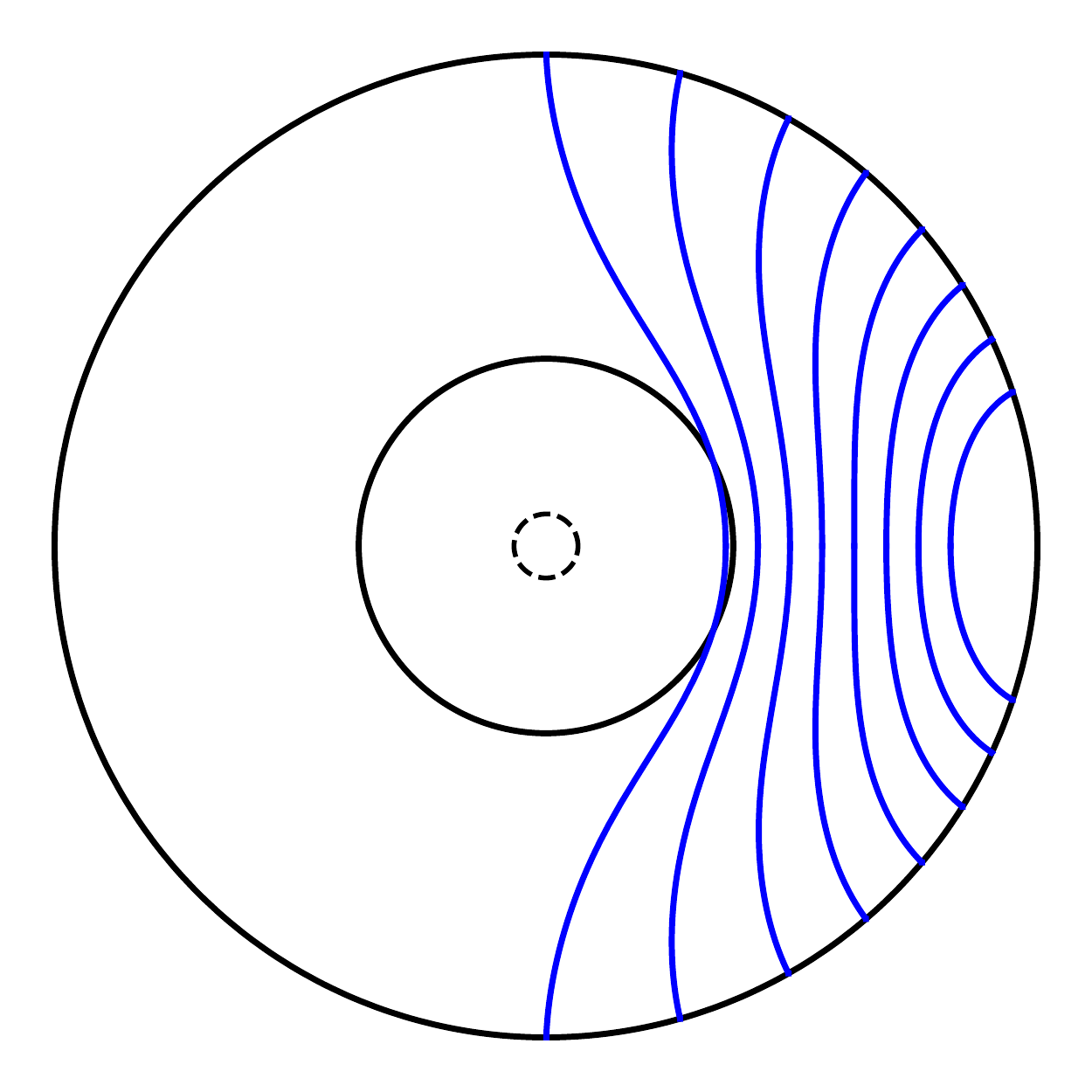}
	\caption{$GM = 1.01$, $R = 0.66 l_{\rm AdS}$: shadow}
\end{subfigure} \\\\
\begin{subfigure}{0.49\textwidth}
	\includegraphics[width=\textwidth]{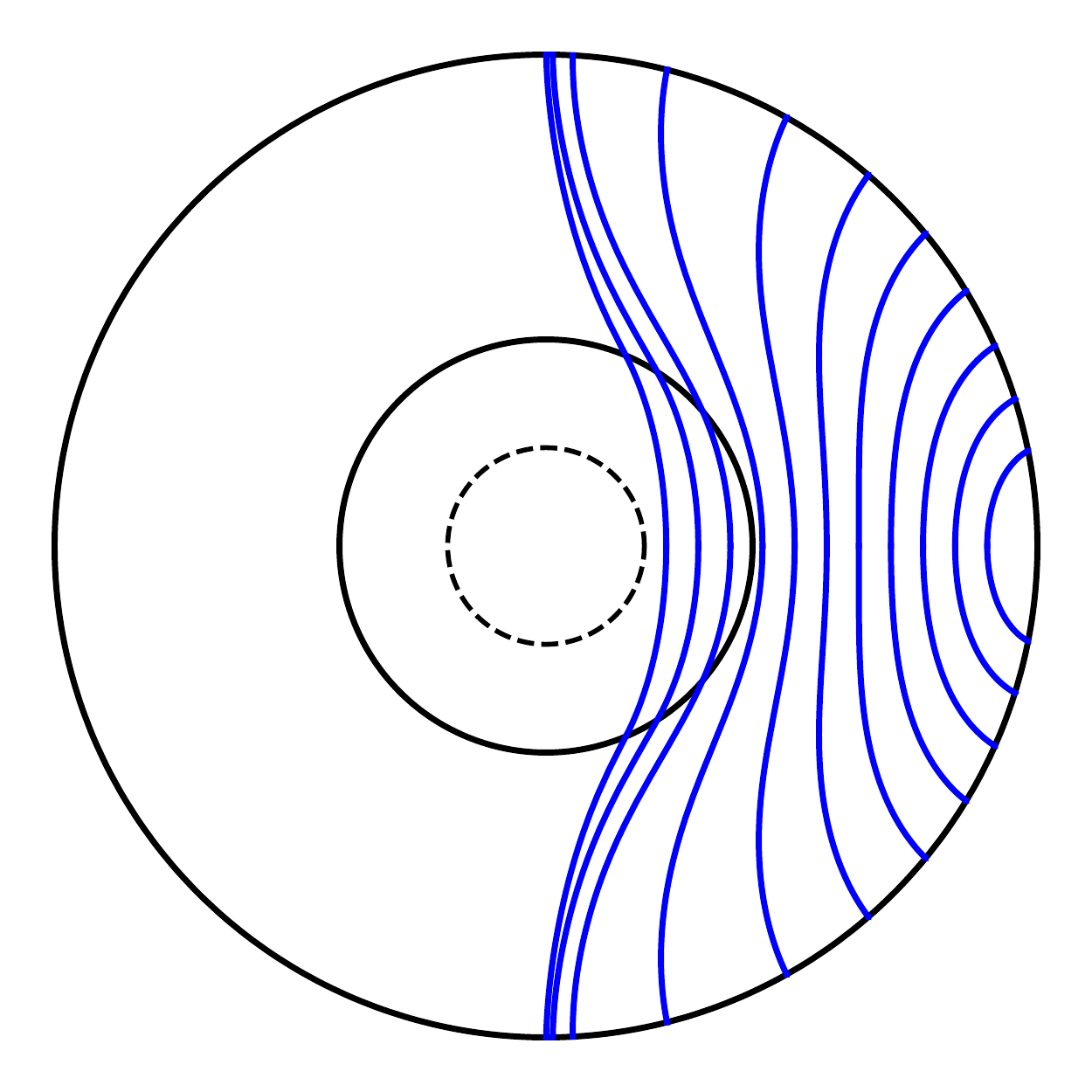}
	\caption{$GM = 1.1$, $R = 0.75 l_{\rm AdS}$: shadow}
\end{subfigure} &
\begin{subfigure}{0.49\textwidth}
	\includegraphics[width=\textwidth]{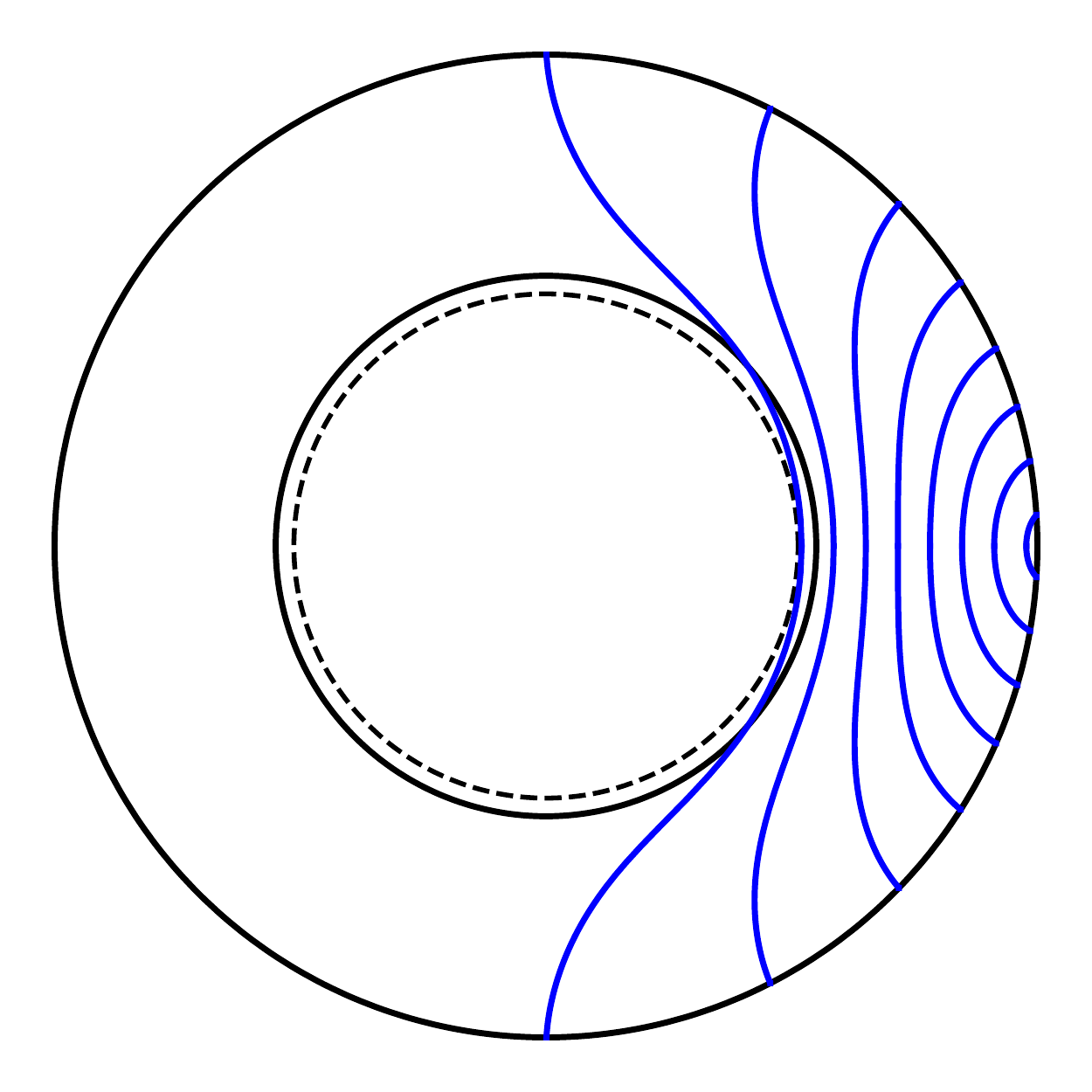}
	\caption{$GM = 2$, $R = 1.12 l_{\rm AdS}$: shadow}
\end{subfigure}\\
\end{tabular}
\caption{Plots of extremal surfaces (blue) for stars of varying density. The solid black circle is the stellar radius $R$; the smaller, dotted black circle is the would-be horizon radius $r_H$. Note that in the first case, which is outside the range \eqref{range}, there is no restriction against covering the entire bulk.}\label{AdSstarGM2R112}
\end{figure} 

This function is plotted for a range of stellar parameters in fig. \ref{StarINOUT}. Clearly, for insufficiently dense stars, $\theta_\infty(r_*)$ is monotonically decreasing, thus SCP is satisfied. However, for sufficiently dense stars, $d\theta_\infty(r_*)/dr_*>0$ as $r_*\rightarrow0$, thus WCP is violated, implying the existence of a shadow within some $r_{\rm min}$. Note that in many cases the shadow extends beyond the stellar boundary, $r_{\rm min}>R$; this is because, for the BTZ geometry in the exterior, we already have $\theta_\infty(\pi/2)$ independent of the stellar mass distribution. In such cases even the assumption of constant density is irrelevant: a shadow will exist as long as enough mass sits within some finite radius $R$. See fig. \ref{AdSstarGM2R112} for plots of the minimal surfaces for a range of stellar densities; the shadow region is easily seen by rotating the surfaces about the center. 

We can obtain an expression for the density range that supports shadows from the condition that $\dd\theta_\infty/\dd r_*>0$ at $r_*=0$, or from demanding the existence of a real solution to $\dd\theta_\infty/\dd r_*=0$. Either condition implies:
\begin{equation}\label{range}
GM-1< R^2 < \frac{GM^2}{GM+1}~.
\end{equation}
Note that the l.h.s. is simply $r_H^2$. Thus the inequality \eqref{range} effectively imposes a lower limit on the density for which one can satisfy SCP: stars of a given mass whose radius falls below the right-hand side will exhibit shadows.

\subsection{Numerical results in AdS$_{3+1}$}
In higher dimensions, one must rely on numerics to solve the second-order differential equation for $\theta_\infty(r_*)$. For $d=3$, we proceed from the following metric:
\begin{equation}
ds^2 = g_{tt}(r) dt^2 + \frac{dr^2}{f(r)} + r^2 d\Omega_2^2~,
\;\;\; f(r) = 1+\frac{r^2}{l_{\rm AdS}^2} -\frac{2M(r)}{r}
\end{equation}
where the mass function $M(r)$ is given by
\begin{equation}
M(r) = m \tanh\left(\frac{r^3}{R^3}\right)~.
\end{equation}
This corresponds to an almost-constant density star, where the mass function has been chosen to be smooth to avoid a step function at the stellar boundary $R$. Note that we have restored the AdS curvature scale in the metric, in order to consider stars much smaller than $l_{\rm AdS}$. In such cases, the scale of the shadow region will be determined exclusively by $m$ and $R$. 

\begin{figure}[h!]
\centering
\includegraphics[width=0.49\textwidth]{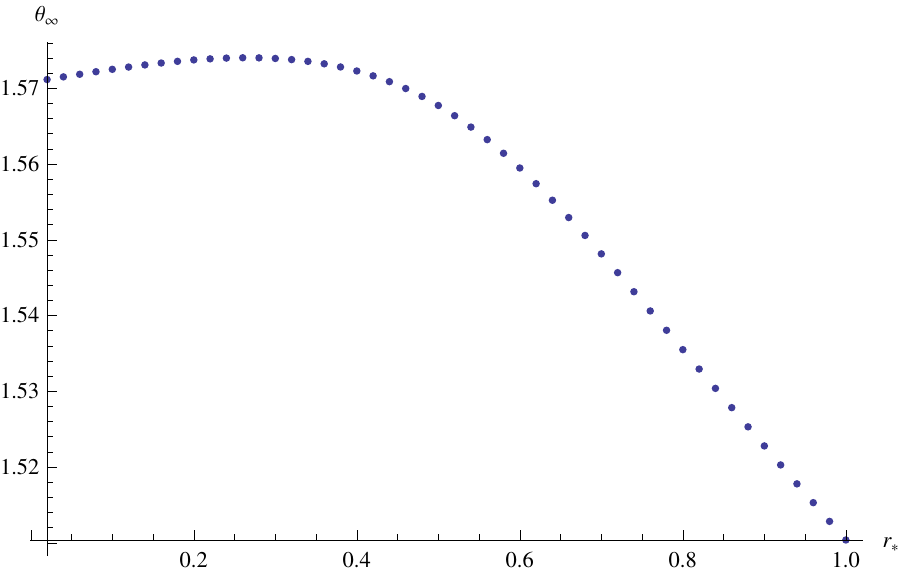}
\caption{$\theta_{\infty}(r_*)$ for an almost-constant density star in AdS$_{3+1}$ with $m=0.1$, $R=0.6$, and $l_{\rm AdS}=10$. The non-monotonicity violates WCP, which implies the existence of an entanglement shadow.}
\label{fig-NS}
\end{figure}

In fig. \ref{fig-NS}, we plot numerical results for the case $m=0.1$, $R=0.6$, $l_{\rm AdS}=10$. We see immediately from the non-monotonicity that, by Theorem 2, this geometry exhibits a shadow. This result remains unchanged if we make $l_{\rm AdS}$ arbitrarily large. Note that this $m/R$ ratio is comparable to that of a typical neutron star in our universe. The shadow persists if we make it slightly less dense, with $R=0.7$, which indicates that the density bound is not particularly severe. 

\section{Minimal area surfaces in Schwarzschild-AdS geometries}\label{sub:BTZEE}
We now turn our attention to singular geometries, in particular AdS with a black hole in the center. Obviously, $\theta_\infty(r_*)$ is undefined if $r_*$ falls within the horizon radius, hence from now on $r_*\geq r_H$ is always implied. 

A key point worth emphasizing is that, for AdS black holes, the phase transition (``switchover'') behaviour is modified. Previously, the global minimum switched solution branches when
\begin{equation}
A(\theta_\infty) = A(\pi - \theta_\infty)
\end{equation}
where $A$ is the area of the associated surface (or set of surfaces). In other words, one switches from a given bulk surface to the complement when the two have equal area, cf. fig. \ref{fig-seg2}. In the case of a black hole however, the complement must include the horizon area \cite{RT1}. This modifies the above area condition to:
\begin{equation}
A(\theta_\infty) = A(\pi-\theta_\infty) + A_{\rm BH}
\label{eq:swithcovercondition}
\end{equation}
where $A_{\rm BH}$ is the portion that wraps the black hole. 

We will present our results in three separate subsections. In sec. \ref{sec:BTZsub} we analytically solve for minimal spacelike geodesics in the BTZ geometry. We then move to higher-dimensional considerations of boundary disks in global AdS in sec. \ref{sec:globalsub}, which we split into large and small black holes to obtain suitable approximations. Although the associated spherically symmetric co-dimension 1 bulk surfaces are the most natural higher-dimensional generalizations of the lower-dimensional geodesics, we also present a similar analysis of boundary strips in planar/Poincar\'e-AdS in sec. \ref{sec:planarSAdS}, as the latter allow for a more straightforward approximation. As we shall see, for large black holes, boundary disks and strips perform almost equally well in the sense that both exhibit exponentially small shadows.\footnote{When referring to AdS-Schwarzschild, we shall speak of the size of the shadow relative to the horizon radius. Thus an exponentially small shadow is one which for which $r_{\rm min}-r_H\sim e^{-\# r_H/l_{\rm AdS}}$, with $\#$ some order one constant.} For small black holes however, strips suffer from more complicated phase transition behaviour that makes them worse boundary shapes than disks, whose associated shadow is of order $r_H$.

\subsection{BTZ black holes}\label{sec:BTZsub}
The bulk quantity dual to the von Neumann entropy of a boundary subregion $\mathcal{A}$ has been conjectured to be given by \cite{RT1,Hubeny:2007xt}:
\begin{align}
S(\mathcal{A}) = \frac{\text{Area}\left(\mathfrak{E}(\partial \mathcal{A}) \right)}{4 G}
\end{align}
where $\mathfrak{E}(\partial \mathcal{A})$ is the extremal bulk surface that ends on $\partial \mathcal{A}$ and has minimal proper area among surfaces continuously deformable to $\mathcal{A}$. When the global state of the boundary is pure, the von Neumann entropy gives a quantitative estimate for the entanglement between the subregion and its complement, called the entanglement entropy. When the global boundary state is mixed, this is no longer necessarily true, although we use the terms von Neumann entropy and entanglement entropy interchangeably in this paper.

A static BTZ black hole is described by the metric
\be
\dd s^2=-(r^2-r_H^2)\dd t^2+\frac{\dd r^2}{r^2-r_H^2}+r^2\dd\theta^2.
\label{eq:BTZmetric}
\ee
To determine the shadow, it is sufficient to consider constant time slices.\footnote{We generalize to $\dd t\neq0$ subregions in appendix \ref{nonconstt} and find that these suffer even larger entanglement shadows.} In $d=2$ the boundary is a circle, and the subsystem $\mathcal{A}$ an interval on the circle. The bulk extremal surface associated with the entanglement entropy is then simply a geodesic anchored at the two points that comprise $\partial\mathcal{A}$. We consider as a boundary region the interval $(-\theta_{\infty},\theta_{\infty})$, where the subscript $\infty$ indicates that the boundary corresponds to $r\rightarrow \infty$ in our coordinates \eqref{eq:BTZmetric}. 

\begin{figure}[h!]
\centering
\includegraphics[width=0.49\textwidth]{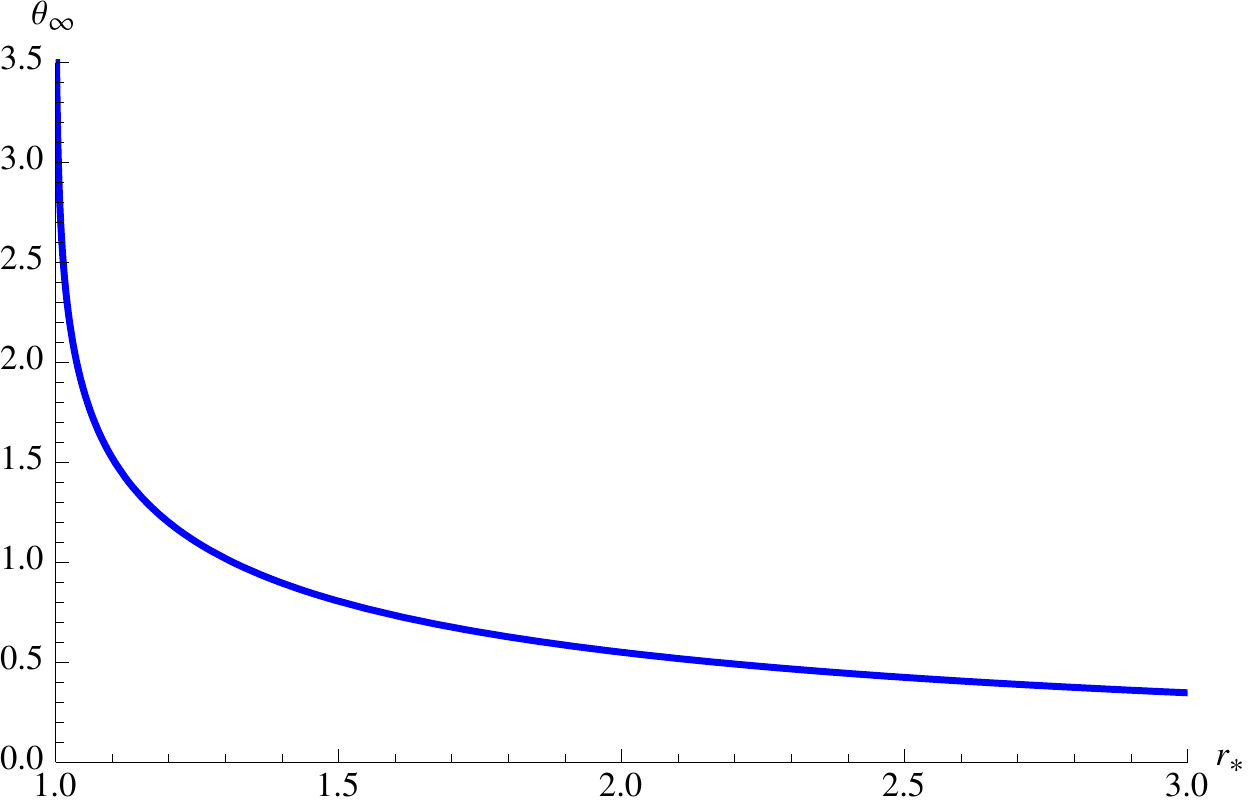}
\caption{$\theta_{\infty}(r_*)$ for a static BTZ black hole with $r_H=1$.}\label{fig:BTZthetainf}
\end{figure}

The Lagrangian describing such a bulk extremal surface is given by
\be
\mathcal{L}=\sqrt{\frac{{r'}^2}{r^2-r_H^2}+r^2}~, \qquad r'\equiv \frac{\dd r}{\dd\theta}~.
\ee
Since the Lagrangian does not depend on $\theta$, there is a conserved momentum due to translation invariance in $\theta$. Hence:
\be
\frac{\delta \mathcal{L}}{\delta r'}r'-\mathcal{L}=\mathrm{constant}~.
\ee
We may fix the constant by the demanding that the surface reaches its minimal value $r_*$ when $r'=0$. This leads to the first-order equation of motion
\be
\frac{\dd r}{\dd\theta}=\frac{r}{r_*}\sqrt{r^2-r_*^2}\sqrt{r^2-r_H^2}
\label{eq:btzdthetadr}
\ee
which may be integrated to obtain
\be
\theta_\infty=\int_{r_*}^\infty\dd r \frac{\dd\theta}{\dd r}=\frac{1}{2 r_H}\cosh^{-1}\left(\frac{r_*^2+r_H^2}{r_*^2-r_H^2} \right)~.
\label{eq:BTZthetainf}
\ee
This curve is plotted in figure \ref{fig:BTZthetainf}. Note that it diverges when $r_*\rightarrow r_H$, and decreases monotonically with increasing $r_*$. 

\begin{figure}[h!]
\centering
\includegraphics[width=0.49\textwidth]{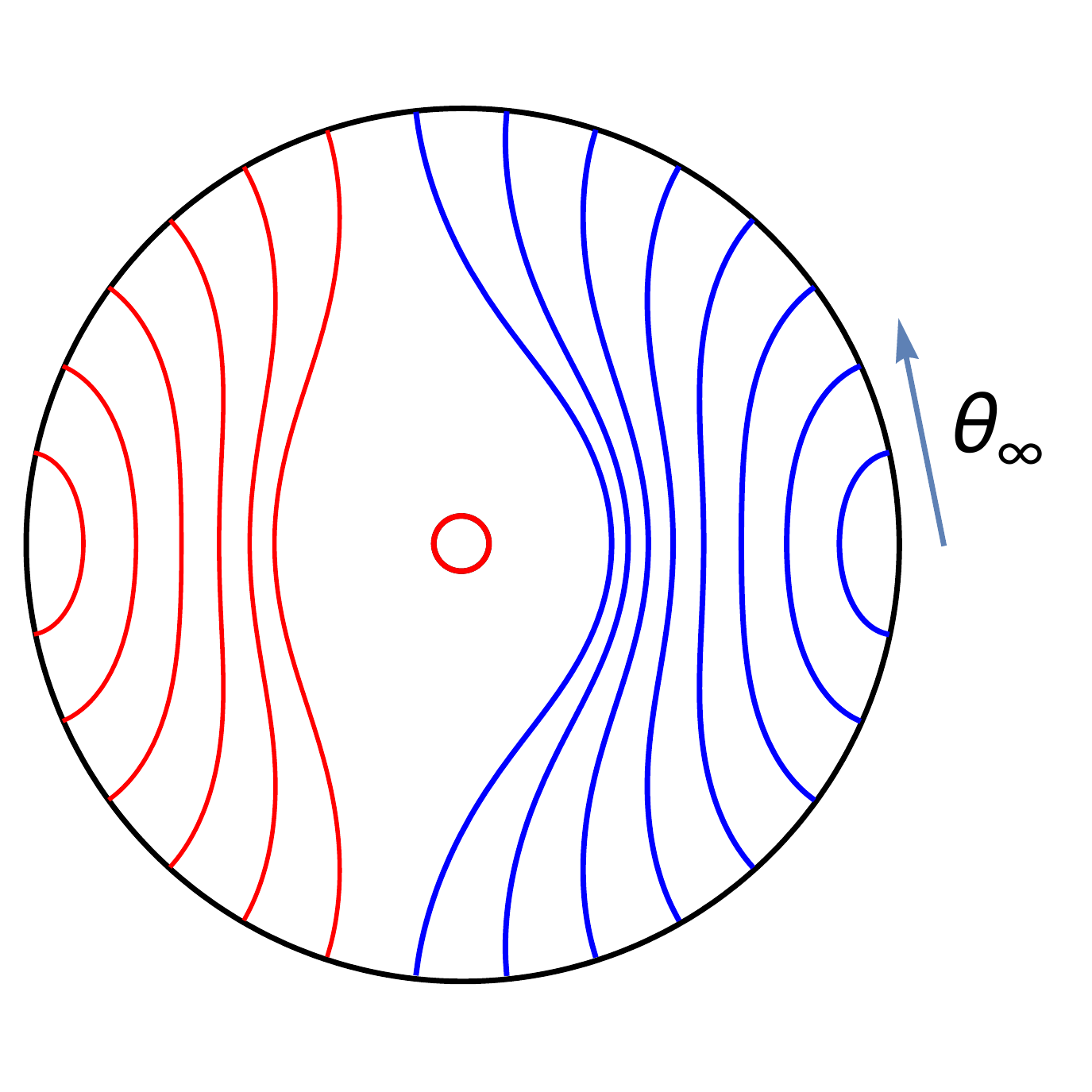}
\includegraphics[width=0.49\textwidth]{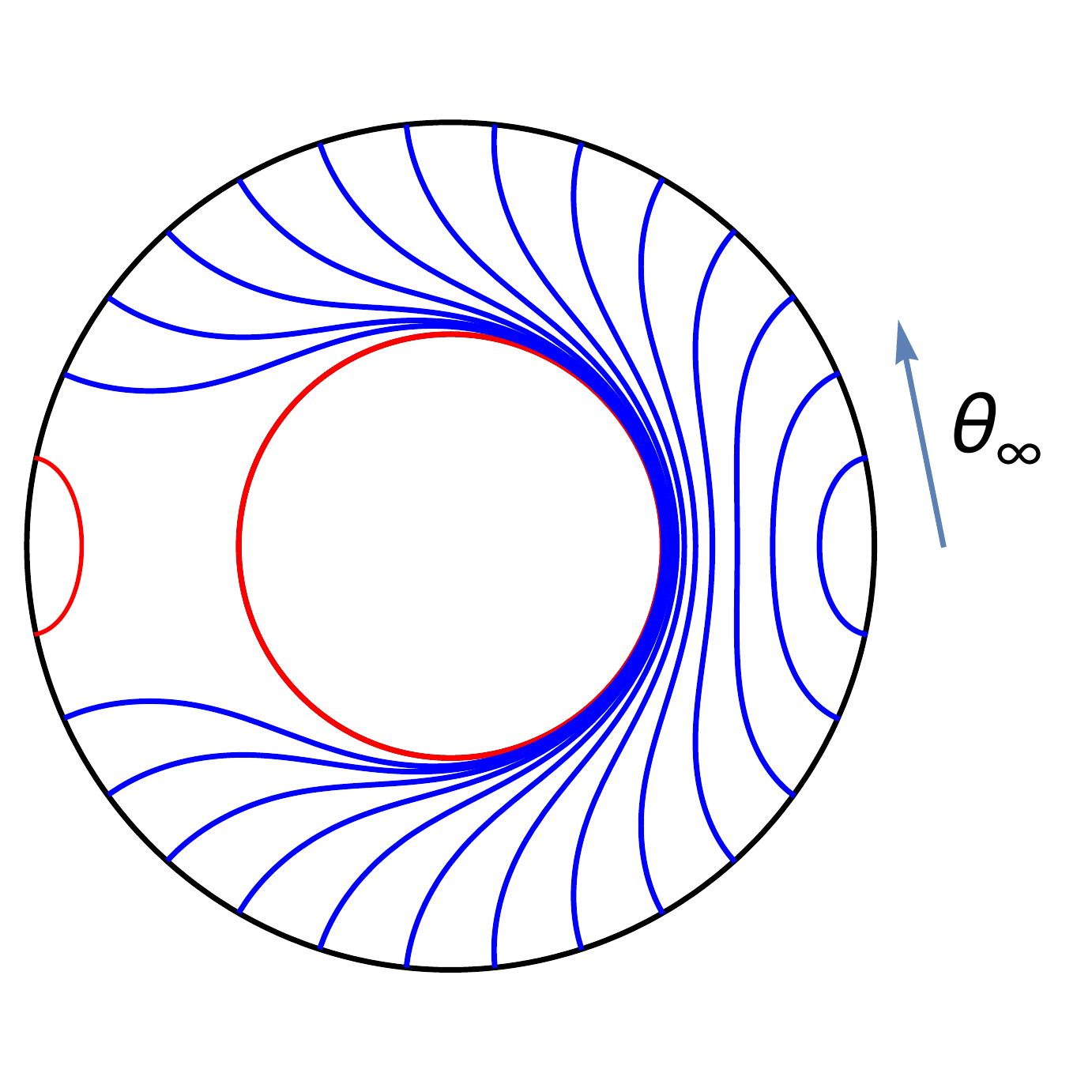}
\caption{Minimal surfaces for boundary intervals of varying size $\theta_\infty$, for a black hole of radius (red circle) $r_H = 0.1l_{\rm AdS}$ (left) and $r_H=l_{\rm AdS}$ (right). The switchover to the disconnected solution (red curves) takes place near $\theta_{\infty}=\pi/2$ for small black holes (left), and approaches $\pi$ for large black holes (right).}\label{BTZmin1}
\end{figure}

We may invert \eqref{eq:BTZthetainf} to obtain:
\be
r_*=\frac{r_H}{\tanh\left(\theta_\infty r_H \right)}.
\label{eq:BTZrstar}
\ee
which is plotted in figure \ref{btzextremealsurfaces}. One clearly sees that that there are geodesics that wind around the black hole one or more times as $r_*$ approaches the horizon. But a surface that intersect itself cannot correspond to a local minimum of the area functional (intuitively, the kinks in the intersection can be infinitesimally smoothed out to reduce the area). Thus for the purpose of identifying the appropriate bulk probe, we only care about the range $\theta_\infty\leq\pi$, since a switchover must occur before $\theta_{\infty}$ reaches this value. The alternative global minimum is then a surface with two disconnected components: a geodesic connecting the endpoints at $\pm\theta_{\infty}$ on the opposite side of the black hole, and a separate part that encircles the horizon; see figure \ref{BTZmin1}. 

We denote the critical angle at which this switchover happens by $\theta_{\rm switch}$, which is given by \eqref{eq:swithcovercondition}: 
\begin{equation}
 l(\theta_{\text{switch}}) = l(\pi -\theta_{\text{switch}})+2\pi r_H,
 \label{eq:btzswitchover}
\end{equation}
where $l(\theta_{\infty})$ is the length of the geodesic connecting the boundary points $\pm \theta_\infty$ and $2\pi r_H$ is the length of the curve that wraps the horizon.

We can compute the length $l(\theta_\infty)$ by integrating the Lagrangian
\be
l(\theta_\infty)=2\int_{r_*}^\infty \sqrt{\frac{1}{r^2-r_H^2}+r^2 \left(\frac{\dd\theta}{\dd r} \right)^2}=2\int_{r_*}^\infty \frac{r\dd r}{\sqrt{r^2-r_H^2}\sqrt{r^2-r_*^2}}
\ee
where we used \eqref{eq:btzdthetadr}, with $r_*$ given by \eqref{eq:BTZrstar}. The integral is divergent, but the divergent parts on the left- and right-hand side of \eqref{eq:btzswitchover} cancel and the finite parts yield:
\begin{equation}\label{thetaswitch}
\theta_{\text{switch}} = \frac{\pi}{2}+\frac{1}{2r_H}\ln\left(\cosh(\pi r_H)\right)~.
\end{equation}

For small black holes ($r_H \ll l_{\rm AdS}$) we have that $\theta_{\text{switch}} \approx \pi/2$, because the area contribution from the black hole in eqn. \eqref{eq:btzswitchover} is close to zero. Conversely, one sees that for large black holes ($r_H \gg l_{\rm AdS}$), $\theta_{\text{switch}} \approx \pi$. See figure \ref{BTZmin1} for an explicit plot of both cases.

\begin{figure}[h!]
\centering
\begin{minipage}[t]{0.48\textwidth}
\centering
\includegraphics[width=\textwidth]{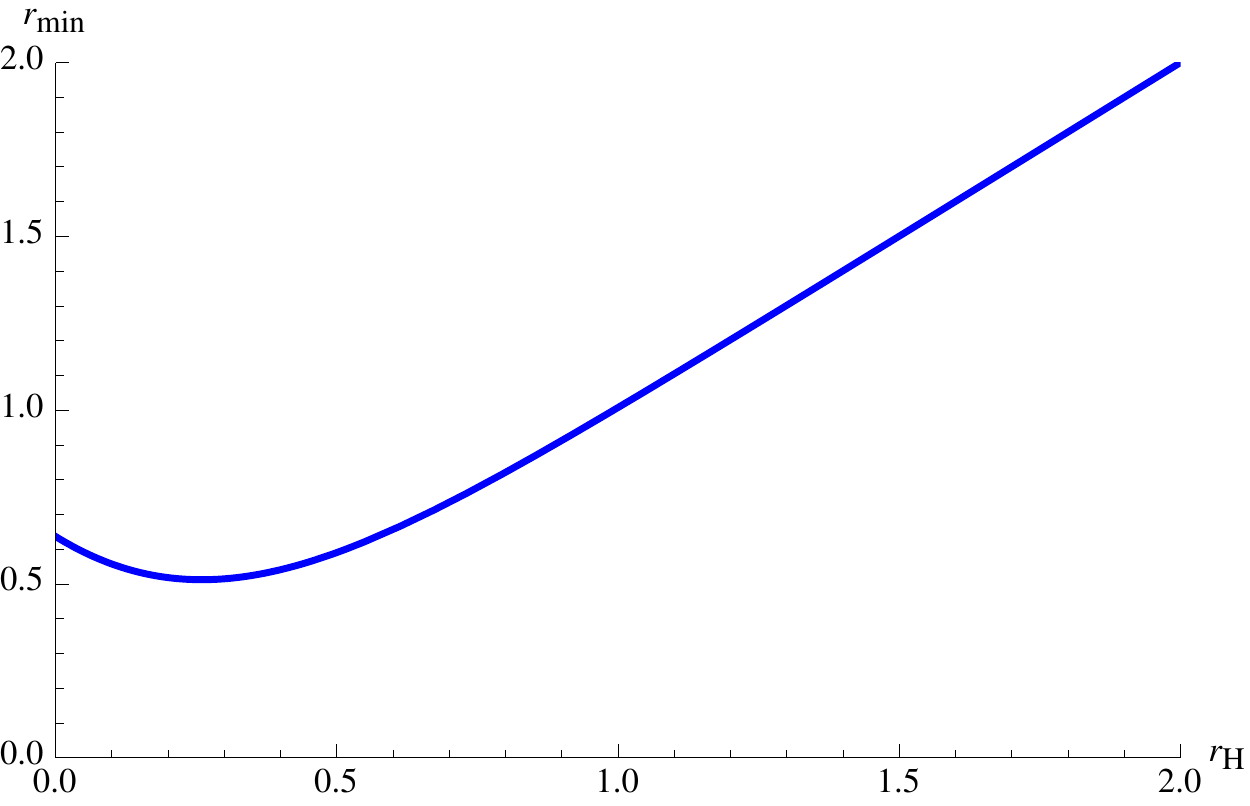}
\caption{Shadow radius $r_{\text{min}}$ as a function of horizon radius $r_H$ for a static BTZ black hole.}\label{fig:BTZfullshadow}
\end{minipage}
\quad
\begin{minipage}[t]{0.48\textwidth}
\centering
\includegraphics[width=\textwidth]{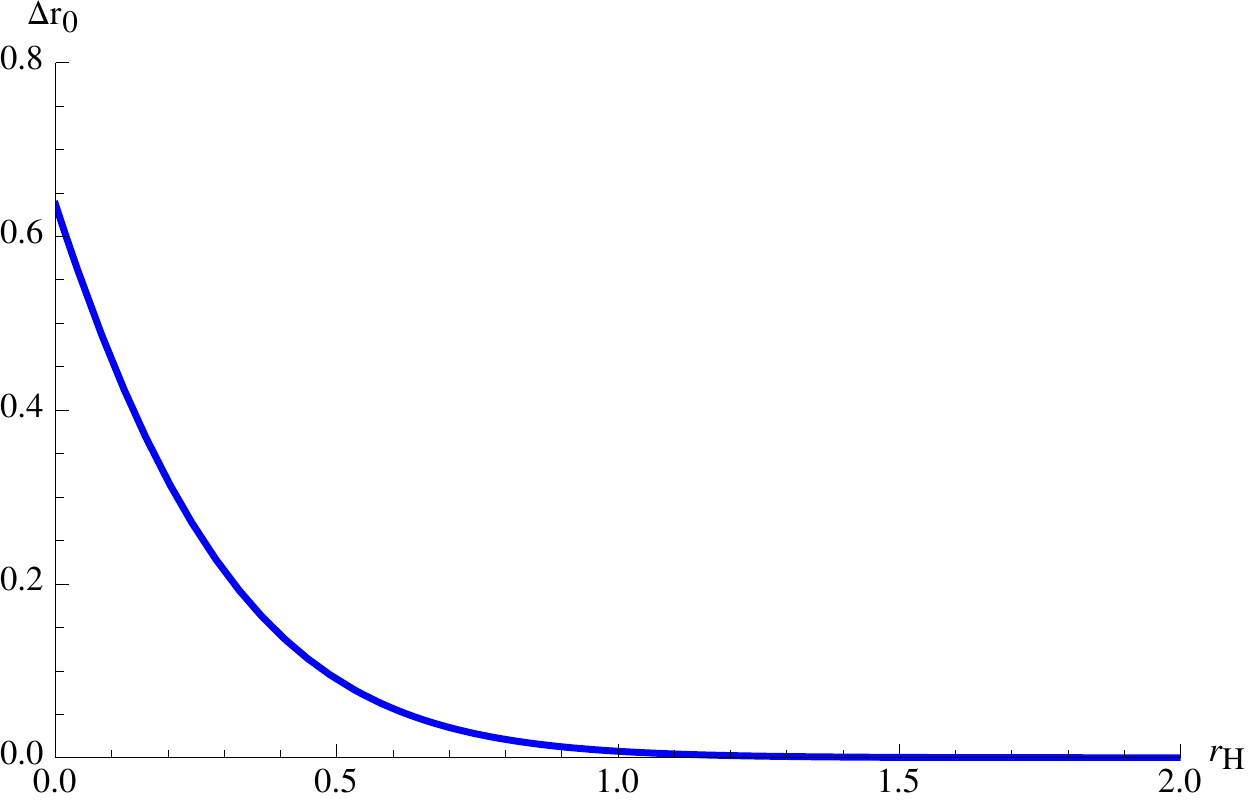}
\caption{Relative shadow size $\Delta r_0$ as a function of horizon radius $r_H$ for a static BTZ black hole.}\label{fig:BTZrelativefullshadow}
\end{minipage}
\end{figure} 

The shadow radius $r_{\text{min}}$, within which no extremal surface associated to entanglement entropy can reach, is finally determined by substituting the value of $\theta_{\text{switch}}$ into \eqref{eq:BTZrstar}:
\begin{equation}
r_{\text{min}} = \frac{r_H}{\tanh(\pi r_H)}+\frac{r_He^{-\pi r_H}}{\sinh(\pi r_H)}~.\label{eq:BTZrmin}
\end{equation}
This curve is plotted in figure \ref{fig:BTZfullshadow}. However, since the black hole is always within the shadow region, the shadow may be more conveniently expressed as
\be
\Delta r_0\equiv r_{\text{min}}-r_H=\frac{2r_He^{-\pi r_H}}{\sinh{\lp\pi r_H\rp}}\label{eq:shadowBTZrel}
\ee
which is plotted in figure \ref{fig:BTZrelativefullshadow}. When referring to the ``size'' of the shadow, we shall implicitly mean the relative quantity \eqref{eq:shadowBTZrel} unless otherwise noted.

From either eqn. \eqref{eq:shadowBTZrel} or fig. \ref{BTZmin1}, one sees that the shadow is exponentially small for large black holes, but remains an order one (AdS radius) distance from the horizon for small black holes. This behavior is easily explained by considering the switchover effect: a large black hole incurs a greater cost from the horizon component in the area condition \eqref{eq:swithcovercondition}, which allows the global minimum to remain on the original (connected) solution branch for larger values of $\theta_\infty$.

It may seem strange that that the shadow radius $r_{\rm min}$ does not go to zero for vanishing horizon radius. This is due to the mass gap in AdS$_3$: letting $r_H\rightarrow 0$ in the BTZ metric \eqref{eq:BTZmetric} will not yield the empty AdS$_3$ metric, but a conical defect. It was previously shown in \cite{BCCB_2014} that the conical defect geometry exhibits entanglement shadows; we shall comment further on this issue in the discussion, sec. \ref{sec:discussionMain}.

\subsection{Global SAdS$_{d+1}$ with $d\geq 3$}\label{sec:globalsub}
We now wish to ask how this result changes for higher-dimensional black holes. Unlike the BTZ case, in which the boundary interval was completely specified by the angle $\theta_\infty$, we will now consider the entanglement entropy of a $(d-1)$-dimensional region in the boundary CFT, which in principle can have an arbitrarily complicated shape (indeed, it does not even need to be simply connected). This allows for much richer phase transition structure when deforming the region. Hence for simplicity, we will generally assume that the boundary region of interest is $O(d-1)$ symmetric, i.e., we consider minimal surfaces of the form $r(\theta)$. 

Note that, among boundary regions of different shapes but equal area, it seems very plausible that these highly symmetric surfaces will maximize the reach into the bulk \cite{Hubeny_2012,Astaneh:2014uba}. However, this does not directly imply that asymmetric regions cannot have minimal surfaces that penetrate the shadows we find herein. This is because, as we have stressed, shadows arise from the switchover behavior, and it is difficult to study such behavior for less symmetric surfaces. Nevertheless, we believe that even if less symmetric surfaces do probe deeper in some cases, it will not eliminate shadows, and probably will not deviate much from the bounds obtained from these highly symmetric surfaces. 

Even when restricting to $O(d-1)$-symmetric surfaces, higher dimensions still allow various interesting new switchover effects. Contrast figures \ref{spheretransition} and \ref{striptransition} below. In figure \ref{spheretransition}, we consider a spherical boundary region, analagous to the BTZ case above. As the radius of this boundary ``disk'' increases, the global minimum will eventually switch to a disconnected bulk solution consisting of the sphereical cap on the far side of the black hole and a part that wraps the horizon. In figure \ref{striptransition}, we instead consider a band around the boundary sphere. As we increase the width of this ``strip'', the dual minimal surface will again undergo a switchover, but now from a single connected piece to two hemispherical caps plus the horion component. 

\begin{figure}[h!]
\centering
\begin{minipage}[t]{0.48\textwidth}
\centering
\includegraphics[width=\textwidth]{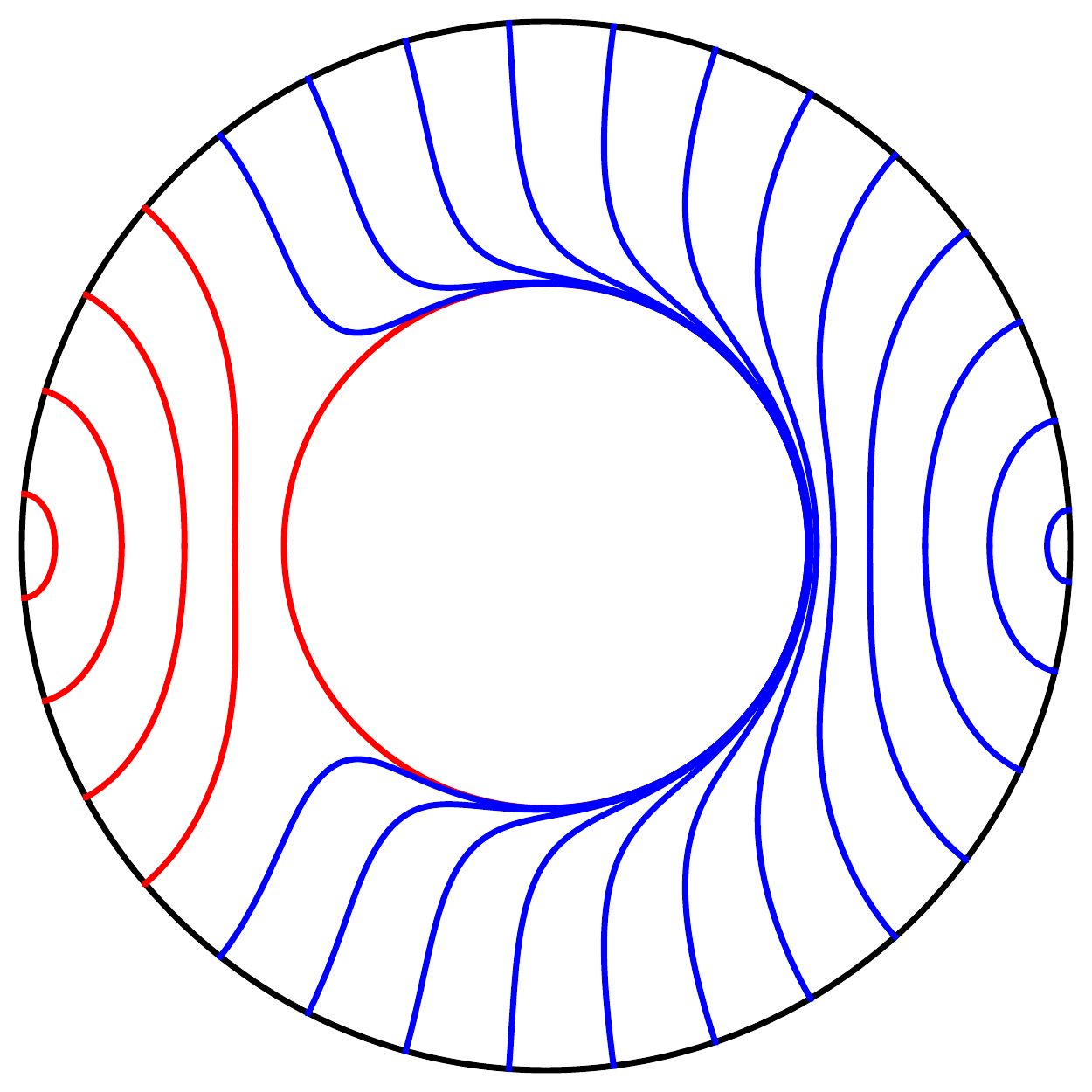}\caption{Transition between two different boundary disks for a black hole with horizon $r_H=l_{\rm AdS}$ in AdS$_5$.}\label{spheretransition}
\end{minipage}
\quad
\begin{minipage}[t]{0.48\textwidth}
\centering
\includegraphics[width=\textwidth]{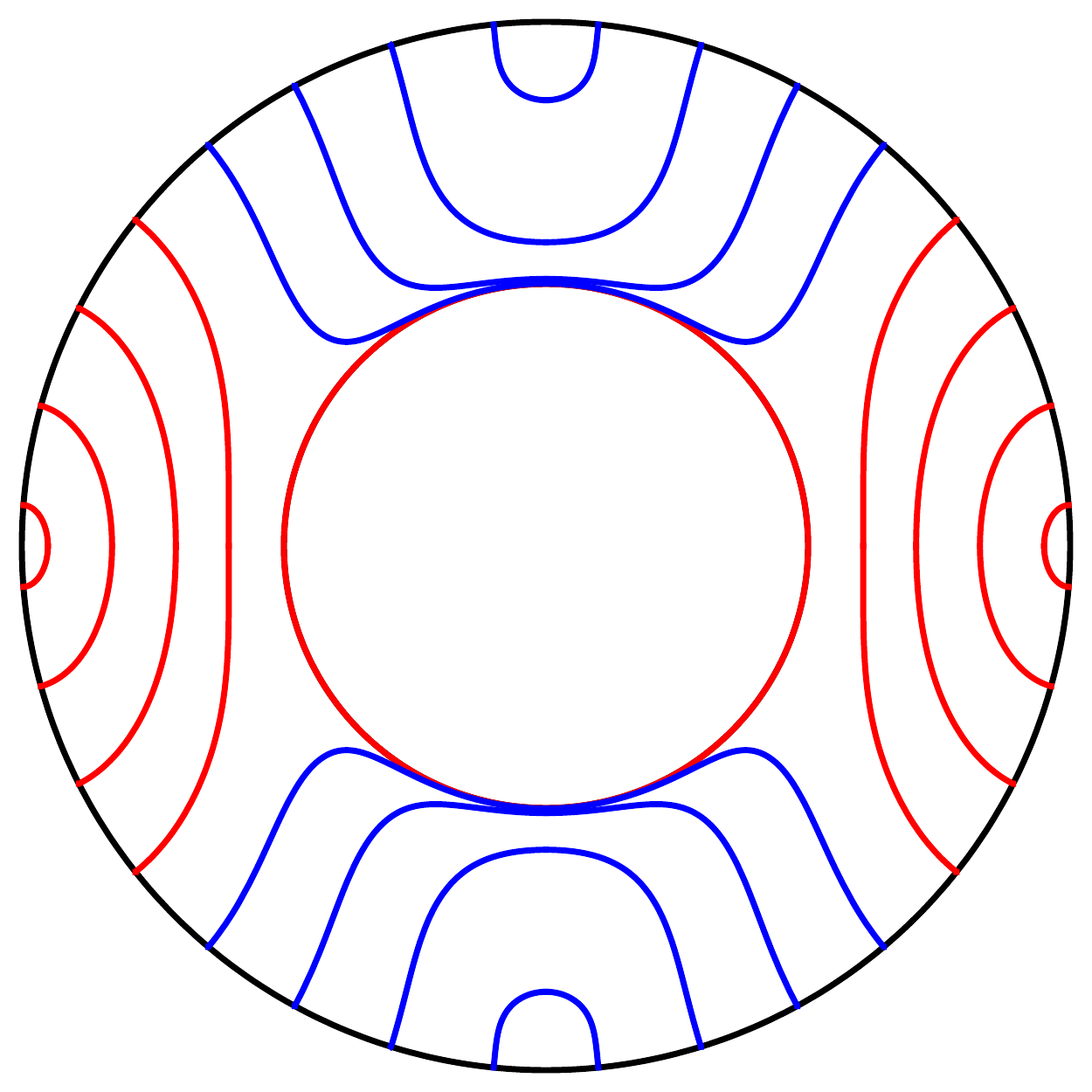}\caption{Transition between a boundary strip and two disks for a black hole with horizon $r_H=l_{\rm AdS}$ in AdS$_5$.}\label{striptransition}
\end{minipage}
\end{figure} 

\begin{figure}[h!]
\centering
\begin{minipage}[t]{0.48\textwidth}
\centering
\includegraphics[width=\textwidth]{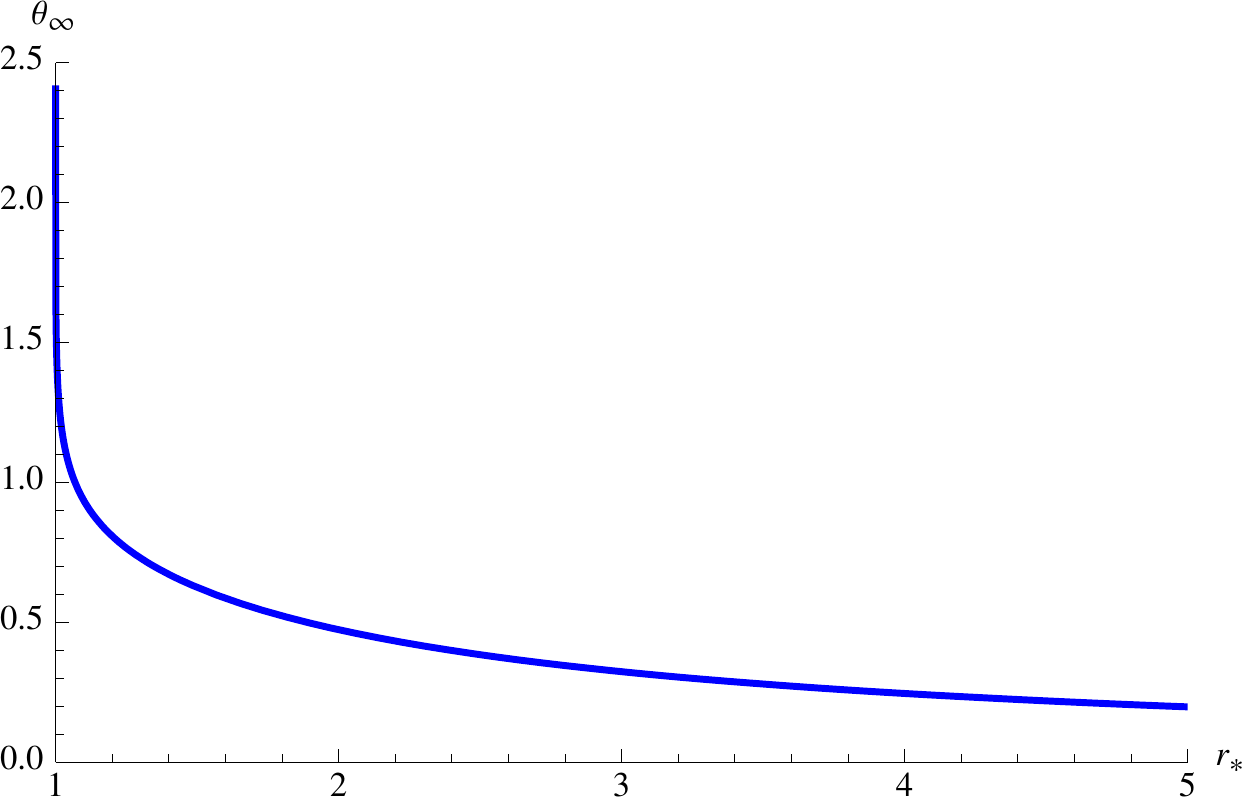}
\caption{$\theta_\infty(r_*)$ for a $SAdS_{5}$ black hole with $r_H=l_{AdS}$.}\label{fig:AdS5thetainfglobal}
\end{minipage}
\quad
\begin{minipage}[t]{0.48\textwidth}
\centering
\includegraphics[width=\textwidth]{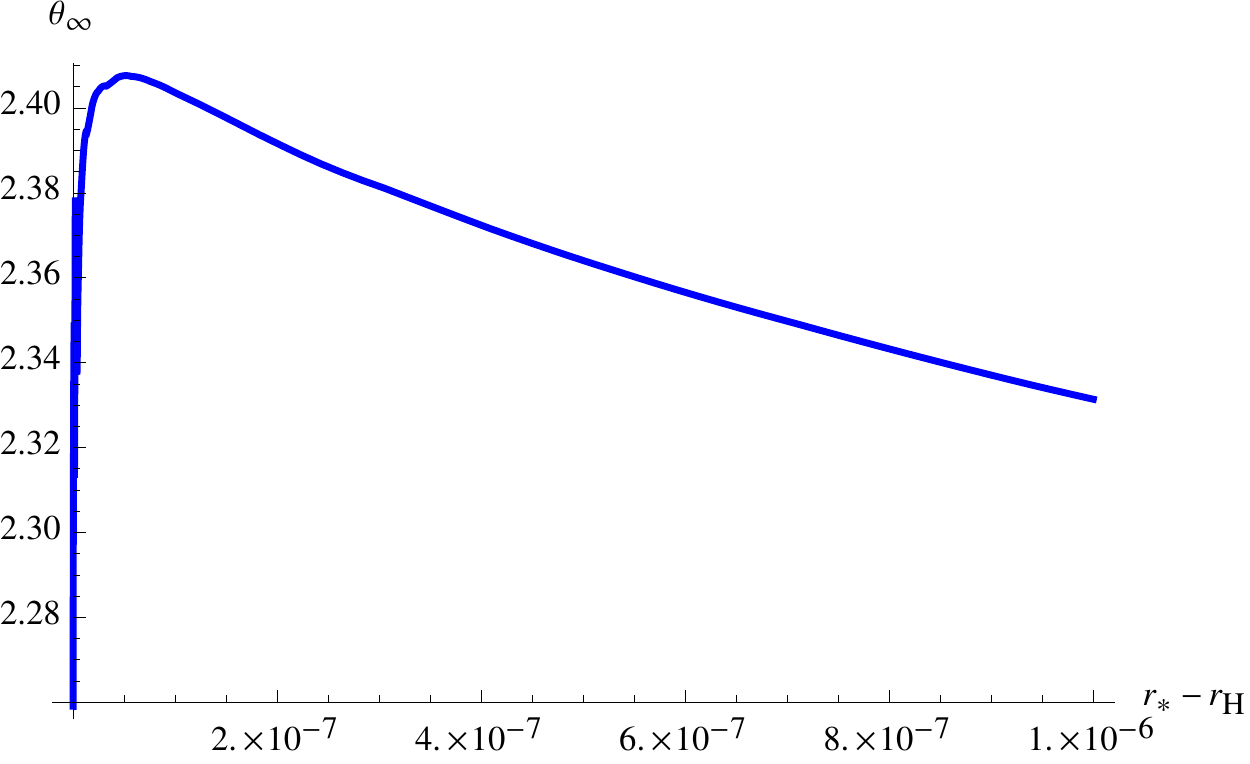}
\caption{Close up of $\theta_\infty(r_*)$ for a $SAdS_{5}$ black hole with $r_H=l_{AdS}$ near $r\approx r_H$.}\label{fig:AdS5thetainfzoom}
\end{minipage}
\end{figure}

\begin{figure}[h!]
\centering
\begin{minipage}[t]{0.48\textwidth}
\centering
\includegraphics[width=\textwidth]{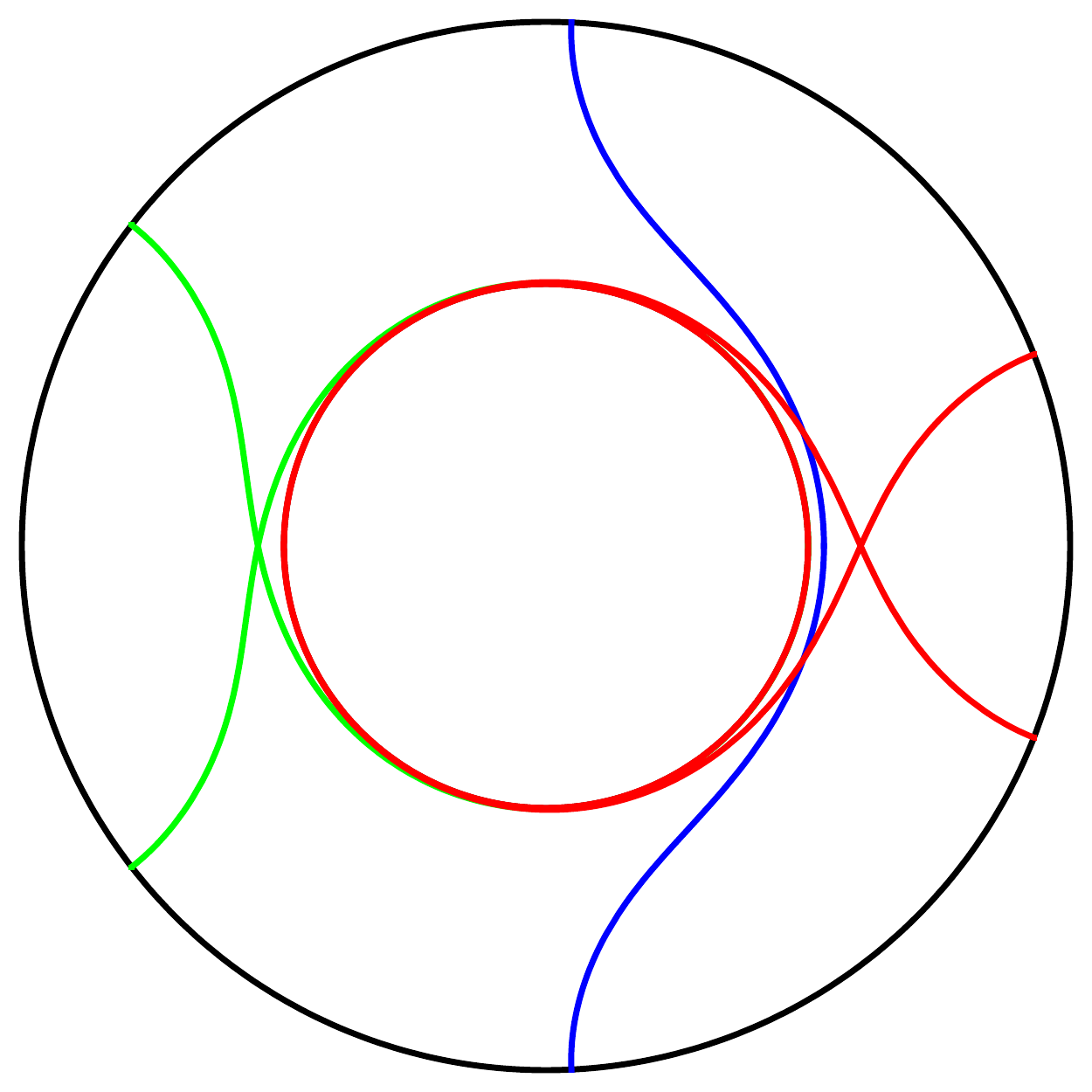}\caption{Extremal entangling surfaces in BTZ with horizon $r_H=l_{\rm AdS}$ and $\Delta r_0=10^{-1}$ (blue), $10^{-3}$ (green), and $10^{-11}$ (red). The red surface wraps the horizon four times. }\label{btzextremealsurfaces}
\end{minipage}
\quad
\begin{minipage}[t]{0.48\textwidth}
\centering
\includegraphics[width=\textwidth]{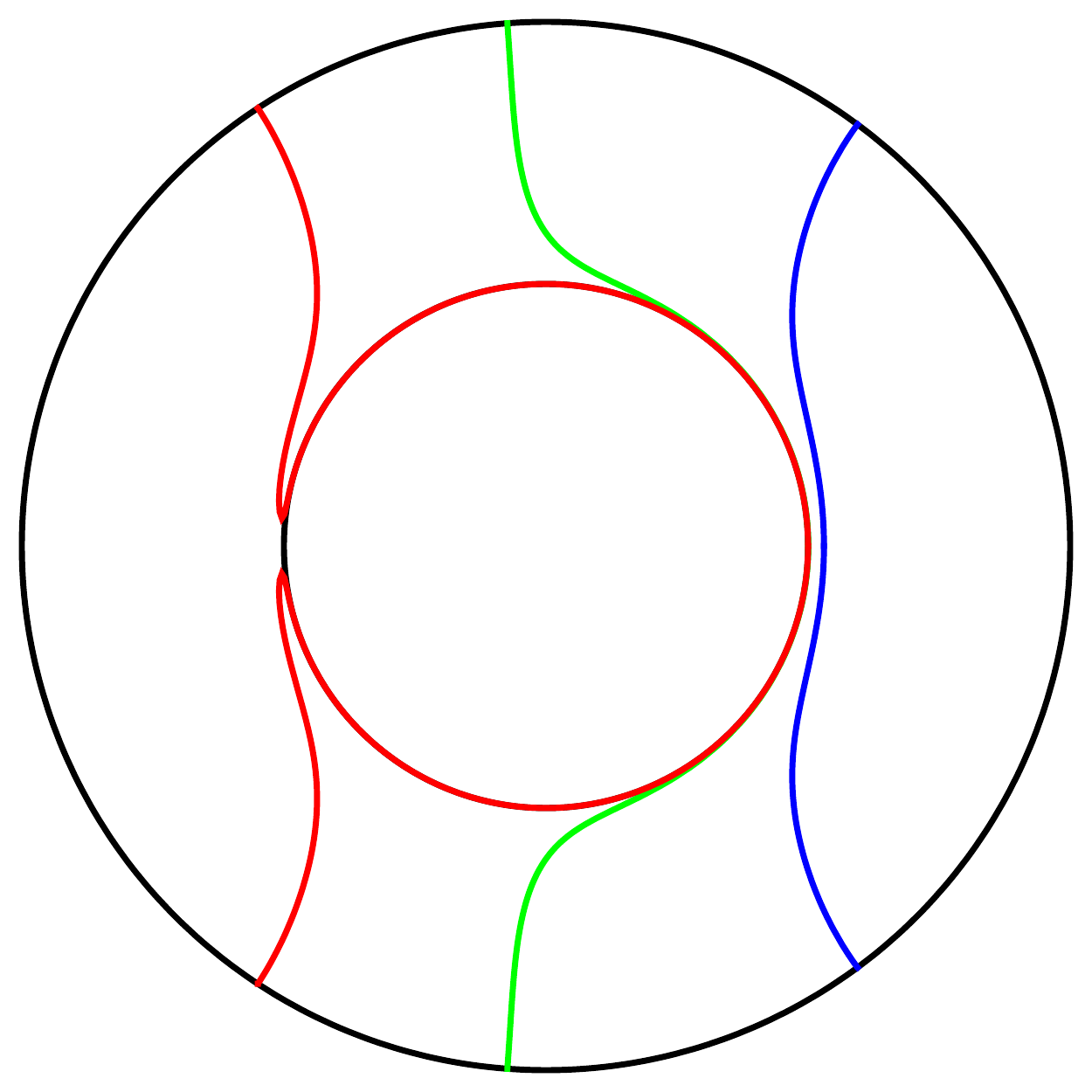}\caption{Extremal entangling surfaces in AdS$_5$ with horizon $r_H=l_{\rm AdS}$ and $\Delta r_0=10^{-1}$ (blue), $10^{-3}$ (green), and $10^{-11}$ (red). Note the folding behaviour in the red surface. }\label{AdS5extremalsurfaces}
\end{minipage}
\end{figure}

In order to study the size of the shadows in these higher dimensional geometries, we will proceed as above, by constructing the function $\theta_\infty(r_*)$ that encodes information about how well the boundary entanglement entropy can reconstruct the bulk. One of the major differences from the AdS$_3$ case is that in higher dimensions the equations of motion describing the minimal surfaces cannot be solved analytically. We rely instead on numerical methods. Results for a black hole with $r_H=l={\rm AdS}$ are displayed in figure \ref{fig:AdS5thetainfglobal}. At first sight, it looks qualitatively very similar to the BTZ case, cf. figure \ref{fig:BTZthetainf}. However, zooming in on the near horizon region, as shown in fig. \ref{fig:AdS5thetainfzoom}, reveals a crucial difference: $\theta_\infty(r_*)$ is \emph{not} actually monotonic. In fact, although not clearly visible in fig. \ref{fig:AdS5thetainfzoom}, it will oscillate an infinite number times as $r_*\rightarrow r_H$ \cite{HMRT_2013}. The difference is due to the fact that in the BTZ geometry the minimal surfaces are geodesics which in principle can self intersect, whereas in higher $d$ the surfaces instead fold into multiple layers around the black hole. See figures \ref{btzextremealsurfaces} and \ref{AdS5extremalsurfaces} for an explicit illustration of these two behaviours. 

To find the shadow, we must study the switchover behavior. Note that while $\theta_\infty$ remains finite as $r_*\rightarrow r_H$, there are values of $r_*$ for which $\theta_\infty(r_*)\geq \frac{\pi}{2}$, which makes switchovers likely. We know from Lemma 3 in appendix \ref{sec-lemma} that values of $r_*$ for which $\dd\theta_\infty/\dd r_*<0$ cannot be minimal surfaces. Additionally, the critical surfaces for which $\theta_\infty$ undergoes oscillations (e.g. the red curve in figure \ref{AdS5extremalsurfaces}) will fold around the black hole and intersect their mirror image. Hence by Lemma 2, they cannot be minimal either. Therefore, we again only need to find the largest value of $r_*$ for which the switchover condition \eqref{eq:swithcovercondition} is satisfied. This $r_*$ then corresponds to the shadow radius $r_{\rm min}$ for the symmetric surfaces under consideration.

\begin{figure}[h!]
\begin{centering}
\includegraphics[scale=0.6]{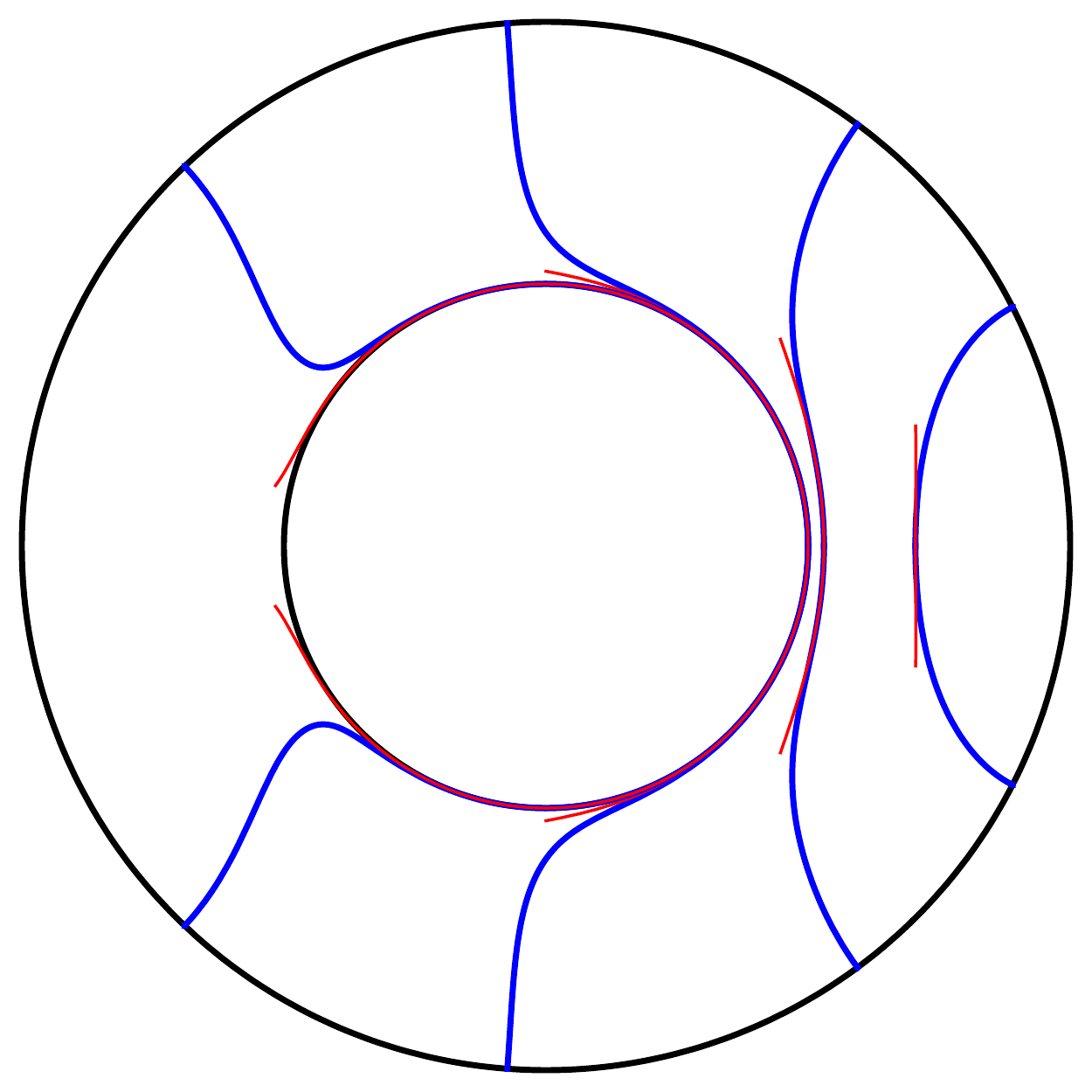}\caption{The approximation \eqref{eq:approximationformula} (red) rendered atop the exact minimal surfaces (blue) for a black hole with horizon $r_H=l_{\rm AdS}$ in AdS$_5$. The surfaces are plotted for $\Delta r_0=10^{-6}$, $10^{-3}$, $10^{-1}$, and 1.}\label{AdS5Approximations}
\end{centering}
\end{figure}

In the limiting case of a large and small\footnote{Although small black black holes have negative heat capacity in $d\geq 3$, they can still describe stable solutions in the microcanonical ensemble for some range of masses \cite{Horowitz:1999uv}.} black hole in AdS, we can analytically approximate the size of the shadow $\Delta r_0$ in arbitrary dimension as follows. The metric for SAdS$_{d+1}$ is given by
\be
\dd s^2=-f(r)\dd t^2+\frac{\dd r^2}{f(r)}+r^2\lp\dd\theta^2+\sin^2\theta \dd\Omega_{d-2}^2\rp
\label{eq:SAdSmetric}
\ee
where
\be
f(r)=r^2+1-\frac{r_H^{d-2}}{r^{d-2}}\lp r_H^2+1\rp~.\label{eq:fSAdS}
\ee
From the Lagrangian describing a $O(d-1)$ minimal surface,
\be
\mathcal{L}=(r(\theta) \sin\theta)^{d-2}\sqrt{\frac{r'(\theta)^2}{f(r)}+r(\theta)^2}
\label{higherdlagrangian}
\ee
we can write down the Euler-Lagrange equation of motion and expand it:
\be
r''(\theta)=(d-1) \left[ r_H^2d+(d-2)\right](r-r_H)-(d-2)\cot(\theta)r'+O(r')^2+O(r-r_H)^2~.
\ee
where as usual the prime denotes differentiation with respect to $\theta$. Assuming we are in a regime where it is permissible to drop the higher order terms (which is near the tip of the surface and close to the horizon), the above may be written
\be
r''(\theta)=-(d-2) \cot (\theta ) r'+ (d-1)r_H\partial_r f(r_H)(r-r_H)~; \qquad r\approx r_H,~r'\ll1.
\ee
This equation can be solved analytically for all $d$, but in $d=4$ it takes the particularly simple form
\be
r(\theta)= r_H + \frac{\Delta r_0}{\lambda} \frac{\sinh \left(\theta  \lambda \right) }{\sin\theta}~,\qquad \lambda\equiv\sqrt{12 r_H^2+5}~.
\label{eq:approximationformula}
\ee
The approximation is plotted on top of the exact solution in figure \ref{AdS5Approximations} for various values of $\Delta r_0$.

\subsubsection{Large black holes: $r_H\gg l_{\rm AdS}$}\label{sec:largesub}
For concreteness, we continue our study of shadows for large black holes in AdS$_5$, but our conclusions will remain valid for large black holes in arbitrary dimension.

In the large black hole limit $r_H\gg 1$, our approximate solution \eqref{eq:approximationformula} reduces to
\be
r(\theta)-r_H\approx  \frac{\Delta r_0}{2\sqrt{12}r_H} \frac{e^{\sqrt{12}r_H \theta} }{\sin\theta}~.
\ee
This solution $r(\theta)$ locally solves the minimal area equations in the near horizon geometry. If we pick the parameter $\Delta r_0$ to be small, this solution will cover an order one angle $\theta$ before the approximation breaks down (see figure \ref{AdS5Approximations}). At this point, the surface is a distance $O(r_H)$ away from the horizon, and one could extend the approximation by matching it to a solution in empty AdS anchored to the boundary. While we don't need to know the exact solution in this regime to estimate the shadow, we can show that the rest of the minimal surface will be quite boring in the sense that it is almost going radially outward towards the boundary. To be more precise, we can show that the amount of angle $\Delta \theta$ that the minimal surface covers when leaving this near horizon regime will be small in the large black hole limit.

We start with Lagrangian \eqref{higherdlagrangian} and approximate $\sin\theta$ to be constant. We then take $f(r)\approx r^2$ since we are relatively far from the black hole. As the Lagrangian no longer depends on $\theta$, there is a conserved quantity $C$ associated to translations in the angular direction, hence:
\be
\frac{\delta\mathcal{L}}{\delta r'} r' - \mathcal{L}=C\implies r^{2d}=C^2 \left(\frac{r'^2}{r^2}+r^2 \right)~.
\ee
The constant $C$ can subsequently be determined by matching, at $r=2r_H$, to our near horizon solution. Specifying henceforth to $d=4$, this yields $C\approx 6 r_H^3$. Plugging this into the above, we find
\be
\Delta \theta = \int_{2 r_H}^\infty \frac{\dd r}{r^2 \sqrt{\frac{r^6}{C^2}-1}}\approx \frac{0.1}{r_H} \ll 1
\ee
which confirms that the minimal surfaces are going approximately radially outward outside $r=2r_H$. Thus we may match our near-horizon solution at a distance $r_H$ from the horizon at some order-one matching angle $\theta_m$ to conclude:
\be
\Delta r_0\approx 2 \sqrt{12}r_H^2   \sin(\theta_m) e^{- \sqrt{12} \theta_m r_H} \propto r_H^2 e^{-\#r_H}
\label{eq:higherdfullshadowsolution}
\ee
where $\#$ is an $O(1)$ number. Thus we find that the shadow region for minimal surfaces is exponentially small for large black holes. Although this particular result has been obtained for SAdS$_5$, we can show that it holds in any dimension.

We must note that in \eqref{eq:approximationformula} we choose as a boundary condition $r_*=r(\theta=0)$, which corresponds to disk-shaped boundary regions. In contrast, the aforementioned boundary strips would require $r_*=r(\theta=\frac{\pi}{2})$. The analysis for the strip is precisely analogous, and also results in an exponentially small shadow. In section \ref{sec:planarSAdS}, we explicitly show that the shadow is exponentially small for all $d$ in planar-SAdS$_{d+1}$, but we first turn to an analysis of small black holes in global SAdS$_5$.

\subsubsection{Small black holes: $r_H\ll l_{\rm AdS}$}\label{sec:smallsub}
For a small black hole, we can make a different argument to estimate the size of the shadow. Since the horizon area is small in eqn. \ref{eq:swithcovercondition}, the switchover angle must be approximately $\pi/2$. Additionally, as explained above, the minimal surface must remain in a single hemisphere, with no folds. The shadow size will therefore be determined by a simple minimal surface at the switchover point. Starting from the boundary at $\theta=\pi/2$, this surface will dive almost radially inward until it is an order $r_H\ll l_{\rm AdS}$ away from the black hole horizon. Here it can be matched to our approximate solution \eqref{eq:approximationformula} in the $r_H\ll l_{\rm AdS}$ limit:
\be
r(\theta)= r_H + \frac{\Delta r_0}{\lambda} \frac{\sinh \left(\theta  \lambda \right) }{\sin\theta}~,\qquad \lambda\approx \sqrt{5}~.
\ee

If we make $\Delta r_0$ too small, the solution will remain in the near horizon regime and the angle traversed will exceed $\pi/2$. Hence, to find the smallest allowed $\Delta r_0$, we must pick it in such way that our approximation breaks down and can be matched onto the radially outward piece at almost $\pi/2$. To estimate (and bound from below) this value of $\Delta r_0$, we let $r(\theta)-r_H\approx r_H$ and take $\theta=\pi/2$ in our approximation \eqref{eq:approximationformula}:
\be
\Delta r_0 \approx \frac{  r_H}{\sinh\lp\sqrt{5} \frac{\pi}{2}\rp}  = \# r_H\label{eq:minimalSmallApprox}
\ee
where $\#$ is again an $O(1)$ number. We conclude that for a small black hole in AdS$_5$, the shadow size is $O(r_H)$. A similar analysis confirms that for every $d\geq 3$ the property $\Delta r_0 \propto r_H$ holds, with the coefficient of proportionality decreasing for increasing $d$. As for the large black hole above, it is important to keep in mind that we presented only disk-shaped boundary regions. It is of course also possible to consider a strip on the boundary, but the small horizon area in this case ensures that the switchover to disconnected surface containing two disks will happen quite soon, which makes strips have even larger shadows. 

While these results conclude our analysis of shadows for small black holes in AdS$_{d+1}$, we would like to end with a parenthetical remark which concerns extending these results to AdS$_{d+1}$ times a compact manifold, as is often the case in concrete realisations of the holographic principle. For example, when considering a small black hole in AdS$_5\times\rm{S}^5$ (smeared uniformly over the S$_5$), one might be inclined to think that the correct minimal surface will be the AdS$_5$ solution as described above, uniformly wrapping the five-sphere. However, when the size of the AdS black hole is small w.r.t the compact manifold, one can show that these black holes are Gregory-Laflamme unstable to localizing on the sphere \cite{Gregory_2011, Hollowood_2006}. This means that the black hole will be an effective ten-dimensional one, and to find the associated minimal surfaces one should analyze it in the appropriate $10d$ background---interpolating between a $10d$ Schwarzschild geometry close to the black hole and an AdS$_5\times\rm{S}^5$ geometry far away. Although we did not analyze this case in detail, we expect that it will not qualitatively alter the above results. 

\subsection{Planar SAdS$_{d+1}$ with $d\geq 3$}\label{sec:planarSAdS}
To show that the shadow for a large black hole is exponentially small in any dimension $d\geq3$, we can perform the analysis in a Poincar\'e patch of Schwarzschild-AdS$_{d+1}$, which is an excellent approximation in the large black hole limit. If we furthermore restrict ourselves to boundary strips, the enhanced symmetry of the problem will allow for an analytical treatment which confirms the exponential size of the shadow in for arbitrary $d\geq3$.

To proceed, we make the change of variables $z = r_H/r$ in the metric \eqref{eq:SAdSmetric} and consider the $r_H\gg 1$ limit: 
\begin{equation}
\dd s^2 \approx (1-z^d)\frac{-\dd t^2}{z^2}+\frac{\dd z^2}{z^2(1-z^d)}+\frac{r_H^2d\Omega_{d-1}^2}{z^2}~.
\end{equation}
For boundary length scales $\theta_\infty\ll r_H$, we can take the boundary metric as approximately flat, $r_H^2\dd\Omega^2_{d-1} \approx\dd\bx_{d-1}^2$. We consider the strip with width $\theta_\infty = ar_H$ with $a\ll 1$ and assume that the strip is sufficiently wide that the deepest point to which the associated bulk minimal reaches, $z_*$, penetrates the near-horizon region, i.e., $z_*-1 \ll 1$. 

The action is given by:
\begin{equation}
S = \int\dd^{d-2}\bx \int\frac{\dd x_1}{z^d}\sqrt{1+\left(\frac{\dd z}{\dd x_1}\right)^2\frac{1}{1-z^d}}~,
\end{equation}
where $x_1$ is the transverse direction. This leads to the equation of motion:
\begin{equation}
\left(\frac{\dd z}{\dd x_1}\right)^2 = (1-z^d)\left(1-\left(\frac{z_*}{z}\right)^{2(d-1)}\right)
\end{equation}
for which the width of the bulk probe is
\begin{equation*}
\frac{\theta_\infty}{2}=\int_0^{\theta_\infty/2}\dd x_1
=\int_0^{z_*}\dd z \left|\frac{\dd x_1}{\dd z}\right|
=\int_0^{z_*}\dd z\lp\frac{z}{z_*}\rp^d\left[\lp1-z^d\rp\lp1-\frac{z^{2(d-1)}}{z_*^{2(d-1)}}\rp\right]^{-1/2}
\end{equation*}
which we may solve approximately by making the change of variables $u\equiv1-z/z_*$ and expanding for small $u$:
\begin{equation*}
\begin{aligned}
\theta_\infty&= 2 z_* \int_0^1\dd u (1-u)^d\left[\lp1-z_*^d (1-u)^d\rp\lp1-(1-u)^{2(d-1)}\rp\right]^{-1/2}\\
&\approx 2z_*\int_0^1\dd u\left[2u(d-1)\lp1-z_*^d +dz_*^d u\rp\right]^{-1/2}
\\&
=\frac{2z_*}{\sqrt{2(d-1)z_*^d}}\cosh^{-1}\left(\frac{2dz_*^d+1-z_*^d}{1-z_*^d}\right)~.
\end{aligned}
\end{equation*}
For ease of comparison with the higher-$d$ solution in global SAdS \eqref{eq:higherdfullshadowsolution}, we make the further approximation $z_* \approx 1$,\footnote{This approximation is valid if $\theta_\infty$ is sufficiently large; this can be accomplished without violating $\theta_\infty\ll r_H$ by taking the large black hole limit, $r_H \gg 1$, which is precisely our current regime.} underwhich the above expression simplifies to:
\begin{equation}\label{zstar}
 z_*\approx1-2d^2\sech\lp \theta_\infty\sqrt{\frac{d-1}{2}}\rp
\implies
 \Delta r_0 \approx 4dr_He^{-ar_H\sqrt{\frac{d-1}{2}}}~.
\end{equation}
We emphasize that this result is only valid for $\theta_\infty=ar_H\gg1$ with $a\ll 1$ and $r_H \gg 1$. Although the calculation was done for a boundary strip and not a disk, the result \eqref{zstar} supports our claim that the shadow is exponentially small for large black holes in all $d\geq3$. 

\section{Wilson loops}\label{sub:BTZWILSON}
In this section, we turn to another bulk probe: static worldsheets arising from certain Wilson loops in the boundary CFT. The bulk dual of the expectation value of a Wilson loop $\mathcal{W}(\mathcal{C})$ evaluated in the supergravity limit is proposed to be \cite{Maldacena:1998im}:
\begin{equation}
\mathcal{W}(\mathcal{C}) \sim e^{-S}
\end{equation}
where $S$ is the proper area of a fundamental string ending on the boundary loop $\mathcal{C}$. To simplify our analysis, we will consider rectangular Wilson loops that extend far into the past and future time-directions. Such a Wilson loop with temporal ``height'' $T$ and spatial width $2\theta_{\infty}$ can be interpreted as the potential between a quark and an anti-quark \cite{Maldacena:1998im,Brandhuber:1998bs}. We assume sufficiently large $T$ that the worldsheet may be considered invariant under time translations. The action for such a static worldsheet is given by
\begin{equation}\label{actionwilson}
S = 2T \int_0^{\theta_{\infty}}\dd\theta \sqrt{\left(\partial_{\theta}r\right)^2+r^2f(r)}~.
\end{equation}
Note that in static spacetimes this quantity takes the standard form of eqn. \eqref{eq-gp} with $F\propto \sqrt{-g_{tt}}$, thus we may treat it as a holographic probe similar to minimal area surfaces.

\begin{figure}[H]
\begin{minipage}[t]{0.475\textwidth}
 \includegraphics[width=\textwidth]{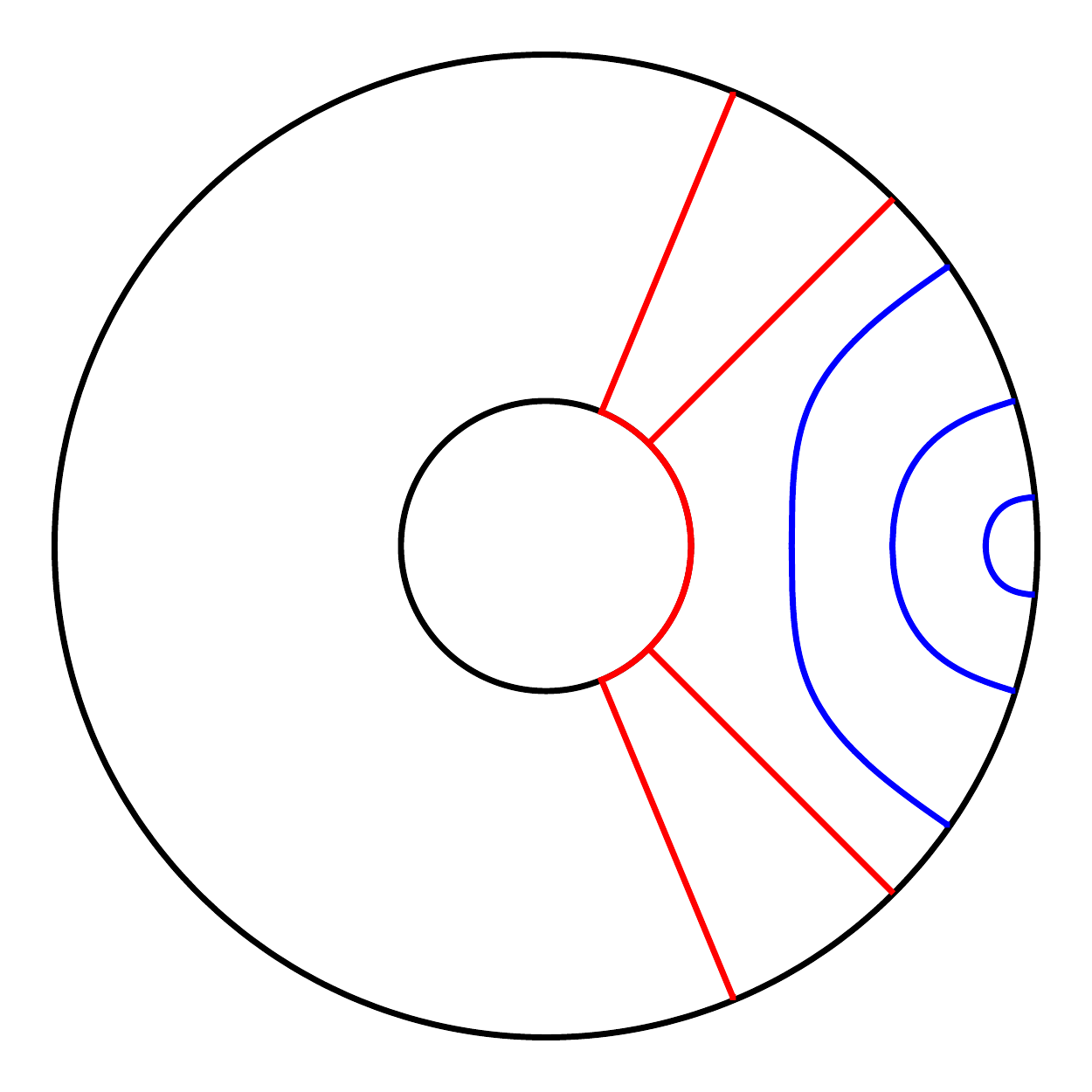}
 \caption{Worldsheets corresponding to different boundary angles for a BTZ black hole of radius $r_H = 0.5l_{\rm AdS}$. The $\cup$-shaped worldsheets are rendered in blue; $\sqcup$-shaped, in red.}\label{BTZwil05}
\end{minipage}
\begin{minipage}{0.05\textwidth}
\quad
\end{minipage}
\begin{minipage}[t]{0.475\textwidth}
 \includegraphics[width=\textwidth]{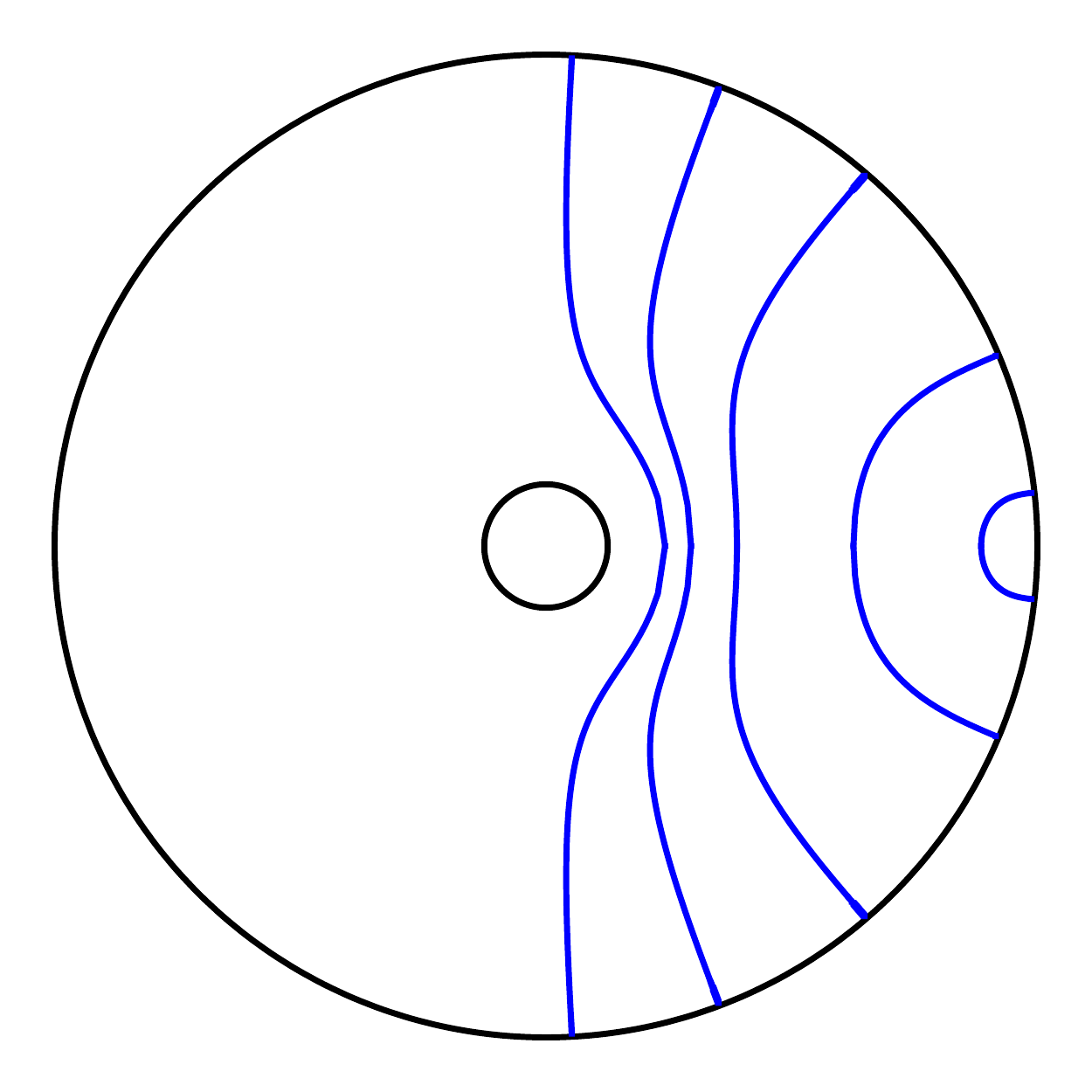}
 \caption{Worldsheets corresponding to different boundary angles for a BTZ black hole of radius $r_H = 0.2l_{\rm AdS}$. Small black holes in $d=2$ are special, because the $\cup$-shaped worldsheet constitutes the leading saddle point for all values of $\theta_\infty$.}\label{BTZwil02}
\end{minipage}
\end{figure}

The action \eqref{actionwilson} does not explicitly depend on $\theta$, so there is a conserved quantity that we shall use to write the equation of motion as a first order differential equation. We will find it convenient to distinguish two types of solutions to this equation:
\begin{description}
\item[$\cup$-shaped worldsheets] are smooth worldsheets anchored on the boundary that do not reach the black hole horizon, instead turning smoothly such that $\left.\partial_{\theta}r\right|_{r=r_*}=0$ at some finite $r_*>r_H$ (see figure \ref{BTZwil05}).
\item[$\sqcup$-shaped worldsheets] consist of two straight segments that extend from the boundary to the black hole, joined discontinuously by a third segment that partially wraps the horizon (see figure \ref{BTZwil05}). 
\end{description}

For a given boundary angle $\theta_{\infty}$, multiple solutions to the equation of motion may exist. Evaluation of the area functional is therefore necessary to determine which worldsheet constitutes the leading saddle point. Generally, we find that a switchover or phase transition occurs from $\cup$-shaped to $\sqcup$-worldsheets, as illustrated in fig. \ref{BTZwil05}. We discuss this behaviour in more detail below. 

\begin{figure}[h!]
\centering
\includegraphics[width=0.49\textwidth]{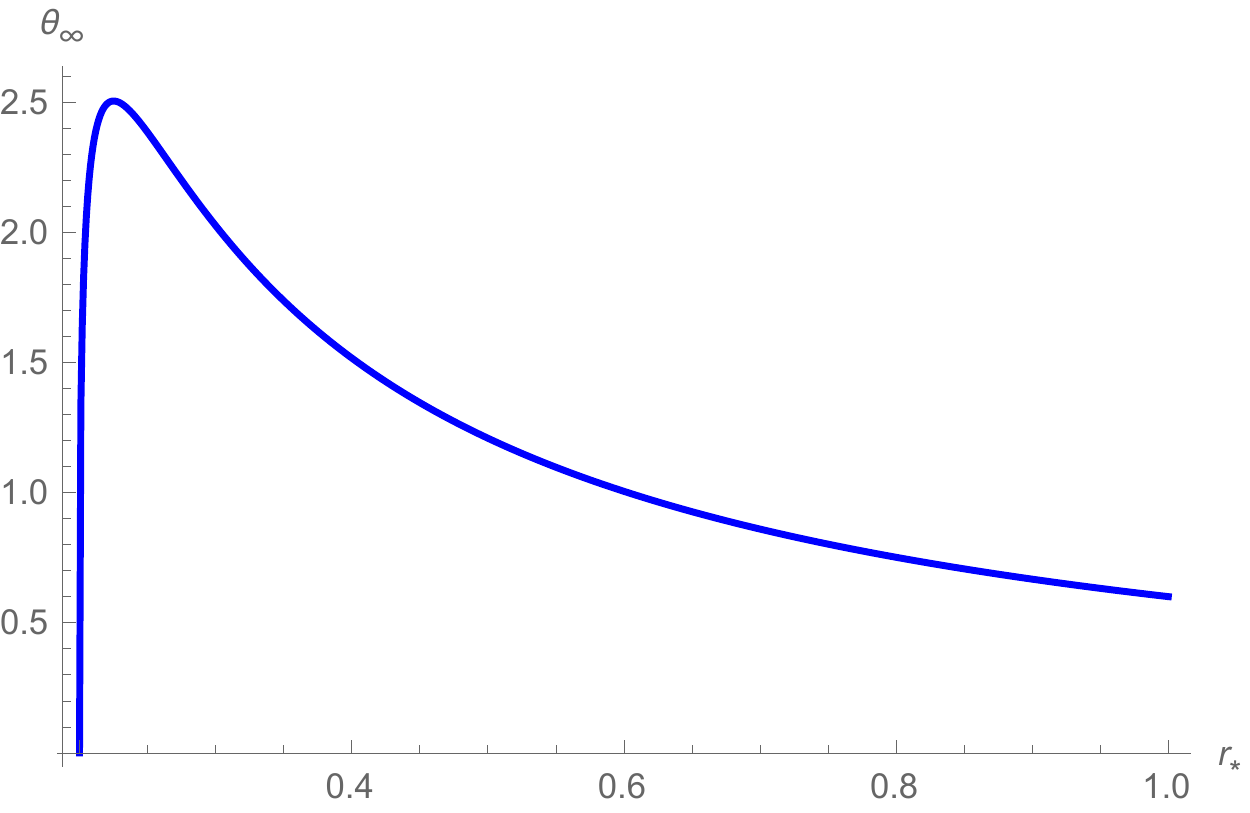}
\includegraphics[width=0.49\textwidth]{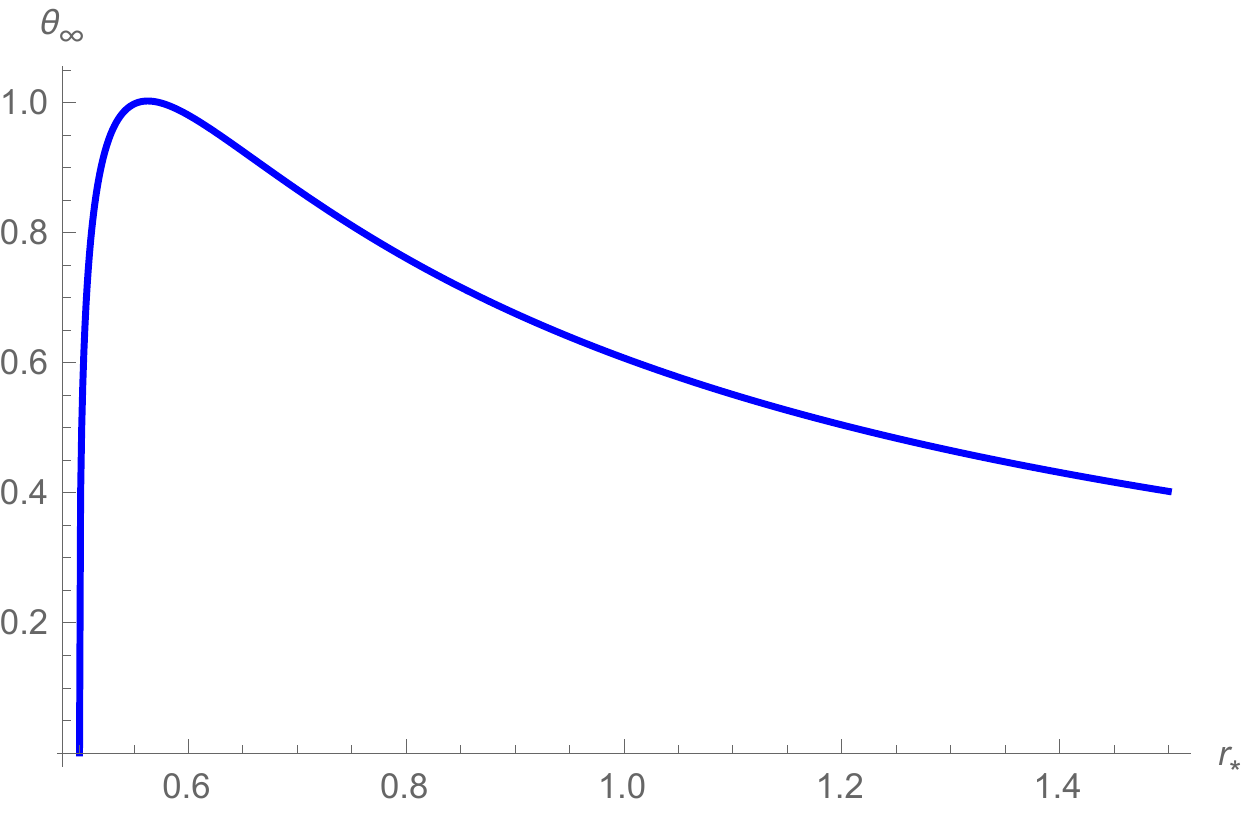}
\caption{$\theta_{\infty}(r_*)$ for Wilson loops for a black hole of radius $r_H = 0.2l_{\rm AdS}$ (left) and $r_h=0.5l_{\rm AdS}$ (right).}\label{BTZwil02graph}
\end{figure}

We first consider the smooth $\cup$-shaped solutions to the equation of motion. We can express the conserved charge in terms of the minimal/turning radius $r_*$. This allows us to find an implicit expression for $\theta_{\infty}$ in terms of $r_*$ by integrating the equation of motion:
\begin{equation}\label{rstar}
\theta_{\infty}(r_*) = \int_{r_*}^{\infty}\dd r \frac{1}{r\sqrt{f(r)}}\frac{1}{\sqrt{\frac{r^2f(r)}{r_*^2f(r_*)}-1}}.
\end{equation}  
Note that this formula only depends on the number of dimensions via $f(r)$, which is given by eqn. \eqref{eq:fSAdS}. $\theta_\infty(r_*)$ is plotted for the BTZ metric (cf. \eqref{eq:BTZmetric}) in figure \ref{BTZwil02graph}. The function is characterized by a single maximum, and decreases monotonically for large $r_*$. Near the horizon however, $\dd\theta_\infty/\dd r_*<0$, and hence by Lemma 3 (see appendix \ref{sec-lemma}) there cannot exist any local minima of the area functional in this range. The $\cup$-shaped worldsheets thus suffer a shadow that extends some finite distance from the horizon, but we postpone further discussion of shadows until after considering $\sqcup$-shaped solutions as well. 

\begin{figure}[h!]
\begin{minipage}[t]{0.55\textwidth}
\includegraphics[width=\textwidth]{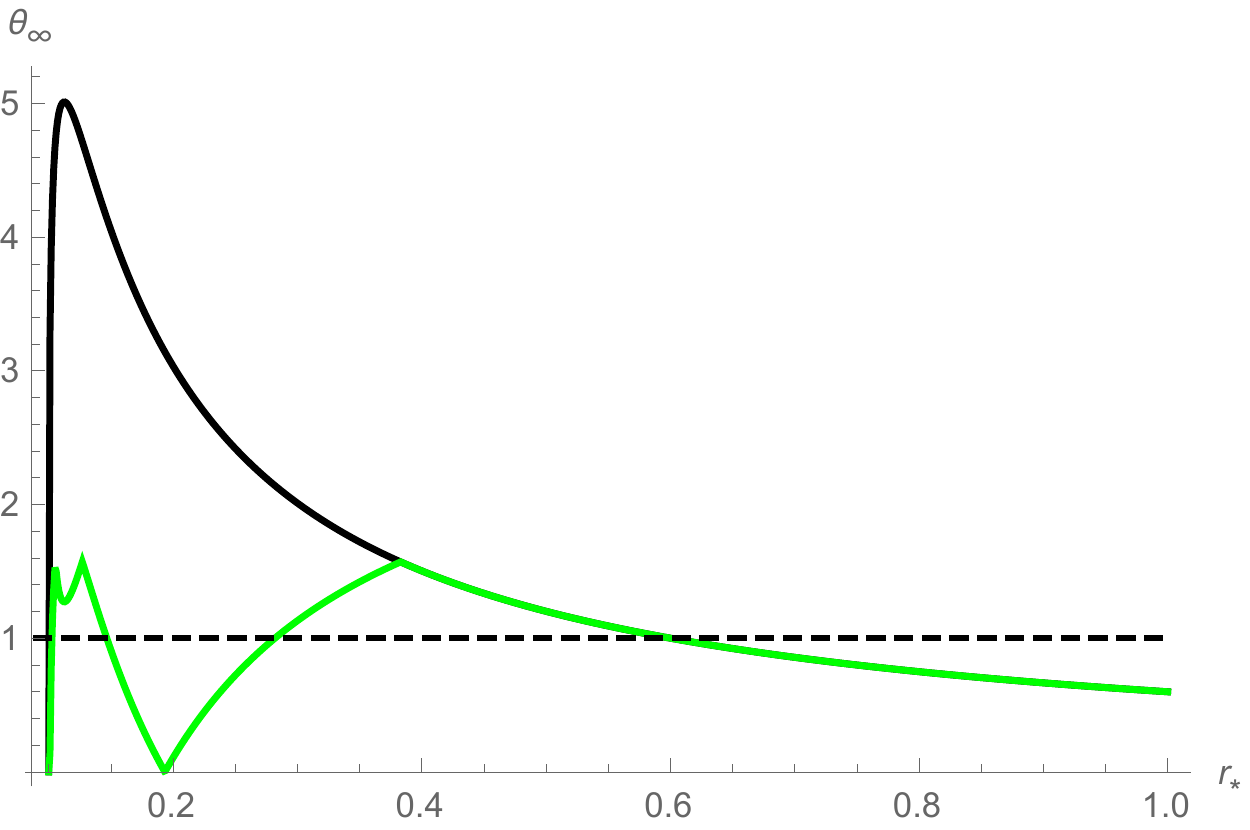}
\caption{$\theta_{\infty}(r_*)$ for a BTZ black hole with radius $r_H = 0.1l_{\rm AdS}$ (black). Solutions with $\theta_{\infty}>\pi/2$ are mapped to the range $\left[0,\pi/2\right]$ (green). The dashed line is at $\theta_{\infty}=1$; every intersection with the green line corresponds to a solution to the equation of motion for this value of $\theta_{\infty}$. These worldsheets are plotted in fig. \ref{Theta1saddles}.}\label{BTZwilanglesTHETA1}
\end{minipage}
\begin{minipage}[t]{0.05\textwidth}
 \quad
\end{minipage}
\begin{minipage}[t]{0.4\textwidth}
\includegraphics[width=\textwidth]{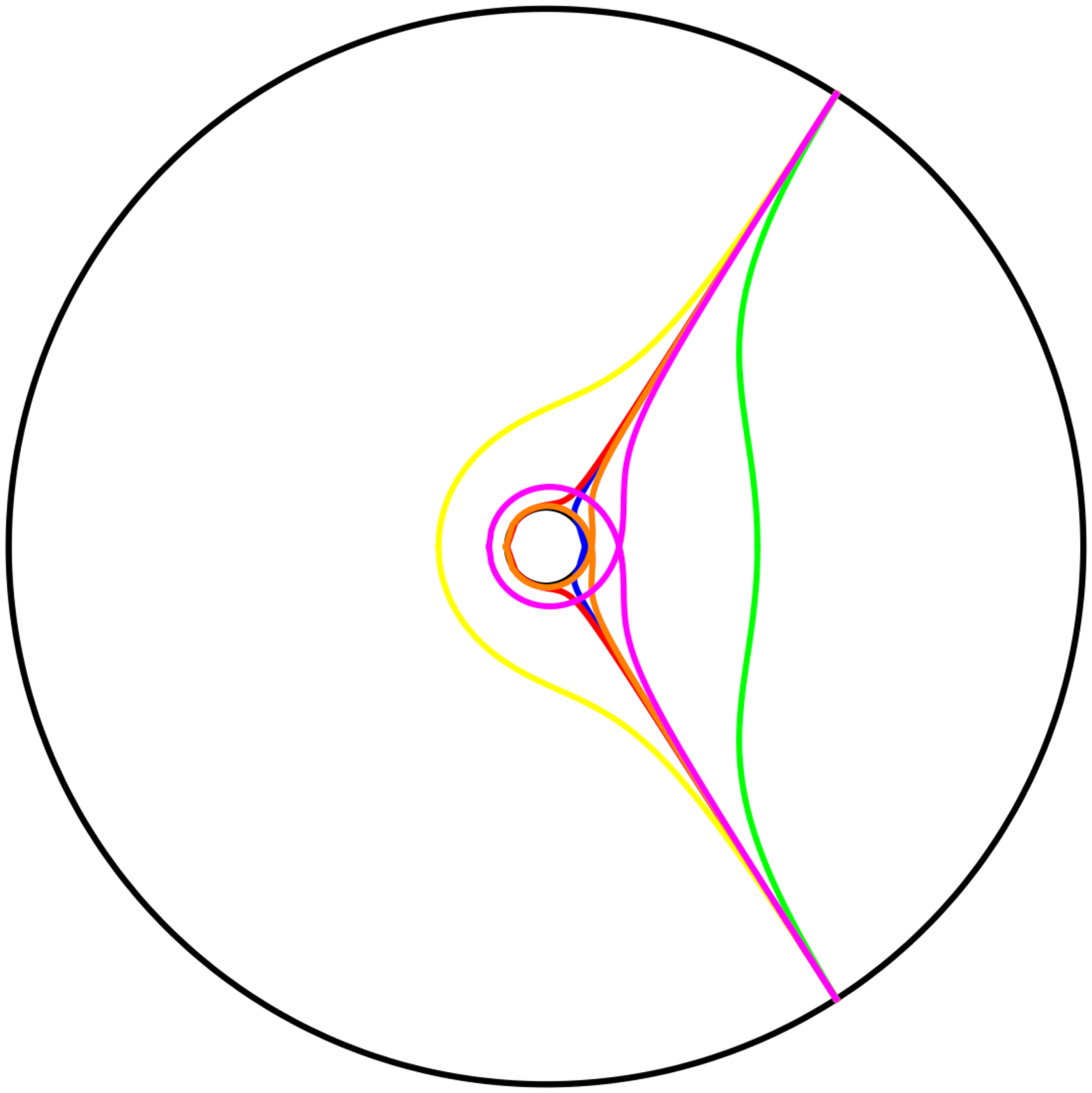}
\caption{Extrema for $\theta_{\infty} = 1$ for a BTZ black hole with horizon radius $r_H = 0.1l_{\rm AdS}$. Only one of these saddle points -- that with zero winding number (green) -- corresponds to a global minimum of the proper area of the worldsheet.}\label{Theta1saddles}
\end{minipage}
\end{figure}

As an aside, we note that for $d=2$, $\theta_\infty$ can be much larger than $\pi/2$. Using the equivalence relations $\theta_{\infty} \sim \theta_{\infty} + n\pi$ and $\theta_{\infty} \sim \pi/2-\theta_{\infty}$, we can map all values of $\theta_{\infty}>\frac{\pi}{2}$ into the range $\left[0,\pi/2\right]$; see figure \ref{BTZwilanglesTHETA1}. The solutions with $\theta_{\infty}>\pi/2$ correspond to strings that wind one or more times around the black hole (see figure \ref{Theta1saddles}). However, strings that cross themselves fail to be minimal, so we can discard these solutions in what follows.

We turn now to the $\sqcup$-shaped solutions, which consist of two radial segments connecting the boundary and the horizon at $\pm\theta_\infty$ and a segment that wraps the horizon (see figure \ref{BTZwil05}). The segment that wraps the horizon does not contribute to the area since the pullback of the metric vanishes. The radial segments have divergent area, which is associated to the unrenormalized self-energy of a quark-antiquark pair. Thus the Wilson loops associated to these $\sqcup$-shaped strings do not encode information about the bulk. Nonetheless, because these $\sqcup$-shaped solutions exist for all boundary angles, evaluation of the area functional is necessary to determine when the $\cup$-shaped solutions constitute the global minimum.

We find that $\cup$-shaped solutions have minimal area up to some critical angle $\theta_{\text{switch}}$, beyond which $\sqcup$-shaped solutions dominate. In general, this switchover will always occur for sufficiently large $\theta_{\infty}<\frac{\pi}{2}$. The only only exception is a small BTZ black hole, for which the minimal area worldsheets are $\cup$-shaped for all $\theta_{\infty}$. In appendix (\ref{wilsonhigherd}) we show that for $d>3$ one always has $\theta_{\text{switch}}<\frac{\pi}{2}$.\footnote{For $d=3$ we can also show that $\theta_{\text{switch}}<\frac{\pi}{2}$ by approximating \eqref{rstar} to find that $\theta_{\infty}(r_*)<\frac{\pi}{2}$, from which it is immediately implied.}

Denote the smallest radius to which the $\cup$-shaped worldsheets reach before the switchover by $r_{\rm s}$. Then the switchover angle $\theta_{\text{switch}}$ and associated switchover radius $r_{\rm s}$ are determined by the equality of the areas of the $\cup$-shaped and $\sqcup$-shaped solutions:
\begin{equation}
S_{\cup}(r_{\rm s}) = S_{\sqcup}~,\qquad
\theta_\infty(r_{\rm s})\equiv\theta_{\rm switch}~.
\label{eq:cupswitch}
\end{equation}
The $\cup$-shaped worldsheet corresponding to the largest possible boundary angle $\theta_{\infty}$ penetrate deepest into the bulk. The switchover angle $\theta_{\text{switch}}$ is the largest angle for which the $\cup$-shaped solutions have minimal area, so the shadow radius $r_{\text{min}}$ is determined by: 
\begin{equation}
r_{\rm min} = {\rm Max}\left[\theta_\infty^{-1}(\pi/2),~r_{\rm s}\right]~.
\end{equation}
We can solve for the value of $r_{\rm s}$ by solving the area condition \eqref{eq:cupswitch}:
\begin{equation}\label{squaretocup}
\begin{aligned}
\int_{r_{\rm s}}^{r_c}\frac{\dd r}{\sqrt{1-\frac{r_{\rm s}^2}{r^2}\frac{f(r_{\rm s})}{f(r)}}}= \int_{r_H}^{r_c}\dd r
\implies \int_{r_{\rm s}}^{\infty}\dd r\left(\frac{1}{\sqrt{1-\frac{r_{\rm s}^2}{r^2}\frac{f(r_{\rm s})}{f(r)}}}-1\right) = r_{\text{s}}-r_H~.
\end{aligned}
\end{equation}
where $r_c$ is a large radial cutoff, necessetated by the fact that both actions are linearly divergent. The dimensional dependence is encapsulated in $f(r)$. For the BTZ metric, we can solve \eqref{squaretocup} exactly by taking $r_{\rm s} = \lambda r_H$:
\begin{equation}
\lambda-1 = \lambda \int_1^{\infty}\dd x\left(\frac{1}{\sqrt{1-\frac{1}{x^2}\frac{\lambda^2-1}{x^2\lambda^2-1}}}-1\right)~,
\end{equation}
which evaluates to $\lambda \approx 1.38$. We emphasize again that the BTZ metric is exceptional in the sense that there is no switchover for small black holes $r_H\lesssim 0.26l_{\rm AdS}$. In this case the $\sqcup$-shaped worldsheets never constitute the leading saddle point of the area functional, even for $\theta_\infty>\pi/2$, and we find numerically that $r_{\rm min}\sim l_{\rm AdS}$.

For $d>2$ we cannot exactly solve \eqref{squaretocup} for the switchover radius, but we can obtain an appromation for large and small black holes. The former is especially well-motivated, since for large black holes there is a natural interpretation of the switchover as a ``confinement-deconfinement'' phase transition \cite{Ageev:2014nva,Brandhuber:1998bs,Rey:1998bq}. In this case, we have $l_{\rm AdS} \ll r_H < r_{\text{s}} \leq r$ so that $f(r) \approx r^2\left(1-r_H^{d-2}/r^{d-2}\right)$. Taking $r_{\text{s}} = \lambda_d r_H$, we have
\begin{equation}
  1-\frac{1}{\lambda_d} = \int_1^{\infty} \dd x\left(\frac{1}{\sqrt{1-\frac{1}{x^4}\frac{\lambda_d^{d-2}-1}{\lambda_d^{d-2}-\frac{1}{x^{d-2}}}}}-1\right)
\end{equation}
from which we conclude that $\lambda_d$ is an order one constant that depends on the dimension.

In the case of small black holes, we cannot exactly solve \eqref{squaretocup}, but we can solve it approximately as follows. We assume that the critical Wilson loop reaches to a location $r_s$ that is much smaller than the AdS radius, but much larger than the black hole radius,
\be
r_H \ll r_s \ll 1
\label{rsregime}
\ee
where we have again set the AdS radius to 1. This approximation will turn out to be self-consistent, and agrees with numerical results. We then approximate the integral in \eqref{squaretocup} in two different regimes: the ``near'' regime, in which $f(r) - f(r_s) \ll 1$, and the ``far'' regime, where $r_s / r \ll 1$. For small black holes, assuming that $r_s$ satisfies (\ref{rsregime}), the near and far regimes have overlapping validity.

\begin{figure}[h!]
\begin{minipage}[t]{0.475\textwidth}
\includegraphics[width=\textwidth]{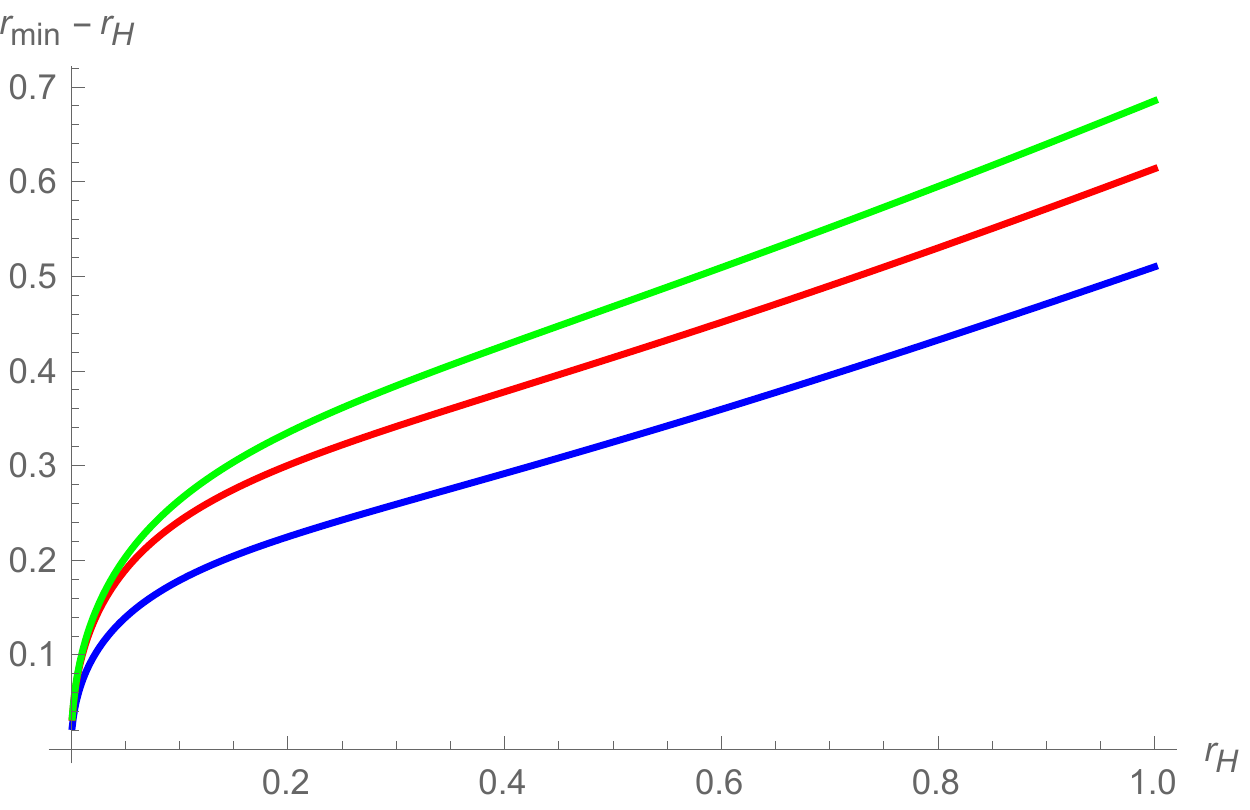}
\caption{Shadow size $r_{\text{min}}-r_H$ for rectangular Wilson loops as a function of $r_H$, for $d=3$ (blue), $d=4$ (red), and $d=5$ (green).}\label{wilsonhd2}
\end{minipage}
\begin{minipage}{0.05\textwidth}
\quad
\end{minipage}
\begin{minipage}[t]{0.475\textwidth}
\includegraphics[width=\textwidth]{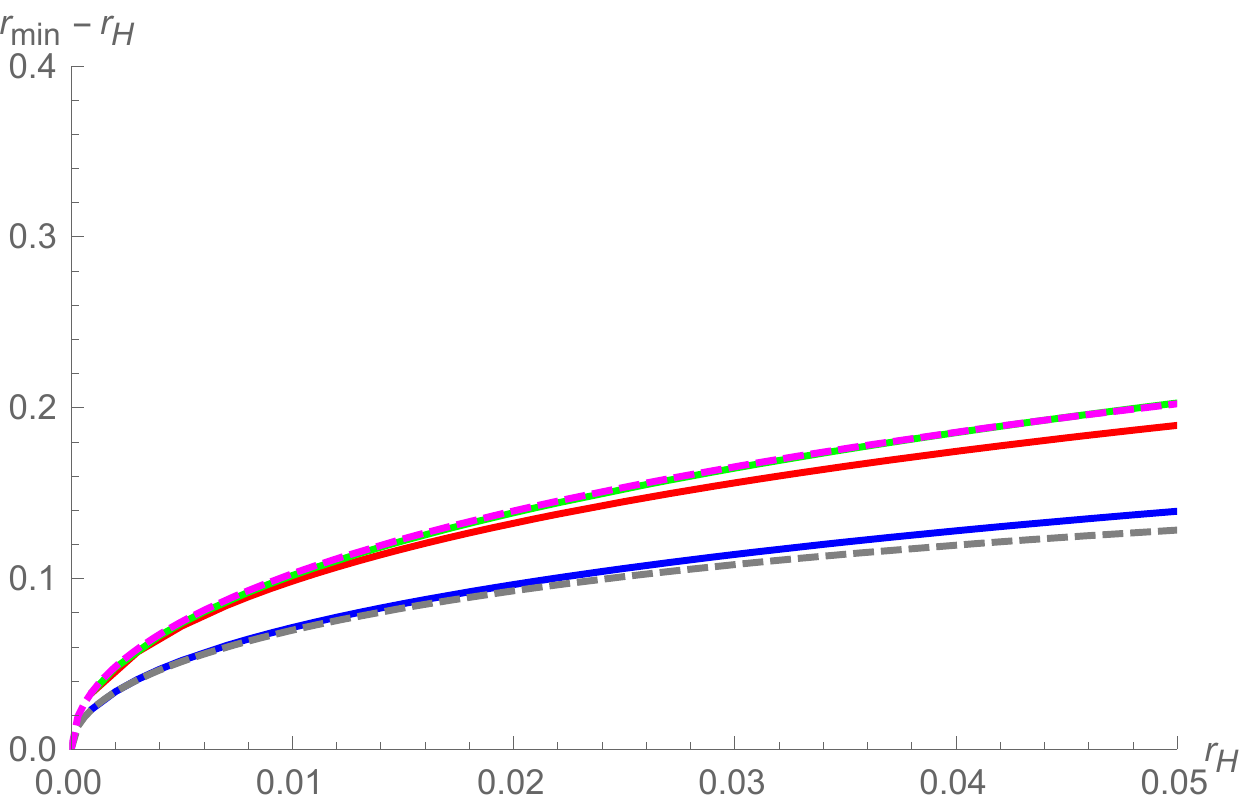}
\caption{The small black hole regime of fig. \ref{wilsonhd2}, superimposed with dotted lines corresponding to our analytical approximation: $r_{\rm min} \sim \sqrt{2r_H/\pi}$ for $d=3$ (grey) and  $r_{\rm min} \sim \sqrt{4r_H/\pi}$ for $d>3$ (magenta).}\label{wilsonhd1}
\end{minipage}
\end{figure}

We can now expand the integrand separately in the near and far regimes,
\be
\begin{aligned}
r_s - r_H &\approx \int_{r_s}^{r_0}  \dd r \left[ \left( {r \over \sqrt{r^2 - r_s^2}} - 1 \right) -  \left( { r_s^2 r \over  2 (r^2 - r_s^2)^{3/2}} {(f(r) - f(r_s) ) \over  f(r_s)} + \ldots \right)  \right] \\
&+\int_{r_0}^\infty \dd r  \left( {r_s^2 f(r_s) \over 2 r^2 f(r)} +\ldots \right)
\end{aligned}
\ee
where we have expanded to first order in the small parameter in each regime. The first term on the right side cancels the $r_s$ on the left side, leading to the equation
\be
2 r_H = \int_{r_s}^{r_0}\dd r { r_s^2 r \over (r^2 - r_s^2)^{3/2}} {f(r) - f(r_s) \over f(r_s)} - \int_{r_0}^\infty {r_s^2 f(r_s) \over r^2 f(r)}~.
\ee
In the near region we use the full formula for $f(r)$, while in the far region we use the pure AdS metric without a black hole.  The integrals can now be performed. Only the term that depends on the black hole mass carries any dimensional dependence. The intermediate scale $r_0$ cancels as it should, and we get
\be
2 r_H = {\pi \over 2} r_s^2 + c_d  r_H  \left( r_H \over r_s \right)^{d-3}
\ee
where $d$ is the spatial dimension and $c_d$ is a positive constant that depends on dimension. For $d>3$, the second term on the right hand side is parametrically smaller than the first and can be dropped; for $d=3$ it must be kept and the constant turns out to be $c_3 = 1$. The final answer is then that the critical Wilson loop reaches to a minimum radius $r_s$ that is related to the horizon radius $r_H$ by
\be
\begin{aligned}
r_s^2 &= {2 \over \pi} r_H \ \ \ \ \ \ &{\rm for} \ d &= 3 \\
r_s^2 &= {4 \over \pi} r_H &{\rm for} \ d &> 3
\end{aligned}
\ee
where $d$ is the spatial dimension. See figure \ref{wilsonhd1} for a comparison of these approximations with our numerical results in the small black hole limit; numerical results for a larger range of black hole radii are plotted in fig. \ref{wilsonhd2}.

Finally, since we are often interested in the boundary angle at which the switchover occurs, a similar computation to the above gives the simple result:
\be
\theta_{\rm switch} \approx {\pi \over 2} (1 - r_s + \ldots)
\ee
valid for all spatial dimension $d \geq 3$.

We close this section with a discussion of how good Wilson loops are, qualitatively, as bulk probes in the context of holographic shadows. We first note that locally, that is, for a given $\theta_{\infty}<\theta_{\text{switch}}$, Wilson loops probe more deeply into the bulk than the corresponding minimal surface due to the extra factor of $\sqrt{-g_{tt}}$ in the action \eqref{actionwilson}. But since the shadow radius $r_{\text{min}}$ is the infimum of the collection of $r_*(\theta_{\infty})$ from $\cup$-shaped worldsheets, we have to take into account the switchover effect in order to make the more appropriate global comparison. In appendix (\ref{wilsonhigherd}), we approximate (\ref{squaretocup}) in higher dimensions ($d\geq3$) for large and small black holes. The results are summarized in table \ref{tablewilson}. 

\begin{table}[h!]
\centering
\begin{tabular}{|c|c|c|c|}
\hline
& $d=2$ & $d=3$ & $d>3$ \\
\hline
$r_H \ll l_{\rm AdS}$ & O(1) & $\sim \sqrt{2r_H/\pi}$ & $\sim \sqrt{4r_H/\pi}$ \\ \hline
$r_H \gg l_{\rm AdS}$ & $\sim \lambda_2 r_H$, $\lambda_2 \approx 1.38$ & $\sim \lambda_3 r_H$, $\lambda_3 \approx 1.46 $ & $\sim \lambda_d r_H$, $\lambda_d \gtrsim 1.52$ \\ \hline
\end{tabular}
\caption{Leading-order approximation of the shadow size $r_{\rm min}-r_H$ for Wilson loops. The proportionality constants are determined numerically via eqn. \eqref{squaretocup}. See also plots in figures \ref{wilsonhd2} and \ref{wilsonhd1}.}\label{tablewilson}
\end{table}

\section{Causal information surfaces}\label{sub:BTZCAUSAL}
The third and final bulk probe we shall consider is the causal information surface \cite{Hubeny:2012wa}, whose associated boundary quantity is dubbed ``causal holographic information''. This differs from the previous two probes in two ways. Firstly, its boundary CFT interpretation is unclear, although suggestions have been made in \cite{Kelly:2013aja,Freivogel:2013zta}. Secondly, it does not take the general form we descirbed in sec. \ref{sec-general} as a minimal geometric object. Nevertheless, it is still natural to define $\theta_\infty(r_*)$ for this probe, and the associated shadow is still effectively due to a phase transition. Thus we can study this probe alongside those above, and later make a comparison of their respective shadows.

The formal definition of causal holographic information is as follows: given a boundary region $a$, we first find its boundary causal diamond $\Diamond_{a}$, defined as the union of the boundary future and past domains of dependence:
\begin{equation}\label{bdiamond}
\Diamond_{a} = D^+(a) \cup D^-(a)~.
\end{equation} 
The causal information surface $\Xi_{\mathcal{A}}$ is then defined as the boundary of the intersection of the bulk future and past domains of influence \cite{Hubeny:2012wa}:
\begin{equation} \label{cis}
\Xi_{a} = \partial J^+(\Diamond_{a}) \cap \partial J^-(\Diamond_{a})~.
\end{equation}

In static, spherically symmetric spacetimes, we can understand this by reversing the construction. Start from a point in the bulk at radial coordinate $r_*$, and construct the two radially outgoing light rays to the future and past. These will end on two boundary points, $p_a^\pm$. The past boundary lightcone from $p_a^+$ and the future boundary lightcone from $p_a^-$ encloses a causal diamond. The waist of diamond is exactly a boundary ball of radius $\theta_\infty$ that sits on the same timeslice as the initial bulk point. In other words,
\begin{equation}
\theta_\infty(r_*)=\int_{r_*}^{\infty}\frac{\dd r}{f(r)}~.
\label{causalradial}
\end{equation}
However, this is only true when $\theta_\infty<\pi$. When $\theta_\infty\geq\pi$, the ball covers the entire asymptotic boundary, and its domain of dependence is the entire spacetime. Therefore, there is an effective phase transition at $\theta_\infty=\pi$, and the shadow radius is given by
\begin{equation}
r_{\rm min} = \theta_\infty^{-1}(\pi)~,
\end{equation}
if this inverse exists. Otherwise there is no shadow.

In spacetimes with a horizon at $r_H$, $f(r)\rightarrow0$ linearly as $r\rightarrow r_H$, thus $\theta_\infty\rightarrow\infty$. So such sapcetimes will always suffer shadows. For example, for the BTZ geometry with $f(r)=r^2-r_H^2$, we have
\begin{equation}
\pi=\int_{r_{\text{min}}}^{\infty}\frac{\dd r}{r^2-r_H^2}
=\frac{1}{r_H}\acoth\lp\frac{r_{\rm min}}{r_H}\rp
\implies r_{\text{min}} = \frac{r_H}{\tanh(r_H\pi)}~.
\end{equation}
Note that this is precisely the first term of \eqref{eq:BTZrmin}! In light of the earlier work by Hubeny \cite{Hubeny:2012wa}, this similarity is not surprising. In the BTZ background, the causal information surface $\Xi_{\mathcal{A}}$ coincides with the extremal surface for a given boundary subregion. The only difference between their respective shadows is that the minimal area surfaces encounter a phase transition at some $\theta_\infty<\pi$ determined by the area matching condition \eqref{eq:swithcovercondition}, while this only happens for the causal information surface at $\theta_\infty=\pi$. In particular, the phase transition for minimal area surfaces with a small black hole occurs when $\theta_\infty\sim\pi/2$, which makes a significant difference from the causal information surfaces. For large black holes, the minimal surface transition occurs at $\theta_\infty\lesssim\pi$, so these two probes agree with each other in this limit.

The situation is similar in higher dimensions. For $d\geq3$ the integral in egn. \eqref{causalradial} is slightly more involved, but since we are primarily interested in knowing how close the surface gets to the black hole, a near-horizon approximation will suffice. Thus we assume $r_*-r_H\ll1$ and expand the integrand in terms of $(r-r_H)$. For large black holes ($r_H\gg1$), the near horizon contribution dominates $\theta_\infty$, so the phase transition happens when 
\begin{equation}
\pi\approx\int_{r_{\rm min}}^{r_{\rm min}+a}\frac{\dd r}{f'(r_H)(r-r_H)}=\frac{1}{f'(r_H)}\ln\lp\frac{r_{\rm min}-r_H+a}{r_{\rm min}-r_H}\rp~.\label{eq:causalLargeBH}
\end{equation}
where $a\lesssim r_H$ is some constant, and $f(r)$ is given by \eqref{eq:fSAdS}. Solving for $r_{\rm min}$, we find
\begin{align}
r_{\rm min}\approx r_H+a e^{-d\pi r_H}~.
\end{align}
Thus for large black holes, the causal information surfaces probe exponentially close to the horizon. 

For small black holes ($r_H\ll1$), the left-most side of \eqref{eq:causalLargeBH} is instead $\pi/2$. This is because far from the horizon, the empty AdS region already contributes almost $\pi/2$ to the integral in \eqref{causalradial}. The solution is then
\begin{align}
r_{\rm min}\approx r_H+a e^{-\frac{\pi (d-2)}{2 r_H}}~.
\end{align}
Thus causal surfaces also probe exponentially close to small black holes, which is dramatically better than minimal area surfaces in this limit (cf. \eqref{eq:minimalSmallApprox}).

\section{Discussion}\label{sec:discussionMain}

\subsection{Comparison of probes in AdS-Schwarzschild}\label{sec:singular}
In this section, we summarize our results by comparing the three probes -- minimal area surfaces, Wilson loops, and causal information surfaces -- for static black holes in asymptotically AdS space. 

\begin{figure}[h!]
\centering
\includegraphics[width=0.49\textwidth]{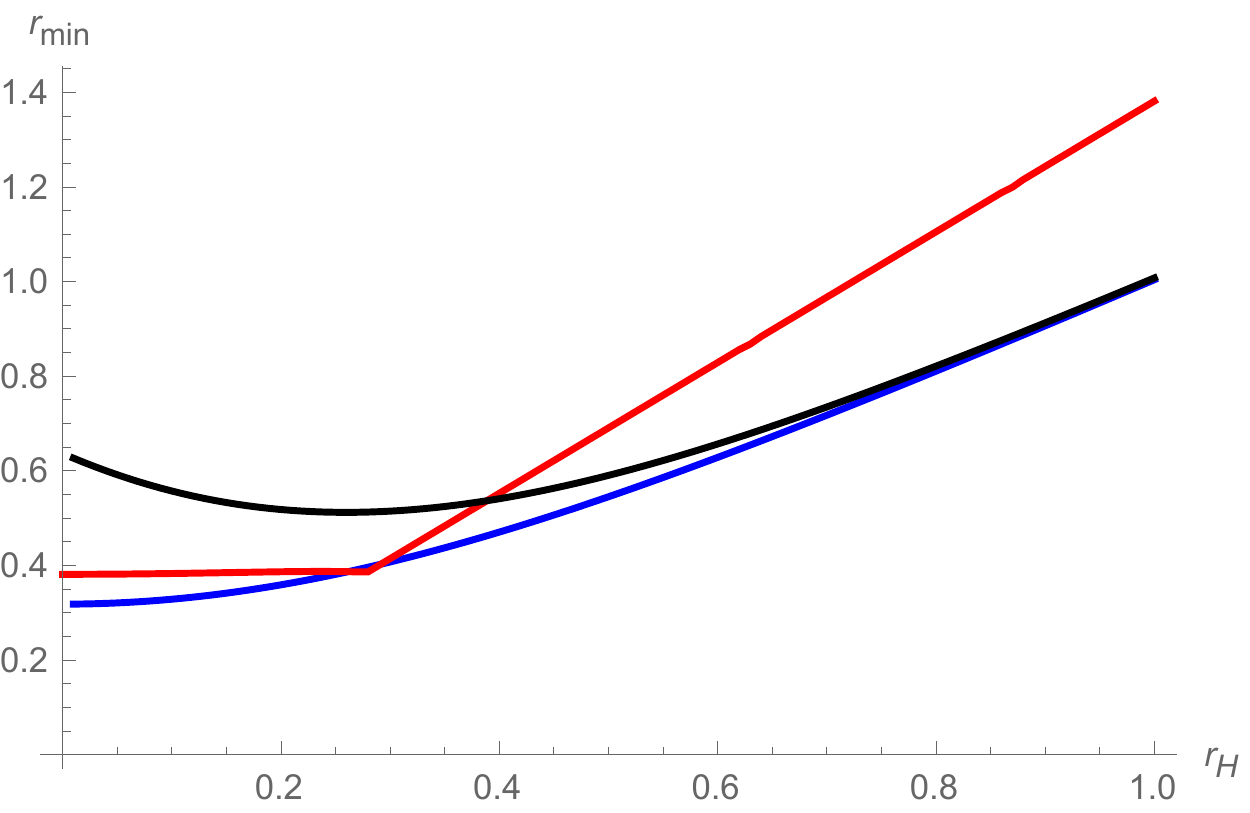}
\includegraphics[width=0.49\textwidth]{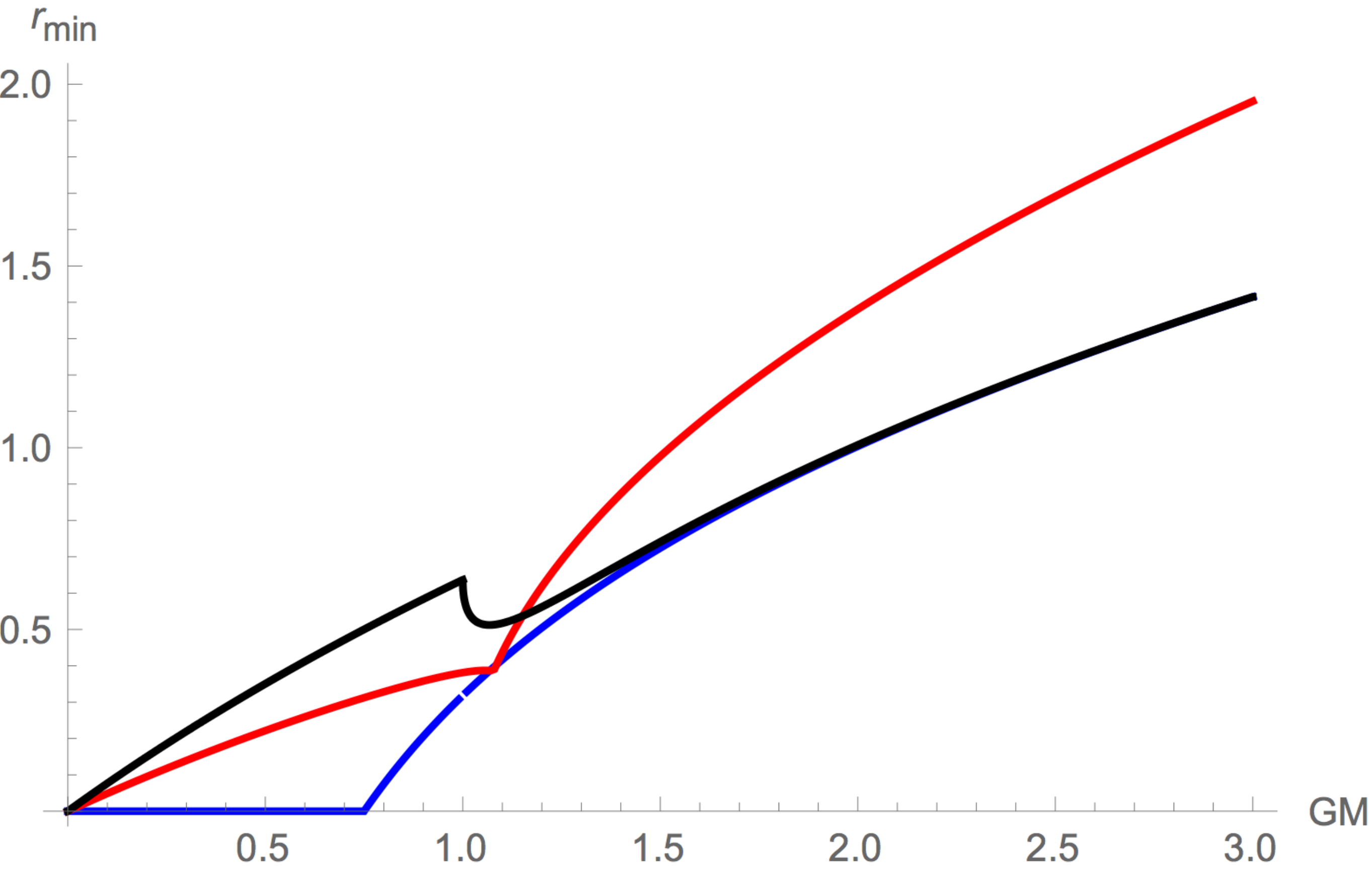}
\caption{Shadow radius $r_{\text{min}}$ as a function of the black hole radius $r_H$ (left) and mass $GM$ (right) for the different bulk probes: entanglement entropy (black), Wilson loops (red), and causal information (blue). The kink in the Wilson loops curves are due to the transition from $\cup$-shaped to $\sqcup$-shaped worldsheets. The kink in the minimal area surface curve in the right panel is exactly at the horizon $r_H=0$, at which point the phase transition angle becomes fixed at $\pi/2$.}\label{BTZcomp}
\end{figure}

For $d=2$, the calculation was sufficiently simple that we were able to obtain exact analytical results in all three cases; see figure \ref{BTZcomp} (left panel). As noted earlier, the shadow persist even when $r_H=0$ due to the mass gap in AdS$_{2+1}$. The horizon radius is related to the ADM mass by $r_H^2 = GM-1$, so a vanishing horizon does not recover empty AdS. In the right panel of figure \ref{BTZcomp}, we extend the parameter range below the mass gap to include the conical defect. Then as $GM\rightarrow0$, all shadows indeed disappear.

We can see clearly that causal information surfaces almost always leave the smallest shadow. This conclusion appears to hold in higher dimensions as well, as indicated by our numerical results and approximations for both small and large black hole. More quantitatively, both causal information and minimal area surfaces can probe exponentially close to the horizon of a large black hole, but the former can also probe exponentially close to a small black hole in $d\geq3$. This fact, and more generally the relative shadow size between probes, can be understood be recalling their respective phase transition behaviours:
\begin{description}
\item[Minimal area surfaces] encounter a phase transition for small black holes when $\theta_\infty\sim\pi/2$, so in this case are significantly worse than causal information surfaces. For large black holes, their phase transition angle approaches $\pi$, so they become comparable to causal information surfaces.
\item[Static Wilson loops] encounter a phase transition at exactly $\pi/2$ for small black holes in $d=2$, and are thus comparable to minimal surfaces in this case. For large black holes or in higher dimensions, they encounter a de-confining phase transition when $\theta_\infty<\pi/2$, and thus suffer a larger shadow. 
\item[Causal information surfaces] only encounter a phase transition when $\theta_\infty=\pi$. This enables them to probe most deeply into the bulk.
\end{description}

It is interesting to note that for a point at radius $r_{\rm min}$, it may be that a given probe can only reach it with a specific orientation, implying a restriction on the accessibility of the bulk tangent space. Empty AdS space satisfies the Strong Coverage Property that the entire tangent space of any point is covered, and indeed this property is necessary for certain reconstruction schemes \cite{BCCBH_2013,BCCB_2014}. It is thus interesting to ask how much of the tangent space once loses due to the presence of a black hole. 

\begin{figure}[h!]
\centering
\includegraphics[width=0.49\textwidth]{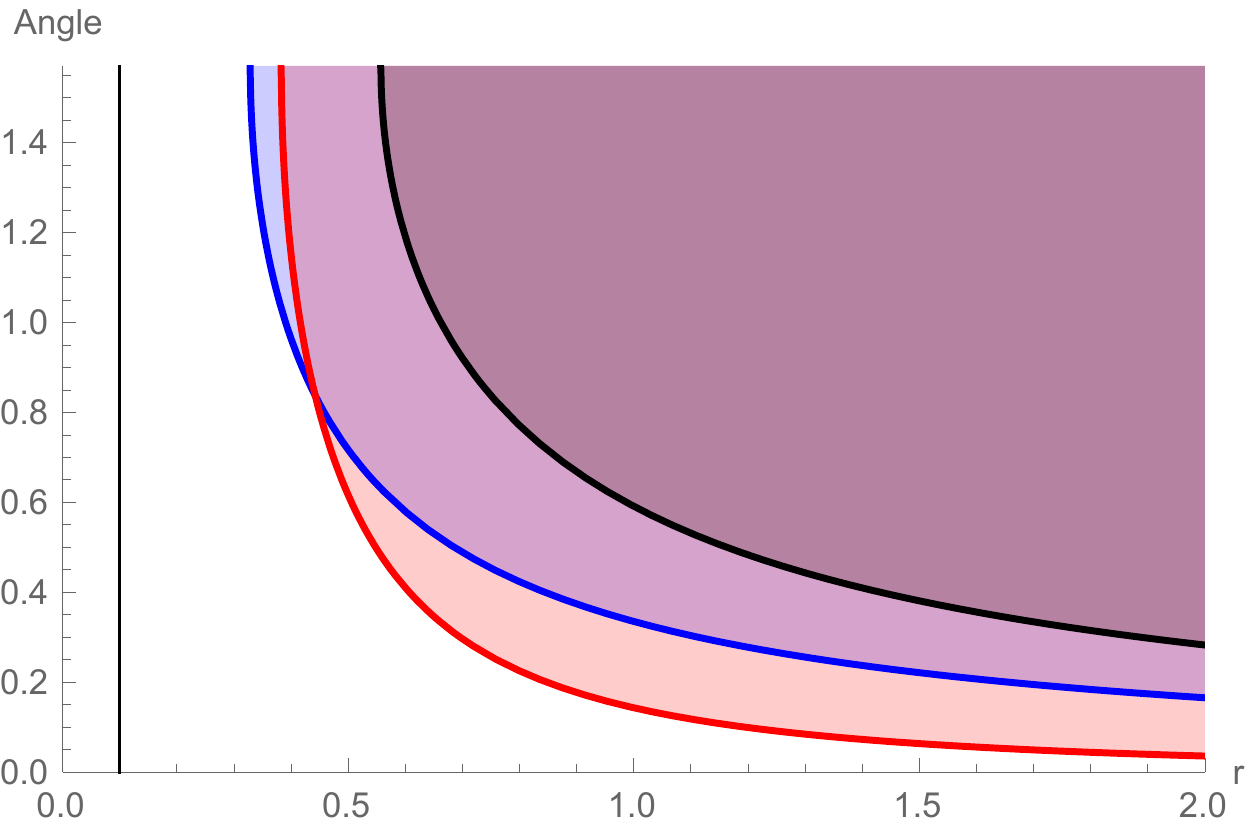}
\includegraphics[width=0.49\textwidth]{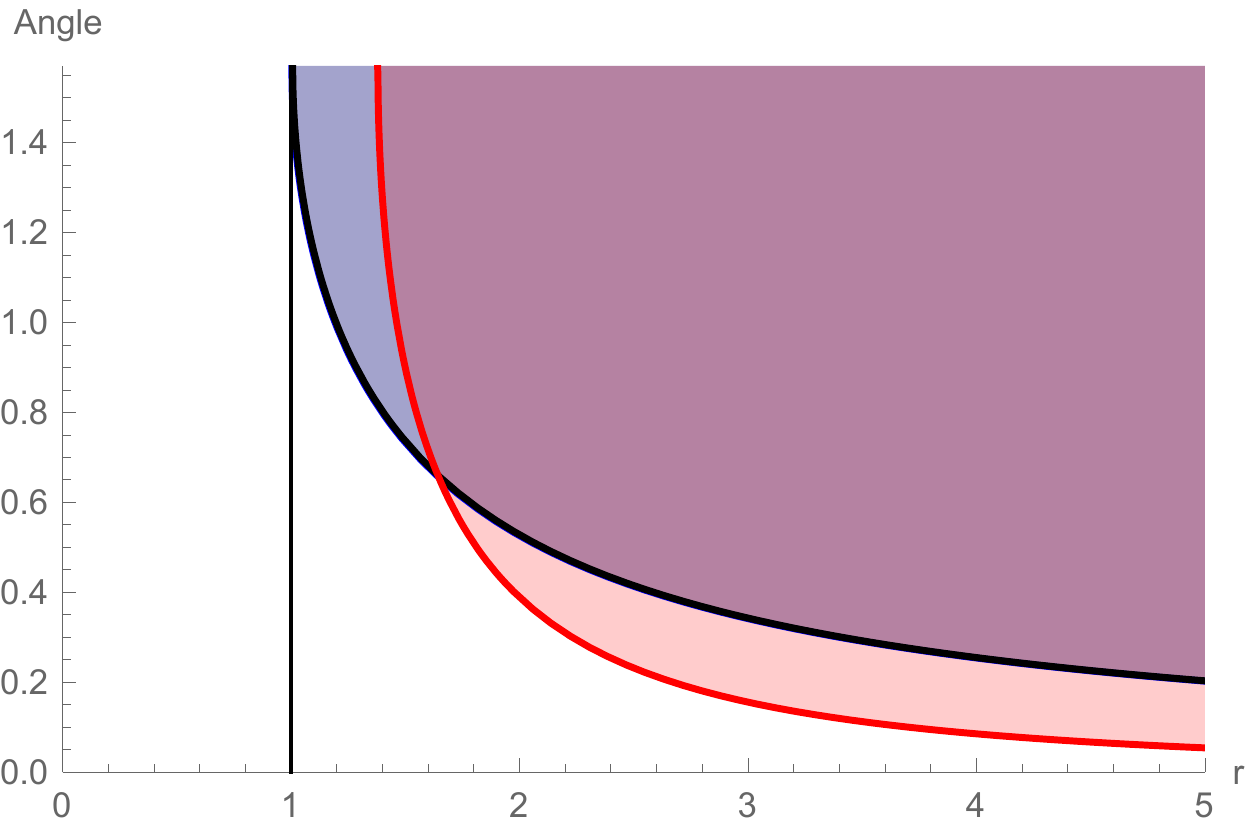}
\caption{The shaded region above each curve represents the part of the tangent space accessible by the associated bulk probe (entanglement (black), Wilson (red), and causal (blue)) as a function of the radial coordinate $r$. $\pi/2$ is purely tangential, and $0$ is purely radial. The horizon radius, $r_H = 0.1l_{\rm AdS}$ (left) and $r_H=l_{\rm AdS}$ (right), is indicated by the vertical line. Note that in the right panel, the blue and black curves almost overlap, reflecting the agreement of minimal and causal information surfaces in the large black hole limit.}\label{halfshadows1b}
\end{figure}

In the BTZ geometry, this question is easy to answer. The deepest probe in any particular family, $b(r_{\rm min})$, also passes through points with $r>r_{\rm min}$ at the steepest angle. Therefore, we need only calculate the slope of this surface to determine the coverage of the tangent space. These ``partial shadows'' are plotted in figure \ref{halfshadows1b}. Somewhat surprisingly, although Wilson loops probe less deeply in general, they exhibit the smallest partial shadows throughout most of the bulk. 

\subsection{Perspectives}\label{sec:discussion}
A holographic duality such as AdS/CFT is an intriguing notion. In principle, every property of the bulk spacetime can be reconstructed from the combination of all boundary data. In practice, one seeks simple properties of the bulk that can be associated with particular observables in some subset of the boundary. The generalized geometric probe we defined in sec. \ref{sec-general} provides a continuous, infinite family of such associations between bulk co-dimension 1 surfaces and boundary regions. Two examples among them -- the area of minimal surfaces and the action of Wilson loops -- are known to have specific boundary observables.

In empty AdS space, these geometric probes faithfully scan through the entire bulk with full coverage of the tangent space at every point. We encapsulated this complete coverage in the Strong Coverage Property, which is a requisite for some specific reconstruction programs, such as recovering Einstein's equations or constructions relying on arbitrary shapes \cite{MvR_2010,BCCBH_2013,LMvR_2013,Faulkner_etal_2013}. However, when coverage of the bulk is incomplete -- either through failure to cover the entire bulk or some portion of the tangent space -- such reconstruction proposals fail.

Black holes are known to create unreachable regions, which we generically referred to as holographic shadows. In particular, these shadows are not limited to the black hole interior, but extend well beyond the horizon. Therefore, even if one replaces the black hole by a sufficiently dense (e.g. neutron) star, such shadows will persist. Proposals to reconstruct the bulk using smearing functions in Lorentzian AdS/CFT encounter similar obstructions in the presence of trapped null geodesics \cite{LR_2013}. In general, it seems that sufficient deviations from pure AdS will pose difficulties for straightforward attempts to completely cover the bulk, even for topologically trivial spacetimes.

It is very interesting to contemplate the implications of these holographic shadows in the context of AdS/CFT. Consider a minimal surface and a bulk field operator $\phi(x)$ inside the region demarcated by the surface, that is, between the surface and the boundary. It is widely believed that this bulk operator $\phi(x)$ can be described in terms of a CFT operator $O(x)$ which has support only in the boundary region defined by the endpoints of this minimal surface. However, if the spacetime exhibits shadows, then the CFT dual of any bulk operators located within the shadow region is less clear.

One can interpret this scenario in various ways. One possibility is that the CFT degrees of freedom that correspond to bulk operators within the shadow region are completely spread out over the boundary sphere. The shadow for a particular geometry would then imply a characteristic nonlocality in the boundary field theory below some IR cutoff. An alternative is that these degrees of freedom are encoded in a quantum secret sharing scheme \cite{Cleve_Gottesman_Lo_1999, Harlow_secret}, an interpretation that follows from the switchover effect.\footnote{We thank Aron Wall for stimulating discussions on this issue.} To see this, let us assume for concreteness that the shadow is caused by the presence of a black hole. The disconnected component that wraps the black hole in principle contains the entire bulk geometry down to the horizon, and one could hope that the CFT must therefore capture all the bulk physics between this surface and the boundary (notably including the shadow). In this picture, the boundary abruptly gains access to all bulk degrees of freedom in the shadow region (the ``secret'') after the phase transition, but contains no information before the switchover. It would be very interesting to make this analogy more precise, but we leave this for future work. Finally, one could conclude that the dual CFT simply does not capture everything that happens in the bulk. This would be the most radical point of view, and also the most unsatisfactory, since it would seem to imply that holographic reconstruction techniques, at least as presently understood, will always be incomplete. 

To our knowledge, the only current proposal that may have no shadows is to use the bulk entwinement surfaces defined in \cite{BCCB_2014}. However, these are dramatically different from the above geometric probes. The boundary data required to reconstruct entwinement surfaces is highly nonlocal, and cannot be associated with a particular subregion of the boundary. Aside from special cases in which the spacetime happens to be an integer quotient of pure AdS, the precise definition of this boundary data is hard to visualize. In light of our results, it seems appropriate to ask whether such explicitly nonlocal observables are necessarily required for holographic reconstruction, or whether there exists some simple geometric probe within our generalized framework that nonetheless leaves no shadow.

The bulk surfaces within this general class are naturally associated with boundary subregions, and hence to observables that are guaranteed to satisfy strong sub-additivity. If there are indeed some probes that cast no shadows in the bulk, then we will have a transparent picture of emergent spacetime in this context. If on the other hand, one can prove that shadows are truly general features of such probes, then we have motivation to conclude that nonlocality will be an intrinsic feature of any successful holographic reconstruction scheme. 

\section*{Acknowledgements}
We thank Jan de Boer, Bartek Czech, Veronika Hubeny, and Aron Wall for stimulating discussions. This work is part of the $\Delta$-ITP consortium and supported in part by the Foundation for Fundamental Research on Matter (FOM); both are parts of the Netherlands Organization for Scientific Research (NWO) funded by the Dutch Ministry of Education, Culture, and Science (OCW). I. S. Y. is supported in part by funding from the European Research Council under the European Union's Seventh Framework Programme (FP7/2007-2013) / ERC Grant agreement no. 268088-EMERGRAV.

\begin{appendix}

\section{Proofs}

In this appendix, we present proofs of the two lemmas used in support of our coverage theorems. Note that Lemma 1 is not limited to globally regular geometries, while the form of Lemma 2 in the main text is. However, we will prove a more general version of Lemma 2 that is applicable to geometries with horizons and/or singularities. We also introduce and prove a third lemma, from which the coverage properties are independent, but which finds utility in the main text.

\label{sec-lemma}

\subsection*{Lemma 1:}
{\it For a boundary sphere $\partial a$, the bulk surface $b$ that minimizes $L$ in eqn. \eqref{eq-gp} with $\partial b=\partial a$ must be spherically symmetric.}

\subsubsection*{Proof:}
\begin{figure}[h!]
\centering
\includegraphics[width=0.49\textwidth]{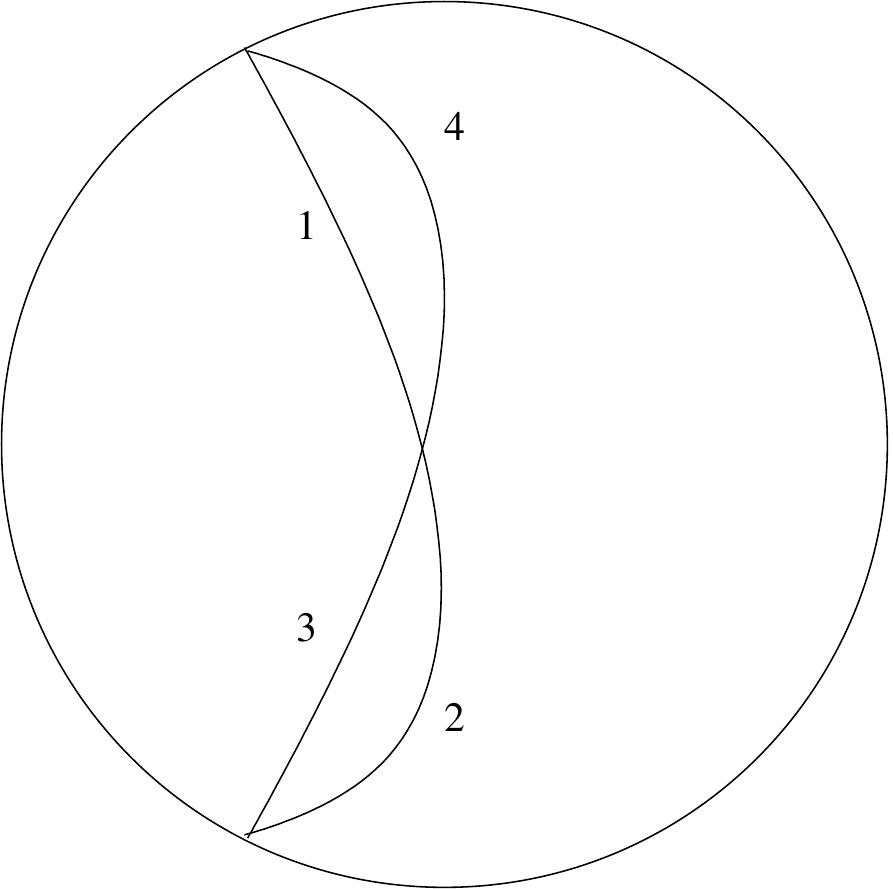}
\includegraphics[width=0.49\textwidth]{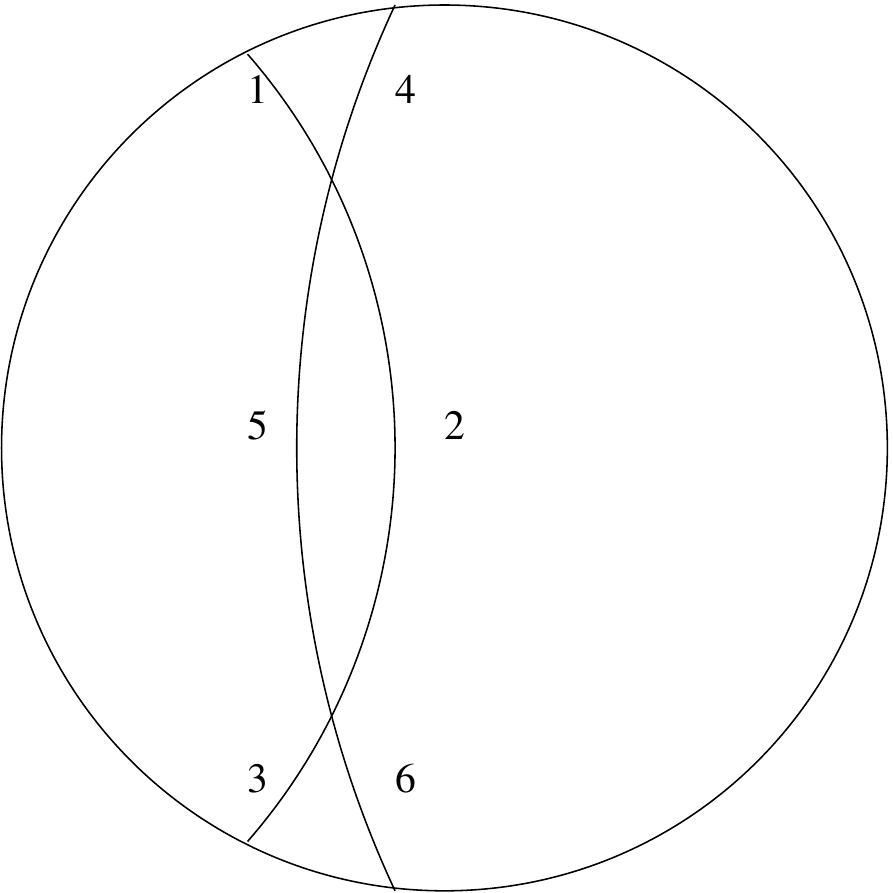}
\caption{The left panel shows two non-spherically symmetric bulk surfaces, $b=(1+2)$ and $b'=(3+4)$, ending on the same spherical boundary, $\partial b = \partial b' = \partial a$. The right panel shows two intersecting bulk surfaces, $b=(1+2+3)$ and $b'=(4+5+6)$, whose corresponding boundary anchors do not intersect. \label{fig-lemma}}
\end{figure}

If the minimal surface $b$ is not spherically symmetric, one can rotate it to get a degenerate minimum $b'$ of the same boundary region, with $\partial b = \partial b' = \partial a$. As shown in the left panel of fig. \ref{fig-lemma}, $b$ and $b'$ must intersect, but it follows from the uniqueness theorem that their normal vectors cannot agree at the intersection. Thus they must intersect with a ``kink''. We assume for simplicity that this kink separates the surfaces into two regions each, but the generalization to multiple intersections is straightforward. Let $b$ be separated into regions 1 and 2, and $b'$ into 3 and 4 as depicted in fig. \ref{fig-lemma}. By symmetry, regions 1 and 3 contribute the same amount to the geometric quantity $L$ in eqn. \eqref{eq-gp}, which we denote $L_{13}$. Similarly, we denote the contribution from regions 2 and 4 by $L_{24}$.

If $L_{24}>L_{13}$, then we could construct a new surface from regions 1 and 3 with the same boundary, thereby contradicting the assumption that both $b$ and $b'$ are minima. Similarly for $L_{13}>L_{24}$. If instead $L_{13}=L_{24}$, then both of the newly constructed surfaces have the same $L$ as $b$ and $b'$. But these new surfaces will not be smooth due to the kink at the intersection, so neither can be a local minimum of $L$. This again contradicts our assumption. QED

\subsection*{Lemma 2:} 
{\it If the boundary anchors $\partial b$ and $\partial b'$ do not cross each other, but the corresponding bulk surfaces $b$ and $b'$ do, and at least one connected region between $b$ and $b'$ does not contain a geometric obstruction, then $b$ and $b'$ cannot both be minimal surfaces.}

\subsubsection*{Proof:}
For this proof, we define a geometric obstruction as any object, defined purely by the metric, through which a bulk surface cannot be deformed without leaving a disconnected piece that wraps the obstruction; this wrapping peice should furthermore have a nonzero contribution to $L$ in \eqref{eq-gp}. (In other words, they are essentially generalizations of the black hole horizon in the case of minimal area surfaces.)

Refer to right panel of figure \ref{fig-lemma}. Let $b=(1+2+3)$, $b'=(4+5+6)$, and assume there is no geometric obstruction within the volume enclosed between 2 and 5. We denote the contribution of region 5 as $L_5$, and the contribution of region 2 as $L_2$. If $L_2>L_5$, then surface $(1+2+3)$ fails to be the minimum since surface $(1+5+3)$ has even smaller $L$. Similarly for $L_5>L_2$. If $L_2=L_5$, the uniqueness theorem again guarantees that the surface $(1+5+3)$ is not smooth, and thus we still arrive at a contradiction. Hence both $b$ and $b'$ cannot be global minima. QED 

\subsection*{\bf Lemma 3:} 
{\it If $\dd\theta_\infty/\dd r_{*}>0$, then the surface $b(r_*)$ cannot be a local minimum.} 

\subsubsection*{Proof:} By continuity, if $b(r_*)$ is a local minimum, there must be an infinitesimal $\delta r$ such that $b(r_*+\delta r)$ is also a local minimum. Since $\dd\theta_\infty/\dd r_{*}>0$, the corresponding boundary regions $a(r_*+\delta r)$ and $a(r_*)$ intersect exactly as in the right panel of fig. \ref{fig-lemma}. Applying Lemma 2 to these two surfaces then implies that they cannot both be local minima. QED

\section{Entanglement surfaces for $\dd t\neq0$}\label{nonconstt}
In this appendix, we consider an entanglement surface with spacelike separated boundary points at $(-t_{\infty},-\theta_{\infty})$ and $(t_{\infty},\theta_{\infty})$. The bulk geodesics between these endpoints are given by:
\begin{equation}\label{btztime}
\begin{aligned}
 r^2(\theta) &= r_H^2\left(\frac{\sinh^2(r_H\theta_\infty)}{\sinh^2(r_H\theta_\infty)-\sinh^2(r_Ht_\infty)}\right)\frac{\cosh^2(r_H\theta_\infty)}{\sinh^2(r_H\theta_\infty)\cosh^2(r_H\theta)-\sinh^2(r_H\theta)\cosh^2(r_H\theta_\infty)}, \\
 r^2(t) &= r_H^2\left(1+\frac{\cosh^2(r_Ht_\infty)}{\sinh^2(r_Ht_\infty)\cosh^2(r_H t)-\sinh^2(r_Ht)\cosh^2(r_Ht_\infty)} \right. \\
 &\times \left. \frac{\sinh^2(r_Ht_\infty)}{\sinh^2(r_H\theta_\infty)\cosh^2(r_H t_\infty)-\sinh^2(r_Ht_\infty)\cosh^2(r_H\theta_\infty)}\right).
\end{aligned}
\end{equation}
For a given boundary region, the minimal radius reached by this geodesic is given by:
\begin{equation}\label{result2}
 \begin{aligned}
  r_*^2 &= \frac{r_H^2\cosh^2(r_H\theta_{\infty})}{\sinh^2(r_H\theta_{\infty})-\sinh^2(r_Ht_{\infty})},
\end{aligned}
\end{equation}
which clearly shows $r_*$ is smallest for $t_{\infty}=0$.

The length of the geodesics (\ref{btztime}) is given by:
\begin{equation}\label{length2}
 l(\theta_{\infty},t_{\infty}) =  2\ln \left(\frac{2 r_c}{r_H}\right)+\ln \left(\sinh^2(r_H\theta_{\infty})-\sinh^2(r_Ht_{\infty})\right)+O\left(r_c^{-2}\right)~,
\end{equation}
where $r_c$ is a radial cut off. As in the case of the constant-time slice analysis, we may determine the switchover angle $\theta_{\text{switch}}$ by the matching condition \eqref{eq:btzswitchover}:
\begin{equation}
\theta_{\text{switch}} = \frac{\pi}{2}+\frac{1}{2r_H}\ln\left(\cosh(\pi r_H)\right)-\frac{1}{2r_H}\ln\left(\cosh(2r_H t_{\infty})\right).
\end{equation}
Thus $\theta_{\text{switch}}$ is indeed smallest for $t_{\infty}=0$.

\section{Wilson loops}\label{wilsonhigherd}

\subsection{Deconfining transition in higher dimensions}
In this section, we will prove that in AdS$_{d+1}$-Schwarzschild geometries with $d \geq 4$, $\theta_\infty(r_*)\leq\pi/2$ for all $r_*$. This implies that a de-confining phase transition completely determines the shadow size independent of the black hole radius $r_H$.

The function $\theta_\infty$ is determined by the metric function $f(r)$ in AdS-Schwarzschild \eqref{eq:fSAdS}:
\begin{equation}
\theta_{\infty}(r_*) = \int_{r_*}^{\infty}\dd r\frac{r_*}{r}\sqrt{\frac{f(r_*)}{f(r)}}\frac{1}{\sqrt{r^2f(r)-r^2_*f(r_*)}}~.
\end{equation}
In contrast, for empty AdS, we have $f_{\rm AdS} = r^2+1\equiv h(r)$, which we call $h$ to avoid confusion in what follows. By symmetry, we have
\begin{equation}
\frac{\pi}{2} = \lim_{r_*\rightarrow 0} \int_{r_*}^{\infty}\dd r\frac{r_*}{r}\sqrt{\frac{h(r_*)}{h(r)}}\frac{1}{\sqrt{r^2h(r)-r^2_*h(r_*)}}~,
\end{equation}
which is essentially the string that cuts through the middle of the space. Thus for finite values of $r_*$:
\begin{equation}
\frac{\pi}{2} \geq \int_{r_*}^{\infty}\dd r\frac{r_*}{r}\sqrt{\frac{h(r_*)}{h(r)}}\frac{1}{\sqrt{r^2h(r)-r^2_*h(r_*)}}~.
\end{equation}

To proceed, we first observe that for $r \geq r_* \geq r_H$, we have:
\begin{equation}
\frac{h(r_*)}{h(r)} \geq  \frac{f(r_*)}{f(r)}~.\label{eq:obs1}
\end{equation}
This is straightforward if we expand both $f$ and $h$ explicitly. Additionally, we shall need the fact that for $d\geq4$ and $r \geq r_*$:
\begin{equation}
 r^2f(r^2)-r_*^2f(r_*^2) \leq r^2 h(r)-r_*^2h(r_*)\label{eq:obs2}
\end{equation}
the proof of which is quite immediate:
\begin{equation*}
\begin{aligned}
r^2f(r^2)-r_*^2f(r_*^2) &= r^2 h(r)-r_*^2h(r_*) 
+(r_H^2+1)r_H^{(d-2)}\left(\frac{1}{r_*^{(d-4)}}-\frac{1}{r^{(d-4)}}\right)\\
&\leq r^2 h(r)-r_*^2h(r_*)
\end{aligned}
\end{equation*}
where we have used the fact that since $r_*\geq r_H$, the second term is negative or zero for $d\geq 4$.

Finally, \eqref{eq:obs1} and \eqref{eq:obs2} allow us to conclude:
\begin{equation}
\begin{aligned}
\theta_{\infty}(r_*) &= \int_{r_*}^{\infty}\dd r\frac{r_*}{r}\sqrt{\frac{f(r_*)}{f(r)}}\frac{1}{\sqrt{r^2f(r)-r^2_*f(r_*)}} \\
  &\leq \int_{r_*}^{\infty}\dd r\frac{r_*}{r}\sqrt{\frac{h(r_*)}{h(r)}}\frac{1}{\sqrt{r^2h(r)-r^2_*h(r_*)}} 
&\leq \frac{\pi}{2}~. 
\end{aligned}
\end{equation}

\subsection{Wilson loops in planar AdS$_5\times \rm{S}^5$}
In this appendix we investigate the possible complications for Wilson loops in the more ``realistic'' (asymptotically) AdS$_5\times \rm{S}^5$ setup. We use notation consistent with Maldacena \cite{Maldacena:1998im}.
\begin{equation}
\dd s^2 = \alpha'\left(-\frac{U^2f(U)}{R^2}\dd t^2+\frac{U^2}{R^2}\dd \vec{x}^2+\frac{R^2}{U^2f(U)}\dd U^2+\dd\Omega_5^2 \right)~,
\end{equation}
where $f(U) = 1$ for pure AdS and $f(U) = 1-U_H^4/U^4$ for the planar black hole with horizon at $U = U_0$. 

\begin{figure}[h!]
\centering
\includegraphics[width=0.55\textwidth]{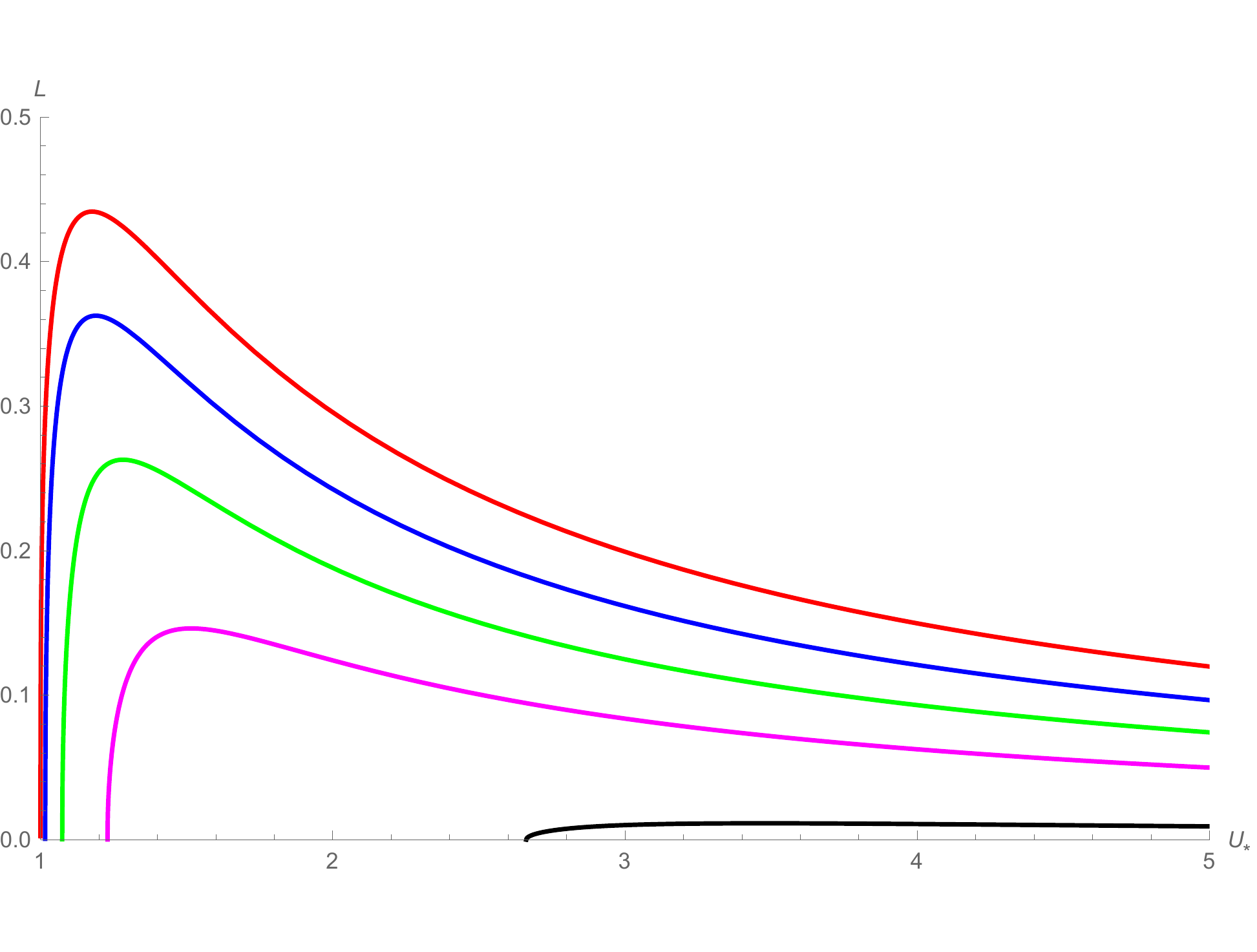}\caption{The boundary width $L$ of the Wilson loop as function of the minimal radial value $U_*$, for a black hole with $U_H = 1$ (in units of the AdS radius $R$) for different angular momenta: $l=0$ (red), $l=0.25$ (blue), $l=0.5$ (green), $l=0.75$ (magenta), and $l=0.99$ (black). }\label{wilsonS5}
\end{figure}

The worldsheet action is given by:
\begin{equation}\label{bbads5}
 S = \frac{T}{2\pi}\int \dd x \sqrt{(\partial_x U)^2+U^2f(U)(\partial_x\theta)^2+\frac{U^4}{R^4}f(U)}~.
\end{equation}
The action \eqref{bbads5} does not explicitly depend on $x$ or $\theta$. In terms of two conserved charges $l$ and $U_*$, the equations of motion are given by:
\begin{equation}
 \begin{aligned}
 (\partial_xU)^2 &= \frac{U^4}{R^4}f(U)\left(\frac{U^2(U^2f(U)-U_*^2l^2)}{U_*^4(f(U_*)-l^2)}-1\right) \\
 (\partial_x\theta)^2 &= \frac{l^2U^4}{U_*^2R^4\left(f(U_*)-l^2\right)} \\
 (\partial_U\theta)^2 &= \frac{l^2}{U_*^2f(U)}\frac{1}{\frac{U^2}{U_*^2}\left(\frac{U^2}{U_*^2}-l^2\right)-\left(1-l^2\right)},
 \end{aligned}
\end{equation}
where $l$ is the angular momentum and $U_*$ the point at which $\partial_xU=0$, related to a second conserved charge associated to the Killing vector $\partial_x$. The quantities $U_*$ and $l$ are related to the loop ``width" $L$ and the angular displacement $\Delta\theta$, respectively, by:
\begin{equation}\label{loopwidth}
 \begin{aligned}
\frac{L}{2} &= \int_{U_*}^{\infty}\dd U \frac{R^2}{U^2\sqrt{f(U)}}\frac{1}{\sqrt{\left(\frac{U^2(U^2f(U)-l^2)}{U_*^4(f(U_*)-l^2)}-1\right)}} \\
 \frac{\Delta\theta}{2} &= \frac{l}{U_*} \int_{U_*}^{\infty} \dd U \frac{1}{\sqrt{f(U)}}\frac{1}{\sqrt{\left(\frac{U^2}{U_*^2}-1\right)\left(\frac{U^2}{U_*^2}-l^2+1\right)}}~.
 \end{aligned}
\end{equation}

From \eqref{loopwidth} we see that for given $U_*$, one can decrease the corresponding boundary loop width $L$ by considering non-zero $l$ and $\Delta\theta$. Numerical evidence shows that $U_{\text{min}}$, the minimum of all $U_*$, is in fact not (significantly) smaller for non-zero ``angular momentum'' $l$ (see figure \ref{wilsonS5}). Firstly, a $\cup$-shaped solution only exists if we have $l \leq \sqrt{f(U_*)}$. Secondly, the $\sqcup$-shaped solutions constitute the dominant saddle points for sufficiently large loop width $L$.

\end{appendix}

\bibliographystyle{arXiv}
\bibliography{article}
\end{document}